\documentclass[10pt, final, journal, letterpaper, oneside, twocolumn]{IEEEtran}
\usepackage{graphicx}
\usepackage{amsmath}
\usepackage{amssymb}
\usepackage{mathrsfs}
\usepackage{stfloats}
\usepackage{dsfont}
\usepackage{setspace}
\begin{document}
\title{Error Performance and Diversity Analysis of Multi--Source Multi--Relay Wireless Networks\\ with Binary Network Coding and Cooperative MRC}
\author{Marco~Di~Renzo,~\IEEEmembership{Member,~IEEE},
        Michela~Iezzi,~\IEEEmembership{Student~Member,~IEEE}, and
        Fabio~Graziosi,~\IEEEmembership{Member,~IEEE}
\thanks{Manuscript received August 12, 2012; revised December 31, 2012. The associate editor coordinating the review of this paper and approving it for publication was W. Su.}
\thanks{M. Di Renzo is with the Laboratoire des Signaux et Syst\`emes, Unit\'e Mixte de Recherche 8506, Centre National de la Recherche Scientifique--\'Ecole
Sup\'erieure d'\'Electricit\'e--Universit\'e Paris--Sud XI, 91192 Gif--sur--Yvette Cedex, France, (e--mail: marco.direnzo@lss.supelec.fr). }
\thanks{M. Iezzi and F. Graziosi are with The University of L'Aquila, Department of Engineering, Computer Science, and Mathematics, Center of Excellence of Research DEWS, via G. Gronchi 18, 67100 L'Aquila, Italy, (e--mail:\{michela.iezzi, fabio.graziosi\}@univaq.it).}
\thanks{This paper was presented in part at the IEEE International Conference on Communications (ICC), Ottawa, Canada, June 2012.}
\thanks{This work was supported in part by the European Commission under the auspices of the FP7--PEOPLE MITN--GREENET project (grant 264759).}
\thanks{Digital Object Identifier XXX.XXX/TWC.XXX.XXX}  }
\markboth{IEEE Transactions on Wireless Communications} {M. Iezzi, M. Di Renzo, and F. Graziosi: Error Performance and Diversity Analysis of Multi--Source Multi--Relay Wireless Networks with Binary Network Coding and Cooperative MRC}
\maketitle
\begin{abstract}
In this paper, we contribute to the theoretical understanding, the design, and the performance evaluation of multi--source multi--relay network--coded cooperative diversity protocols. These protocols are useful to counteract the spectral inefficiency of repetition--based cooperation. We provide a general analytical framework for analysis and design of wireless networks using the Demodulate--and--Forward (DemF) protocol with binary Network Coding (NC) at the relays and Cooperative Maximal Ratio Combining (C--MRC) at the destination. Our system model encompasses an arbitrary number of relays which offer two cooperation levels: i) \emph{full--cooperative relays}, which postpone the transmission of their own data frames to help the transmission of the sources via DemF relaying and binary NC; and ii) \emph{partial--cooperative relays}, which exploit NC to transmit their own data frames along with the packets received from the sources. The relays can apply NC on different subsets of sources, which is shown to provide the sources with unequal diversity orders. Guidelines to choose the packets to be combined, \emph{i.e.}, the network code, to achieve the desired diversity order are given. Our study shows that partial--cooperative relays provide no contribution to the diversity order of the sources. Theoretical findings and design guidelines are validated through extensive Monte Carlo simulations.
\end{abstract}
\section{Introduction} \label{Introduction}
\PARstart{R}{elaying} and distributed cooperation have recently emerged as potential candidate technologies for many future wireless applications and standards \cite{Nosratinia}. Fundamental design objective of these systems is to maximally protect the transmission of some network nodes (usually known as sources) while minimizing: i) the extra bandwidth demanded to accomplish this protection; and ii) the resources of the network nodes (usually known as relays) willing to help the sources \cite{Krikidis}. Early transmission protocols for cooperative diversity were mainly based on the repetition coding principle with transmissions over orthogonal channels \cite{Laneman}, \cite{DiRenzoSep2009}. The main limitation of these protocols is that the diversity gain comes at the cost of low spectral efficiency. Motivated by these considerations, during the last few years many solutions have been proposed to overcome, at least in part, the throughput reduction of repetition--based orthogonal relaying protocols. Some examples are distributed space--time coded protocols \cite{LanemanDSTC}, non--orthogonal protocols \cite{Azarian}, successive relaying protocols \cite{Belfiore}, \cite{Thompson2007}, shifted successive relaying protocols \cite{Thompson2010}, two--way relay protocols \cite{ANALOG_NC}, and cognitive cooperation \cite{Sadek}. Advantages and disadvantage of these cooperative diversity protocols are discussed in \cite{Krikidis}.

More recently, a new family of cooperative diversity protocols has been introduced to overcome the throughput limitations of repetition--based protocols, while still keeping their affordable implementation complexity. They are today known as network--coded cooperative diversity protocols, as they exploit the emerging concept of Network Coding (NC) \cite{Ahlswede} for a better diversity and throughput tradeoff \cite{Zorzi}. Some examples where the achievable diversity of network--coded cooperative diversity protocols is studied are \cite{RayLiuGLOBE2010}--\cite{Rebelatto} for erasure channel models and \cite{SchoberGLOBE2010}--\cite{DiRenzo} for error channel models. These papers have all shown that NC can be especially beneficial to improve the spectral efficiency of multi--user cooperative networks, where the available relays must simultaneously serve many independent sources. For example: i) in \cite{RayLiuGLOBE2010} and \cite{RayLiu}, it is shown that NC--aided cooperative protocols are useful to overcome the accurate frequency and timing synchronization requirements of distributed space--time coded protocols; ii) in \cite{Topakkaya}, it is shown that network--coded cooperation provides a better diversity--multiplexing tradeoff than distributed space--time coded, repetition--based, and selection relaying protocols; iii) in \cite{SchoberGLOBE2010} and \cite{Nasri}, it is shown that NC can reduce the number of channel uses from $2N_S$ to $N_S+1$ in multiple--access single relay networks with $N_S$ sources, while still achieving second--order diversity for every source; and iv) in \cite{DiRenzo}, it is shown that, in a multi--source multi--relay network setup, the network code can be adequately chosen to provide each source a different diversity order, and, thus, a different robustness to multipath fading. A comprehensive state--of--the--art survey of advantages and disadvantages of these protocols is available in \cite{DiRenzo}.

Motivated by these potential advantages of network--coded cooperative diversity protocols against repetition--based cooperative diversity protocols, in the present paper we contribute to the theoretical understanding and the design of multi--source multi--relay cooperative networks with generic binary NC and non--identically distributed fading over all the wireless links. In our system model, $N_S$ sources are assumed to broadcast, in orthogonal time--slots, their data to the available $N_R$ relays and to the single destination. The relays first demodulate and then perform NC on the estimated data by applying a Demodulate--and--Forward (DemF) relaying protocol \cite{Laneman2007}, \cite{Morgado}. More specifically, NC is applied to all received packets regardless of correct or incorrect data demodulation (error channel model). Each relay can apply a different binary encoding vector to the received packets, thus being able to apply NC only on the packets received from a subset of sources. In the present paper, it is shown that this degree of freedom enables the sources to achieve unequal diversity orders, which can be useful for application to heterogenous networks \cite{Iezzi_ICC2011} and green communications \cite{RayLiu_Loc2009}. From our analytical framework, guidelines for the design of network codes to allow each source to achieve the desired distributed diversity order are derived.

Two classes of relay nodes are considered in the present paper: \emph{full--cooperative} and \emph{partial--cooperative} relays. The relays in the first class are willing to help the sources by delaying the transmission of their own data frames. These relays demodulate the packets received from the sources and then apply binary NC on a subset of received packets. The relays in the second class are willing to relaying the packets received from the sources if and only if the transmission of their data frames is not delayed. To this end, these relays demodulate the packets received from the sources and then apply binary NC on both a subset of these received packets and the first packet available in their own buffer. Thus, full--cooperative relays transmit only redundant bits, while partial--cooperative relays transmit, on the same channel use, redundant bits and new information packets. As a consequence, partial--cooperative relays entail no reduction of the network rate. The investigation of how these two classes of relays contribute to the diversity order of each source is motivated by the recent results in \cite{Fasolo} and \cite{Zorzi_Phoenix}, where partial--cooperative relays are shown to be useful to avoid dedicated resources to forward only the packets of the sources, as well as to reduce the transmission delay of their own data frames. However, in \cite{Zorzi_Phoenix} the impact of partial--cooperative relays on the achievable diversity of the sources is not explicitly investigated. Furthermore, our study is motivated by \cite{Ahlswede} and \cite{Koetter}, where, to achieve the network capacity, all network nodes are assumed to encode the data available in their buffers with the incoming data transmitted from other nodes. However, \cite{Ahlswede} and \cite{Koetter} focus their attention on the solvability of the NC problem, and, on the other hand, diversity is not investigated. Finally, we mention that the data frames available in the buffers of the partial--cooperative relays can be the data frames received, in proceeding time--slots, from other (than the $N_S$ of the cooperative network) sources. Understanding how the partial--cooperative relays contribute to the diversity order of the $N_S$ sources is important to understand whether they can help these sources without being dedicated network elements. Finally, the destination combines all the packets received from sources and relays by using a generalized Cooperative Maximal Ratio Combining (C--MRC) receiver \cite{Nasri}, \cite{Laneman2007}, which accounts for both the DemF relaying protocol and the binary encoding vectors used at the relays. For analytical tractability, it is assumed that sources and relays use binary modulation, \emph{i.e.}, Binary Phase Shift Keying (BPSK). Even though this assumption might seem restrictive, it is worth mentioning that BPSK modulation is currently used in many wireless standards, such as WiMax (Worldwide Interoperability for Microwave Access) and LTE (Long Term Evolution) for control channels. Furthermore, it is commonly used for first analytical investigations of very complex communication systems. The generalization of mathematical framework and analytical diversity assessment to non--binary modulations and non--binary encoding vectors is currently under investigation. Some comments are provided throughout the manuscript. Some preliminary Monte Carlo simulation results are available in \cite{Thang_ACT2012}. Furthermore, recent simulation results including channel coding are available in \cite{Poulliat}--\cite{Marium_CAMAD2012}. In particular, recent results in \cite{Marium_CAMAD2012} have shown that channel coding does not contribute to the diversity order of the sources if the channel is assumed to be quasi--static over all the wireless links.

The main findings of our analysis can be summarized as follows: i) the diversity order of each source is given by the separation vector \cite{Boyarinov_Marc1981} of the distributed network code, and linear block codes with Unequal Error Protection (UEP) properties \cite{VanGils} can be used as network codes even in the presence of demodulation errors at the relays; ii) only full--cooperative relays contribute to the diversity order of the sources. On the other hand, partial--cooperative relays contribute neither to the diversity order nor to the coding gain of the sources, irrespective of the binary encoding vectors; iii) properly designed (deterministic) binary network codes provide greater diversity orders than random binary network codes, which achieve only first--order diversity; iv) the C--MRC receiver provides error performance very close to the Maximum--Likelihood (ML) optimum receiver, but with reduced signal processing complexity; v) the C--MRC receiver with BPSK modulation is shown to satisfy the so--called uniform error property \cite{Raphaeli}, which greatly simplifies the analysis of the error performance and the design of binary network codes with UEP properties; and vi) compared with repetition--based cooperative relaying, network--coded cooperative diversity protocols are shown to achieve a larger range of diversity orders for the same number of total channel uses.

Finally, we emphasize that compared with state--of--the-art papers, which investigate the design and the analysis of network--coded cooperative diversity protocols under an error channel model \cite{SchoberGLOBE2010}--\cite{DiRenzo}, the novelty and contributions of the present paper are as follows: i) in \cite{SchoberGLOBE2010}--\cite{Nasri}, the analysis is restricted to a single relay node that combines the packets of all the sources. In the present paper, we consider an arbitrary number of relays, each of them using a different encoding vector to selectively applying NC on the packets received from the sources according to the desired diversity order. In particular, the system model in \cite{SchoberGLOBE2010}, \cite{Nasri} can provide only second--order diversity, while our system model guarantees a larger range of diversity orders for each source. Furthermore, two classes of relays are considered, and their role on the end--to--end diversity is discussed; and ii) in \cite{DiRenzo}, a different and sub--optimal demodulator is investigated, as well as only full--cooperative relays are considered. In the present paper, we show that the C--MRC receiver not only provides better performance than the demodulator in \cite{DiRenzo}, but it also needs lower signal processing complexity.

The remainder of the present paper is organized as follows. In Section \ref{SystemModel}, system model, transmission protocol, and C--MRC receiver are introduced. In Section \ref{ErrorPerformance}, the error performance of network--coded cooperative diversity protocols is studied, and an asymptotically--tight analytical framework is developed. In Section \ref{Diversity_NCdesign}, the diversity order is studied and guidelines for network code design are given. Furthermore, the role played by full-- and partial--cooperative relays is discussed. In Section \ref{NumericalResults}, numerical results are shown to substantiate our analytical derivation and findings. Finally, Section \ref{Conclusion} concludes this paper.
\section{System Model and Transmission Protocol} \label{SystemModel}
A multi--source multi--relay network with $N_S$ sources ($S_t$ for $t = 1,2, \ldots ,N_S$), $N_R$ relays ($R_q$ for $q = 1,2, \ldots ,N_R$), and a single destination ($D$) is considered. Transmissions of sources and relays occur in orthogonal time--slots. In time--slot $T_t$, the source $S_t$ broadcasts its data packet to $D$ and to the $N_R$ relays. This transmission (broadcast) phase lasts $N_S$ time--slots. The signals received at $R_q$ and $D$ are $y_{S_t R_q }  = \sqrt {E_{S_t} } h_{S_t R_q } x_{S_t }  + n_{S_t R_q }$ and $y_{S_t D}  = \sqrt {E_{S_t} } h_{S_t D} x_{S_t }  + n_{S_t D}$, respectively, where: i) $x_{S_t } = {1 - 2b_{S_t } }$ is the BPSK--modulated signal transmitted by $S_t$; ii) $b_{S_t }  \in \left\{ {0,1} \right\}$ is the bit emitted by $S_t$, iii) ${E_{S_t}}$ is the transmitted energy per bit of $S_t$; iv) $h_{XY}$ is the fading coefficient from node $X$ to node $Y$, which is a circular symmetric complex Gaussian Random Variable (RV) with zero mean and variance ${{\sigma _{XY}^2 } \mathord{\left/ {\vphantom {{\sigma _{XY}^2 } 2}} \right. \kern-\nulldelimiterspace} 2}$ per real dimension (Rayleigh fading). Independent but non--identically distributed (i.n.i.d.) fading is considered; and v) $n_{XY}$ is the complex Additive White Gaussian Noise (AWGN) at the input of node $Y$ and related to the transmission from node $X$. The AWGN is independent and identically distributed (i.i.d.) with zero mean and variance ${{N_0 } \mathord{\left/{\vphantom {{N_0 } 2}} \right. \kern-\nulldelimiterspace} 2}$ per real dimension.

Upon reception of $y_{S_t R_q }$ in time--slot $T_t$, the relay $R_q$ applies ML--optimum demodulation:
\setcounter{equation}{0}
\begin{equation}
\label{Eq_1} \hat b_{S_t}^{ (R_q) }  = \mathop {\arg \min }\limits_{\tilde b_{S_t }  \in \left\{ {0,1} \right\}} \left\{ {\left| {y_{S_t R_q }  - \sqrt {E_{S_t} } h_{S_t R_q } \left( {1 - 2\tilde b_{S_t } } \right)} \right|^2 } \right\}
\end{equation}
\noindent where $\hat b_{S_t}^{ (R_q) }$ denotes the estimate of $b_{S_t}$ at $R_q$.

After $N_S$ time--slots, $R_q$ takes turn transmitting, in time--slot $T_{N_S+q}$, a data packet to $D$. This transmission (relaying) phase lasts $N_R$ time--slots. Two classes of relays are considered: \emph{full--cooperative} and \emph{partial--cooperative} relays. Full--cooperative relays transmit a linear combination of the data packets received from a subset of sources and delay the transmission of their own data packets. Partial--cooperative relays transmit a linear combination of the data packets received from a subset of sources and the first available data packet in their own buffer. Thus, full--cooperative relays transmit only redundant data packets (parity bits), while partial--cooperative relays transmit redundant and new data packets. Among the $N_R$ relays, $N_R^{FC}$ and $N_R^{PC}$ act as full-- and partial--cooperative relays, respectively, with $N_R^{FC} + N_R^{PC} = N_R$. The sets of full-- and partial--cooperative relays are denoted by $\mathcal{N}_R^{FC}$ and $\mathcal{N}_R^{PC}$, respectively. It is worth noticing that in our system model the relays do not listen, in the relaying phase, to the transmission of other relays. In other words, relay--to--relay transmissions are neglected in our protocol, and, thus, in our performance and diversity analysis. This makes our communication protocol sub--optimal. In fact, the parity bits transmitted by the relays may be exploited by other relays to provide better estimates of the data transmitted from the sources. This option is, however, beyond the scope of this paper and its analysis is postponed to future research.

Let $\hat x_{R_q}^{(\rm{NC})}  = {1 - 2 \hat b_{R_q}^{(\rm{NC})} }$ be the BPSK--modulated signal transmitted from $R_q$ in time--slot $T_{N_S+q}$, and $\hat b_{R_q}^{(\rm{NC})}$ be the network--coded bit estimated at $R_q$. This latter coded bit is defined as $\hat b_{R_q}^{(\rm{NC})} = g_{S_1 R_q } \hat b_{S_1}^{ (R_q)}  \oplus g_{S_2 R_q } \hat b_{S_2}^{ (R_q)}  \oplus  \ldots  \oplus g_{S_{N_S } R_q } \hat b_{S_{N_S } }^{ (R_q)} \oplus g_{R_{q } R_q } b_{R_q}$, where: i) $\oplus$ denotes exclusive OR (XOR) operations; ii) ${\bf{g}}_{R_q }  = \left[ {g_{S_1 R_q } ,g_{S_2 R_q } , \ldots ,g_{S_{N_S } R_q } } \right]$ is the $1 \times N_S$ binary encoding vector at $R_q$ \cite{Ahlswede}, with $g_{S_t R_q } \in \left\{ {0,1} \right\}$; iii) $b_{R_q}$ is the data available in the buffer of $R_q$; and iv) $g_{R_q R_q } = 0$ if $R_q \in \mathcal{N}_R^{FC}$, while $g_{R_q R_q } = 1$ if $R_q \in \mathcal{N}_R^{PC}$. Full--cooperative relays are allowed to transmit their own data packets in the first available time--slot at the end of the cooperation (broadcast--plus--relaying) phase, which lasts $N_S + N_R$ time--slots. Thus, the signal received at $D$ is $y_{R_q D}  = \sqrt {E_{R_q} } h_{R_q D} \hat x_{R_q}^{(\rm{NC})}  + n_{R_q D}$, where ${E_{R_q}}$ is the transmitted energy per bit of $R_q$.

In our system model, Channel State Information (CSI) is available at the receiver but not at the transmitter. Thus, a uniform energy allocation scheme at the sources and at the relays is assumed. In formulas, $E_{S_t} = E_m$ for $t = 1,2, \ldots ,N_S$, $E_{R_q} = E_m$ if $R_q \in \mathcal{N}_R^{PC}$, and $E_{R_q} = E_m / 2$ if $R_q \in \mathcal{N}_R^{FC}$ for $q = 1,2, \ldots ,N_R$. This energy allocation scheme takes into account that full--cooperative relays must split their available energy to help the sources and to transmit their own data at the end of the cooperation phase.
\begin{figure*}[!t]
\setcounter{equation}{2}
\begin{equation} \footnotesize
\label{Eq_3} \begin{split}
 & \Lambda ^{\left( {{\rm{ML}}} \right)} \left( {{\bf{\tilde b}}_S ,{\bf{\tilde b}}_R } \right) = \prod\limits_{t = 1}^{N_S } {\exp \left( { - \frac{{\left| {y_{S_t D}  - \sqrt {E_{S_t } } h_{S_t D} \left( {1 - 2\tilde b_{S_t } } \right)} \right|^2 }}{{N_0 }}} \right)}  \\
   &\hspace{0.55cm} \times \prod\limits_{q = 1}^{N_R } {\left[ {P_{R_q }^{\left( {{\rm{NC}}} \right)} \exp \left( { - \frac{{\left| {y_{R_q D}  + \sqrt {E_{R_q } } h_{R_q D} \left( {1 - 2\tilde b_{R_q }^{\left( {{\rm{NC}}} \right)} } \right)} \right|^2 }}{{N_0 }}} \right) + \left( {1 - P_{R_q }^{\left( {{\rm{NC}}} \right)} } \right)\exp \left( { - \frac{{\left| {y_{R_q D}  - \sqrt {E_{R_q } } h_{R_q D} \left( {1 - 2\tilde b_{R_q }^{\left( {{\rm{NC}}} \right)} } \right)} \right|^2 }}{{N_0 }}} \right)} \right]}  \\
 \end{split}
\end{equation}
\normalsize \hrulefill \vspace*{0pt}
\end{figure*}
\begin{figure*}[!t]
\setcounter{equation}{3}
\begin{equation} \footnotesize
\label{Eq_4}
\begin{split}
 P_{R_q }^{\left( {{\rm{NC}}} \right)}  &= \Pr \left\{ {g_{S_1 R_q } \hat b_{S_1 }  \oplus g_{S_2 R_q } \hat b_{S_2 }  \oplus  \ldots  \oplus g_{S_{N_S } R_q } \hat b_{S_{N_S } }  \oplus g_{R_q R_q } b_{R_q }  \ne g_{S_1 R_q } b_{S_1 }  \oplus g_{S_2 R_q } b_{S_2 }  \oplus  \ldots  \oplus g_{S_{N_S } R_q } b_{S_{N_S } }  \oplus g_{R_q R_q } b_{R_q } } \right\} \\
 &\mathop  = \limits^{\left( a \right)} \Pr \left\{ {g_{S_1 R_q } \hat b_{S_1 }  \oplus g_{S_2 R_q } \hat b_{S_2 }  \oplus  \ldots  \oplus g_{S_{N_S } R_q } \hat b_{S_{N_S } }  \ne g_{S_1 R_q } b_{S_1 }  \oplus g_{S_2 R_q } b_{S_2 }  \oplus  \ldots  \oplus g_{S_{N_S } R_q } b_{S_{N_S } } } \right\} \\
 &\mathop  = \limits^{\left( b \right)} \sum\limits_{t = 1}^{N_S } {\left\{ {g_{S_t R_q } Q\left( {\sqrt {2\gamma _{S_t R_q } } } \right)\prod\limits_{r = t + 1}^{N_S } {\left[ {1 - 2g_{S_r R_q } Q\left( {\sqrt {2\gamma _{S_r R_q } } } \right)} \right]} } \right\}} \mathop  \approx \limits^{\left( c \right)} \sum\limits_{t = 1}^{N_S } {g_{S_t R_q } Q\left( {\sqrt {2\gamma _{S_t R_q } } } \right)}  \\
 \end{split}
\end{equation}
\normalsize \hrulefill \vspace*{0pt}
\end{figure*}
\subsection{Diversity Combining at the Destination} \label{ML_and_CMRC}
Upon reception of all signals $y_{S_t D}$ and $y_{R_q D}$ in time--slots $T_t$ and $T_{N_S+q}$, respectively, $D$ jointly estimates the data transmitted from the sources and the $N_R^{PC}$ relays. The data of the $N_R^{FC}$ relays is transmitted at the end of the cooperation phase, \emph{i.e.}, from time--slot
$T_{N_S+N_R+1}$. Since the data of the full--cooperative relays is independent of the data transmitted by the sources and by the partial--cooperative relays, it can be estimated individually, without loss of optimality, by using a demodulator similar to (\ref{Eq_1}). Demodulation of full--cooperative relays is not considered in the present paper, as it can be found in textbooks \cite{Simon}. Let ${\bf{\hat b}}_S^{\left( D \right)}$ and ${\bf{\hat b}}_R^{\left( D \right)}$ be $1 \times N_S$ and $1 \times {N}_R^{\left( {{{PC}}}\right)}$ vectors whose entries are $\hat b_{S_t }^{\left( D \right)}$ for $t = 1,2, \ldots ,N_S$ and $\hat b_{R_q }^{\left( D \right)}$ for $R_q  \in \mathcal{N}_R^{\left( {{{PC}}} \right)}$ and $q = 1,2, \ldots ,N_R$, respectively. More specifically, ${\bf{\hat b}}_S^{\left( D \right)}$ and ${\bf{\hat b}}_R^{\left( D \right)}$ denote the estimates at $D$ of the information bits, ${\bf{ b}}_S$ and ${\bf{ b}}_R$, actually transmitted from the $N_S$ sources and the $N_R^{PC}$ relays. The vectors ${\bf{ b}}_S$ and ${\bf{ b}}_R$ have similar definition as ${\bf{\hat b}}_S^{\left( D \right)}$ and ${\bf{\hat b}}_R^{\left( D \right)}$, respectively. By assuming an ML--optimum diversity combiner, they can be computed as:
\setcounter{equation}{1}
\begin{equation}
\label{Eq_2}
\left\{ {{\bf{\hat b}}_S^{\left( D \right)} ,{\bf{\hat b}}_R^{\left( D \right)} } \right\} = \mathop {{\rm{argmax}}}\limits_{\scriptstyle \tilde b_{S_t }  \in \left\{ {0,1} \right\},\;t = 1,2, \ldots ,N_S  \hfill \atop
  \scriptstyle \tilde b_{R_q }  \in \left\{ {0,1} \right\},\;R_q  \in \mathcal{N}_R^{\left( {{\rm{PC}}} \right)}  \hfill} \left\{ {\Lambda ^{\left( {{\rm{ML}}} \right)} \left( {{\bf{\tilde b}}_S ,{\bf{\tilde b}}_R } \right)} \right\}
\end{equation}
\noindent where ${\bf{\tilde b}}_S$ and ${\bf{\tilde b}}_R$ are $1 \times N_S$ and $1 \times {N}_R^{\left( {{{PC}}}\right)}$ vectors whose entries are $\tilde b_{S_t }$ and $\tilde b_{R_q }$, respectively, and $\Lambda ^{\left( {{\rm{ML}}} \right)} \left( \cdot, \cdot \right)$ is shown in (\ref{Eq_3}) at the top of this page, where $\tilde b_{R_q }^{\left( {{\rm{NC}}} \right)}  = g_{S_1 R_q } \tilde b_{S_1 }  \oplus g_{S_2 R_q } \tilde b_{S_2 }  \oplus  \ldots  \oplus g_{S_{N_S } R_q } \tilde b_{S_{N_S } }  \oplus g_{R_q R_q } \tilde b_{R_q }$, $b_{R_q }^{\left( {{\rm{NC}}} \right)}  = g_{S_1 R_q } b_{S_1 }  \oplus g_{S_2 R_q } b_{S_2 }  \oplus  \ldots  \oplus g_{S_{N_S } R_q } b_{S_{N_S } }  \oplus g_{R_q R_q } b_{R_q }$ is the network--coded bits at $R_q$ in the absence of demodulation errors, and $P_{R_q }^{\left( {{\rm{NC}}} \right)}  = \Pr \left\{ {\hat b_{R_q }^{\left( {{\rm{NC}}} \right)}  \ne b_{R_q }^{\left( {{\rm{NC}}} \right)} } \right\}$ is the probability that $R_q$ forwards a wrong network--coded bit to $D$, which is shown in (\ref{Eq_4}) at the top of this page, where: i) $\gamma _{S_t R_q }  = \left| {h_{S_t R_q } } \right|^2 \left( {{{E_{S_t } } \mathord{\left/ {\vphantom {{E_{S_t } } {N_0 }}} \right. \kern-\nulldelimiterspace} {N_0 }}} \right)$ is an exponential RV with parameter $\bar \gamma _{S_t R_q }  = \sigma _{S_t R_q }^2 \left( {{{E_{S_t } } \mathord{\left/{\vphantom {{E_{S_t } } {N_0 }}} \right. \kern-\nulldelimiterspace} {N_0 }}} \right)$ and probability density function $f_{\gamma _{S_t R_q } } \left( \xi  \right) = \left( {{1 \mathord{\left/ {\vphantom {1 {\bar \gamma _{S_t R_q } }}} \right. \kern-\nulldelimiterspace} {\bar \gamma _{S_t R_q } }}} \right)\exp \left( { - {\xi  \mathord{\left/ {\vphantom {\xi  {\bar \gamma _{S_t R_q } }}} \right. \kern-\nulldelimiterspace} {\bar \gamma _{S_t R_q } }}} \right)$; ii) $Q\left( x \right) = \left( {{1 \mathord{\left/ {\vphantom {1 {\sqrt {2\pi } }}} \right. \kern-\nulldelimiterspace} {\sqrt {2\pi } }}} \right)\int_x^{ + \infty } {\exp \left( { - {{t^2 } \mathord{\left/ {\vphantom {{t^2 } 2}} \right. \kern-\nulldelimiterspace} 2}} \right)dt}$ is the Q--function; iii) $\mathop  = \limits^{\left( a \right)}$ follows because the data of the partial--cooperative relays are not affected by demodulation errors; iv) $\mathop  = \limits^{\left( b \right)}$ follows from \cite[Proposition 1]{DiRenzo}; and v) $\mathop  \approx \limits^{\left( c \right)}$ holds for high Signal--to--Noise--Ratio (SNR), where ${\prod\nolimits_{r = t + 1}^{N_S } {\left[ {1 - 2g_{S_r R_q } Q\left( {\sqrt {2\gamma _{S_r R_q } } } \right)} \right]} } \to 1$.

Even though ML--optimum, the diversity combiner in (\ref{Eq_3}) is computationally complex. Thus, similar to \cite{Nasri} and \cite{Laneman2007}, low--complexity diversity combiners are needed. We consider a low--complexity diversity combiner that is best known as C--MRC \cite{Laneman2007}, and provide the main steps of its derivation for the specific transmission protocol under analysis. According to \cite[Section II--B]{Laneman2007}, the C--MRC can be derived by regarding $P_{R_q }^{\left( {{\rm{NC}}} \right)}$ in (\ref{Eq_4}) as the error probability of an equivalent point--to--point, rather than multipoint--to--point, channel with ${b_{R_q }^{\left( {{\rm{NC}}} \right)} }$ at its input and ${\hat b_{R_q }^{\left( {{\rm{NC}}} \right)} }$ at its output. In formulas, $P_{R_q }^{\left( {{\rm{NC}}} \right)}  \approx \sum\nolimits_{t = 1}^{N_S } {g_{S_t R_q } Q\left( {\sqrt {2\gamma _{S_t R_q } } } \right)}  \approx Q\left( {\sqrt {2\gamma _{eqR_q } } } \right)$, where ${\gamma _{eqR_q } }$ is the SNR of the point--to--point equivalent channel at $R_q$. ${\gamma _{eqR_q } }$ can be obtained as:
\setcounter{equation}{4}
\begin{equation}
\label{Eq_5}
\begin{split} & \sum\limits_{t = 1}^{N_S } {g_{S_t R_q } Q\left( {\sqrt {2\gamma _{S_t R_q } } } \right)} \\ & \hspace{2cm} \mathop  \ge \limits^{\left( a \right)} \max _{t = 1,2, \ldots ,N_S } \left\{ {Q\left( {\sqrt {2g_{S_t R_q }^{ - 1} \gamma _{S_t R_q } } } \right)} \right\} \\ & \hspace{2cm} \mathop  = \limits^{\left( b \right)} Q\left( {\sqrt {2\min _{t = 1,2, \ldots ,N_S } \left\{ {g_{S_t R_q }^{ - 1} \gamma _{S_t R_q } } \right\}} } \right) \end{split}
\end{equation}
\noindent where: i) $\mathop  \ge \limits^{\left( a \right)}$ holds because the Q--function is a non--negative function; and ii) $\mathop  = \limits^{\left( b \right)}$ holds because the Q--function is monotonically decreasing. Thus, from $P_{R_q }^{\left( {{\rm{NC}}} \right)} \approx Q\left( {\sqrt {2\gamma _{eqR_q } } } \right)$, we obtain $\gamma _{eqR_q }  = \min _{t = 1,2, \ldots ,N_S } \left\{ {g_{S_t R_q }^{ - 1} \gamma _{S_t R_q } } \right\}$, which is an exponential RV with parameter $\bar \gamma _{eqR_q }  = \left( {\sum\nolimits_{t = 1}^{N_S } {g_{S_t R_q } \bar \gamma _{S_t R_q }^{ - 1} } } \right)^{ - 1}$. We note that $g_{S_t R_q }^{ - 1} = \infty$ if $g_{S_t R_q }=0$. This implies that $\hat b_{S_t }$ is not network--coded at relay $R_q$, and, thus, the related demodulation error probability has no impact in (\ref{Eq_5}).
\begin{figure}[!t]
\centering
\includegraphics [width=\columnwidth] {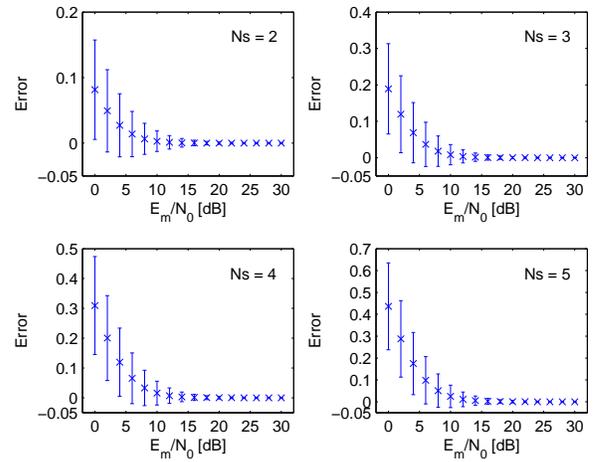}
\caption{Mean and standard deviation of the error function in (\ref{Eq_5bis}). Setup: i) $g_{S_t R_q }  = 1$ for $t=1, 2, \ldots, N_S$; and ii) channel fading is i.i.d. with $\sigma _{S_t R_q }^2  = 1$ for $t=1, 2, \ldots, N_S$.} \label{Fig0}
\end{figure}
\begin{figure*}[!t]
\setcounter{equation}{7}
\begin{equation}
\label{Eq_7}
\Lambda ^{\left( {{\rm{C - MRC}}} \right)} \left( {{\bf{\tilde b}}_S ,{\bf{\tilde b}}_R } \right) = \sum\limits_{t = 1}^{N_S } {\frac{{\left| {y_{S_t D}  - \sqrt {E_{S_t } } h_{S_t D} \left( {1 - 2\tilde b_{S_t } } \right)} \right|^2 }}{{N_0 }}}  + \sum\limits_{q = 1}^{N_R } {\lambda _{R_q } \frac{{\left| {y_{R_q D}  - \sqrt {E_{R_q } } h_{R_q D} \left( {1 - 2\tilde b_{R_q }^{\left( {{\rm{NC}}} \right)} } \right)} \right|^2 }}{{N_0 }}}
\end{equation}
\normalsize \hrulefill \vspace*{0pt}
\end{figure*}
\begin{figure*}[!t]
\setcounter{equation}{8}
\begin{equation}
\label{Eq_8}
\begin{split}
 {\rm{PEP}}\left( {\left. {{\bf{c}} \to {\bf{\tilde c}}} \right|{\bf{h}}} \right) &= \Pr \left\{ {\left. {\Lambda ^{\left( {{\rm{C - MRC}}} \right)} \left( {{\bf{\tilde b}}_S ,{\bf{\tilde b}}_R } \right) < \Lambda ^{\left( {{\rm{C - MRC}}} \right)} \left( {{\bf{b}}_S ,{\bf{b}}_R } \right)} \right|{\bf{h}}} \right\} = \Pr \left\{ {\left. {\Delta ^{\left( {{\rm{C - MRC}}} \right)} \left( {{\bf{c}},{\bf{\tilde c}}} \right) < 0} \right|{\bf{h}}} \right\} \\
  & \mathop  = \limits^{\left( a \right)} \sum\limits_{m = 0}^{2^{N_R }  - 1} {\Pr \left\{ {\left. {\Delta ^{\left( {{\rm{C - MRC}}} \right)} \left( {{\bf{c}},{\bf{\tilde c}}} \right) < 0} \right|\mathcal{E}_m ,{\bf{h}}} \right\}\Pr \left\{ {\left. {\mathcal{E}_m } \right|{\bf{h}}} \right\}} \\ & \mathop  = \limits^{\left( b \right)} \sum\limits_{m = 0}^{2^{N_R }  - 1} {\left[ {\frac{1}{{2\pi j}}\int\nolimits_{\delta  - j\infty }^{\delta  + j\infty } {\frac{{{\mathbb{E}}_{\bf{n}} \left\{ {\exp \left( { - s\Delta ^{\left( {{\rm{C - MRC}}} \right)} \left( {\left. {{\bf{c}},{\bf{\tilde c}}} \right|\mathcal{E}_m ,{\bf{h}}} \right)} \right)} \right\}}}{s}ds} } \right]\Pr \left\{ {\left. {\mathcal{E}_m } \right|{\bf{h}}} \right\}}  \\
 \end{split}
\end{equation}
\normalsize \hrulefill \vspace*{0pt}
\end{figure*}
\begin{figure*}[!t]
\setcounter{equation}{9}
\begin{equation}
\label{Eq_9}
\left\{ \begin{array}{l}
 \mathcal{E}_m^{\left( {{\rm{ok}}} \right)}  = \left\{ {\left. {\hat b_{R_q }^{\left( {{\rm{NC}}} \right)}  = b_{R_q }^{\left( {{\rm{NC}}} \right)} } \right|m = 2^0 \mu _0  + 2^1 \mu _1  +  \ldots  + 2^{N_R  - 1} \mu _{N_R  - 1} \;{\rm{and}}\;\mu _{q - 1}  = 0\;{\rm{for}}\;q = 1,2, \ldots ,N_R } \right\} \\
 \mathcal{E}_m^{\left( {{\rm{nok}}} \right)}  = \left\{ {\left. {\hat b_{R_q }^{\left( {{\rm{NC}}} \right)}  \ne b_{R_q }^{\left( {{\rm{NC}}} \right)} } \right|m = 2^0 \mu _0  + 2^1 \mu _1  +  \ldots  + 2^{N_R  - 1} \mu _{N_R  - 1} \;{\rm{and}}\;\mu _{q - 1}  = 1\;{\rm{for}}\;q = 1,2, \ldots ,N_R } \right\} \\
 \end{array} \right.
\end{equation}
\normalsize \hrulefill \vspace*{0pt}
\end{figure*}

The accuracy of the bound in (\ref{Eq_5}) is studied in Fig. \ref{Fig0} as a function of $N_S$ and SNR. In particular, Fig. \ref{Fig0} shows mean and standard deviation of the error function defined as follows:
\setcounter{equation}{5}
\begin{equation}
\label{Eq_5bis}
\begin{split} {\rm{Error}} & = \sum\limits_{t = 1}^{N_S } {g_{S_t R_q } Q\left( {\sqrt {2\gamma _{S_t R_q } } } \right)} \\ & - Q\left( {\sqrt {2\mathop {\min }\limits_{t = 1,2, \ldots ,N_S } \left\{ {g_{S_t R_q }^{ - 1} \gamma _{S_t R_q } } \right\}} } \right) \ge 0 \end{split}
\end{equation}

Figure \ref{Fig0} clearly shows that (\ref{Eq_5}) is increasing tight with the SNR, and, thus, it is useful for high--SNR analysis, \emph{e.g.}, to study the diversity order. Furthermore, we note that, as expected, the accuracy of the bound decreases with the number of network--coded sources. The error function for $N_S = 1$ is not shown as it would be equal to zero.

By using the same arguments as in \cite[Section II--C]{Laneman2007}, the C--MRC with binary NC at the relays is:
\setcounter{equation}{6}
\begin{equation}
\label{Eq_6}
\left\{ {{\bf{\hat b}}_S^{\left( D \right)} ,{\bf{\hat b}}_R^{\left( {{\rm{D}}} \right)} } \right\} = \mathop {{\rm{argmin}}}\limits_{\scriptstyle \tilde b_{S_t }  \in \left\{ {0,1} \right\},\;t = 1,2, \ldots ,N_S  \hfill \atop
  \scriptstyle \tilde b_{R_q }  \in \left\{ {0,1} \right\},\;R_q  \in \mathcal{N}_R^{\left( {{\rm{PC}}} \right)}  \hfill} \left\{ {\Lambda ^{\left( {{\rm{C - MRC}}} \right)} \left( {{\bf{\tilde b}}_S ,{\bf{\tilde b}}_R } \right)} \right\}
\end{equation}
\noindent with $\Lambda ^{\left( {{\rm{C - MRC}}} \right)} \left( \cdot, \cdot \right)$ shown in (\ref{Eq_7}) at the top of this page, where $\lambda _{R_q }  = {{\min \left\{ {\gamma _{eqR_q } ,\gamma _{R_q D} } \right\}} \mathord{\left/ {\vphantom {{\min \left\{ {\gamma _{eqR_q } ,\gamma _{R_q D} } \right\}} {\gamma _{R_q D} }}} \right. \kern-\nulldelimiterspace} {\gamma _{R_q D} }}$, and $\gamma _{R_q D}  = \left| {h_{R_q D} } \right|^2 \left( {{{E_{R_q } } \mathord{\left/ {\vphantom {{E_{R_q } } {N_0 }}} \right. \kern-\nulldelimiterspace} {N_0 }}} \right)$ is an exponential RV with parameter $\bar \gamma _{R_q D}  = \sigma _{R_q D}^2 \left( {{{E_{R_q } } \mathord{\left/ {\vphantom {{E_{R_q } } {N_0 }}} \right. \kern-\nulldelimiterspace} {N_0 }}} \right)$. In what follows, we use the notation $\tilde x_{S_t }  = 1 - 2\tilde b_{S_t }$ and $\tilde x_{R_q }  = 1 - 2\tilde b_{R_q }^{\left( {{\rm{NC}}} \right)}$.

It is worth noticing that the C--MRC in (\ref{Eq_7}) is different from and less computationally complex than the two--step demodulator in \cite[Eq. (6)]{DiRenzo}. In Section \ref{NumericalResults}, it is shown that the C--MRC in (\ref{Eq_7}) also outperforms the two--step demodulator in \cite[Eq. (6)]{DiRenzo}. Furthermore, the C--MRC in (\ref{Eq_7}) is similar to the diversity combiner in \cite[Eq. (4)]{Nasri}. The difference is that (\ref{Eq_7}) accounts for arbitrary binary encoding vectors, and, thus, each relay can apply NC on a different subset of packets received from the sources.
\begin{figure*}[!t]
\setcounter{equation}{10}
\begin{equation}
\label{Eq_10}
\Pr \left\{ {\left. {\mathcal{E}_m } \right|{\bf{h}}} \right\} = \begin{cases}
   \prod\limits_{q = 1}^{N_R } {\left( {1 - P_{R_q }^{\left( {{\rm{NC}}} \right)} } \right)}  \approx \prod\limits_{q = 1}^{N_R } {\left( {1 - \sum\limits_{t = 1}^{N_S } {g_{S_t R_q } Q\left( {\sqrt {2\gamma _{S_t R_q } } } \right)} } \right)} & \; {\rm{if}} \quad m = 0 \hfill  \\
   \prod\limits_{q = 1}^{N_R } {P_{R_q }^{\left( {{\rm{NC}}} \right)} }  \approx \prod\limits_{q = 1}^{N_R } {\left( {\sum\limits_{t = 1}^{N_S } {g_{S_t R_q } Q\left( {\sqrt {2\gamma _{S_t R_q } } } \right)} } \right)} & \; {\rm{if}}\quad m = 2^{N_R }  - 1 \hfill  \\
   \approx \prod\limits_{\scriptstyle q = 1 \atop
  \scriptstyle \mu _{q - 1}  = 0}^{N_R } {\left( {1 - \sum\limits_{t = 1}^{N_S } {g_{S_t R_q } Q\left( {\sqrt {2\gamma _{S_t R_q } } } \right)} } \right)}  \times \prod\limits_{\scriptstyle q = 1 \atop
  \scriptstyle \mu _{q - 1}  = 1}^{N_R } {\left( {\sum\limits_{t = 1}^{N_S } {g_{S_t R_q } Q\left( {\sqrt {2\gamma _{S_t R_q } } } \right)} } \right)} & \; {\rm{if}}\quad 0 < m < 2^{N_R }  - 1 \hfill  \\
\end{cases}
\end{equation}
\normalsize \hrulefill \vspace*{-5pt}
\end{figure*}
\begin{figure*}[!t]
\setcounter{equation}{11}
\begin{equation}
\label{Eq_11}
\Delta ^{\left( {{\rm{C - MRC}}} \right)} \left( {\left. {{\bf{c}},{\bf{\tilde c}}} \right|\mathcal{E}_m ,{\bf{h}}} \right) = \sum\limits_{t = 1}^{N_S } {\left( 4{\gamma _{S_t D} d_{S_t }^2  + 4\sqrt {\gamma _{S_t D} } d_{S_t } {\mathop{\rm Re}\nolimits} \left\{ {\eta _{S_t D}^* } \right\}} \right)}  + \sum\limits_{q = 1}^{N_R } {\lambda _{R_q } \left( 4{\gamma _{R_q D} d_{R_q }^{\left( m \right)}  + 4\sqrt {\gamma _{R_q D} } d_{R_q } {\mathop{\rm Re}\nolimits} \left\{ {\eta _{R_q D}^* } \right\}} \right)}
\end{equation}
\normalsize \hrulefill \vspace*{-5pt}
\end{figure*}
\begin{figure*}[!t]
\setcounter{equation}{12}
\begin{equation}
\label{Eq_12} \begin{split}
\mathcal{M}_{\Delta ^{\left( {{\rm{C - MRC}}} \right)} } \left( {\left. s \right|{\bf{c}},{\bf{\tilde c}},\mathcal{E}_m ,{\bf{h}}} \right) & = \prod\limits_{t = 1}^{N_S } {\left[ {\exp \left( { - 4\gamma _{S_t D} d_{S_t }^2 s} \right)\exp \left( {4\gamma _{S_t D} d_{S_t }^2 s^2 } \right)} \right]} \\ & \times \prod\limits_{q = 1}^{N_R } {\left[ {\exp \left( { - 4\lambda _{R_q } \gamma _{R_q D} d_{R_q }^{\left( m \right)} s} \right)\exp \left( {4\lambda _{R_q }^2 \gamma _{R_q D} d_{R_q }^2 s^2 } \right)} \right]} \end{split}
\end{equation}
\normalsize \hrulefill \vspace*{0pt}
\end{figure*}
\section{Asymptotic End--to--End Error Performance of C--MRC} \label{ErrorPerformance}
In this section, the end--to--end error performance of the C--MRC in (\ref{Eq_7}) is studied. The analytical development comprises two steps: i) first, the Average Pairwise Error Probability (APEP) is introduced and computed in Section \ref{APEP}; and ii) then, the Average Bit Error Probability (ABEP) of the data transmitted from the sources and the partial--cooperative relays is estimated in Section \ref{ABEP_Sources} and in Section \ref{ABEP_Relays}, respectively.
\subsection{Computation of the Average Pairwise Error Probability} \label{APEP}
Let ${\bf{c}} = \left[ {b_{S_1 } ,b_{S_2 } , \ldots ,b_{S_{N_S } } ,b_{R_1 }^{\left( {{\rm{NC}}} \right)} ,b_{R_2 }^{\left( {{\rm{NC}}} \right)} , \ldots ,b_{R_{N_R } }^{\left( {{\rm{NC}}} \right)} } \right]$ be the actual bits transmitted from the $N_S$ sources and the $N_R$ relays, and ${\bf{\tilde c}} = \left[ {\tilde b_{S_1 } ,\tilde b_{S_2 } , \ldots ,\tilde b_{S_{N_S } } ,\tilde b_{R_1 }^{\left( {{\rm{NC}}} \right)} ,\tilde b_{R_2 }^{\left( {{\rm{NC}}} \right)} , \ldots ,\tilde b_{R_{N_R } }^{\left( {{\rm{NC}}} \right)} } \right]$ be the hypothesis in (\ref{Eq_7}). Codewords ${\bf{c}}$ and ${\bf{\tilde c}}$ are $1 \times (N_S+N_R)$ binary vectors, and their entries are $c_p$ and $\tilde c_p$ for $p=1, 2, \ldots, N_S+N_R$. The APEP is the probability of detecting ${\bf{\tilde c}}$ in lieu of ${\bf{c}}$ under the assumption that they are the only two codewords possibly being transmitted. In formulas, ${\rm{APEP}}\left( {{\bf{c}} \to {\bf{\tilde c}}} \right) = {\mathbb{E}}_{\bf{h}} \left\{ {{\rm{PEP}}\left( {\left. {{\bf{c}} \to {\bf{\tilde c}}} \right|{\bf{h}}} \right)} \right\}$, where: i) ${\mathbb{E}}_X \left\{  \cdot  \right\}$ is the expectation operator computed over the RV $X$; ii) ${\bf{h}}$ is a short--hand that denotes all channel gains; and iii) ${{\rm{PEP}}\left( {\left. {{\bf{c}} \to {\bf{\tilde c}}} \right|{\bf{h}}} \right)}$ is the PEP conditioned upon channel fading. From (\ref{Eq_7}), ${{\rm{PEP}}\left( {\left. {{\bf{c}} \to {\bf{\tilde c}}} \right|{\bf{h}}} \right)}$ is given in (\ref{Eq_8}) at the top of this page, where: i) $\Delta ^{\left( {{\rm{C - MRC}}} \right)} \left( {{\bf{c}},{\bf{\tilde c}}} \right) = \Lambda ^{\left( {{\rm{C - MRC}}} \right)} \left( {{\bf{\tilde b}}_S ,{\bf{\tilde b}}_R } \right) - \Lambda ^{\left( {{\rm{C - MRC}}} \right)} \left( {{\bf{b}}_S ,{\bf{b}}_R } \right)$; ii) $\mathcal{E}_m$, with $m = 2^0 \mu _0  + 2^1 \mu _1  +  \ldots  + 2^{N_R  - 1} \mu _{N_R  - 1}$ and $\mu _{q-1}  \in \left\{ {0,1} \right\}$ for $q=1,2,\ldots,N_R$, denotes the joint event that the relays with index $q$ such that $\mu _{q - 1}  = 0$ transmit correct bits, \emph{i.e.}, $\hat b_{R_q }^{\left( {{\rm{NC}}} \right)}  = b_{R_q }^{\left( {{\rm{NC}}} \right)}$, and the relays with index $q$ such that $\mu _{q - 1}  = 1$ transmit wrong bits, \emph{i.e.}, $\hat b_{R_q }^{\left( {{\rm{NC}}} \right)}  \ne b_{R_q }^{\left( {{\rm{NC}}} \right)}$. In formulas, $\mathcal{E}_m  = \mathcal{E}_m^{\left( {{\rm{ok}}} \right)}  \cup \mathcal{E}_m^{\left( {{\rm{nok}}} \right)}$ shown in (\ref{Eq_9}) at the top of this page; iii) $\mathop  = \limits^{\left( a \right)}$ follows from the total probability theorem by taking into account the ${2^{N_R } }$ independent demodulation events $\mathcal{E}_m$ at the $N_R$ relays; iv) $j = \sqrt { - 1}$ is the imaginary unit; v) ${\bf{n}}$ is a short--hand to denote all AWGNs at $D$; and vi) $\mathop  = \limits^{\left( b \right)}$ follows from the inverse Laplace transform \cite[Eq. (5)]{Biglieri} with $\delta$ defined in \cite[Sec. 2]{Biglieri}.

From (\ref{Eq_4}) and the independence of the demodulation outcomes at the relays, ${\Pr \left\{ {\left. {\mathcal{E}_m } \right|{\bf{h}}} \right\}}$ in (\ref{Eq_9}) is explicitly given in (\ref{Eq_10}) shown at the top of this page. From (\ref{Eq_7}) and some algebra, $\Delta ^{\left( {{\rm{C - MRC}}} \right)} \left( {\left. {{\bf{c}},{\bf{\tilde c}}} \right|\mathcal{E}_m ,{\bf{h}}} \right)$ in (\ref{Eq_8}) can be simplified as shown in (\ref{Eq_11}) at the top of this page, where: i) ${\mathop{\rm Re}\nolimits} \left\{  \cdot  \right\}$ is the real part operator; ii) $\angle \left(  \cdot  \right)$ denotes the phase of a complex number; iii) $\left(  \cdot  \right)^*$ is the complex conjugate operator; iv) $\eta _{S_t D}^*  = {{\left( {n_{S_t D}^* \angle h_{S_t D} } \right)} \mathord{\left/ {\vphantom {{\left( {n_{S_t D}^* \angle h_{S_t D} } \right)} {\sqrt {N_0 } }}} \right. \kern-\nulldelimiterspace} {\sqrt {N_0 } }}$ and $\eta _{R_q D}^*  = {{\left( {n_{R_q D}^* \angle h_{R_q D} } \right)} \mathord{\left/ {\vphantom {{\left( {n_{R_q D}^* \angle h_{R_q D} } \right)} {\sqrt {N_0 } }}} \right. \kern-\nulldelimiterspace} {\sqrt {N_0 } }}$; v) $d_{S_t }  = \tilde b_{S_t }  - b_{S_t }  \in \left\{ { - 1,0,1} \right\}$ and $d_{R_q }  = \tilde b_{R_q }^{\left( {{\rm{NC}}} \right)}  - b_{R_q }^{\left( {{\rm{NC}}} \right)}  \in \left\{ { - 1,0,1} \right\}$; and vi) $d_{R_q }^{\left( m \right)}  = \hat b_{R_q }^{\left( {{\rm{NC}}} \right)} d_{R_q }  = \left| {\tilde b_{R_q }^{\left( {{\rm{NC}}} \right)}  - b_{R_q }^{\left( {{\rm{NC}}} \right)} } \right| \in \left\{ {0,1} \right\}$ if $\mu _{q - 1}  = 0$ and $d_{R_q }^{\left( m \right)}  = \hat b_{R_q }^{\left( {{\rm{NC}}} \right)} d_{R_q }  =  - \left| {\tilde b_{R_q }^{\left( {{\rm{NC}}} \right)}  - b_{R_q }^{\left( {{\rm{NC}}} \right)} } \right| \in \left\{ { - 1,0} \right\}$ if $\mu _{q - 1}  = 1$.

From (\ref{Eq_11}), $\mathcal{M}_{\Delta ^{\left( {{\rm{C - MRC}}} \right)} } \left( {\left. s \right|{\bf{c}},{\bf{\tilde c}},\mathcal{E}_m ,{\bf{h}}} \right) = {\mathbb{E}}_{\bf{n}} \left\{ {\exp \left( { - s\Delta ^{\left( {{\rm{C - MRC}}} \right)} \left( {\left. {{\bf{c}},{\bf{\tilde c}}} \right|\mathcal{E}_m ,{\bf{h}}} \right)} \right)} \right\}$ in (\ref{Eq_8}) reduces to (\ref{Eq_12}), shown at the top of this page, by using the identity ${\mathbb{E}}_{\bf{n}} \left\{ {\exp \left( { - K{\mathop{\rm Re}\nolimits} \left\{ {\eta _{XY}^* } \right\}} \right)} \right\} = \exp \left( {{{K^2 } \mathord{\left/ {\vphantom {{K^2 } 4}} \right. \kern-\nulldelimiterspace} 4}} \right)$ \cite[Eq. (19)]{Biglieri}.

\begin{figure*}[!t]
\setcounter{equation}{13}
\begin{equation}
\label{Eq_13}
\left\{ \begin{array}{l}
 {\rm{APEP}}\left( {{\bf{c}} \to {\bf{\tilde c}}} \right) = \sum\limits_{m = 0}^{2^{N_R }  - 1} {\left( {\frac{1}{{2\pi j}}\int\nolimits_{\delta  - j\infty }^{\delta  + j\infty } {\frac{{\mathcal{\bar M}_{\Delta ^{\left( {{\rm{C - MRC}}} \right)} } \left( {\left. s \right|{\bf{c}},{\bf{\tilde c}},\mathcal{E}_m } \right)}}{s}ds} } \right)}  \\
 \mathcal{\bar M}_{\Delta ^{\left( {{\rm{C - MRC}}} \right)} } \left( {\left. s \right|{\bf{c}},{\bf{\tilde c}},\mathcal{E}_m } \right) = {\mathbb{E}}_{\bf{h}} \left\{ {\mathcal{M}_{\Delta ^{\left( {{\rm{C - MRC}}} \right)} } \left( {\left. s \right|{\bf{c}},{\bf{\tilde c}},\mathcal{E}_m ,{\bf{h}}} \right)\Pr \left\{ {\left. {\mathcal{E}_m } \right|{\bf{h}}} \right\}} \right\} = \prod\limits_{t = 1}^{N_S } {\mathcal{F}_t \left( {\left. s \right|{\bf{c}},{\bf{\tilde c}}} \right)}  \times \mathcal{G}\left( {\left. s \right|{\bf{c}},{\bf{\tilde c}},\mathcal{E}_m } \right) \\
 \mathcal{F}_t \left( {\left. s \right|{\bf{c}},{\bf{\tilde c}}} \right) = {\mathbb{E}}_{\gamma _{S_t D} } \left\{ {\exp \left( { - 4\gamma _{S_t D} d_{S_t }^2 s} \right)\exp \left( {4\gamma _{S_t D} d_{S_t }^2 s^2 } \right)} \right\} \\
 \mathcal{G}\left( {\left. s \right|{\bf{c}},{\bf{\tilde c}},\mathcal{E}_m } \right) = {\mathbb{E}}_{\left\{ {{\pmb{\gamma }}_{SR} ,{\pmb{\gamma }}_{RD} } \right\}} \left\{ {\prod\limits_{q = 1}^{N_R } {\left[ {\exp \left( { - 4\lambda _{R_q } \gamma _{R_q D} d_{R_q }^{\left( m \right)} s} \right)\exp \left( {4\lambda _{R_q }^2 \gamma _{R_q D} d_{R_q }^2 s^2 } \right)} \right]}  \times \Pr \left\{ {\left. {\mathcal{E}_m } \right|{\pmb{\gamma }}_{SR} } \right\}} \right\} \\
 \end{array} \right.
\end{equation}
\normalsize \hrulefill \vspace*{0pt}
\end{figure*}
\begin{figure*}[!t]
\setcounter{equation}{15}
\begin{equation}
\label{Eq_15}
\mathcal{G}_q \left( {\left. s \right|{\bf{c}},{\bf{\tilde c}},\mathcal{E}_m } \right) \approx \begin{cases}
 \left[ {\left( {4\bar \gamma _{eqR_q } s} \right)^{ - 1}  + \left( {4\bar \gamma _{R_q D} s\left( {1 - s} \right)} \right)^{ - 1}  - \left( {4\bar \gamma _{eqR_q } } \right)^{ - 1} \mathcal{I}\left( {s;\mu _{q - 1}  = 0} \right)} \right]^{\left| {\tilde b_{R_q }^{\left( {{\rm{NC}}} \right)}  - b_{R_q }^{\left( {{\rm{NC}}} \right)} } \right|} & \quad {\rm{if}}\quad \mu _{q - 1}  = 0 \\
 \left( {4\bar \gamma _{eqR_q } } \right)^{ - 1} \left[ {\mathcal{I}\left( {s;\mu _{q - 1}  = 1} \right)} \right]^{\left| {\tilde b_{R_q }^{\left( {{\rm{NC}}} \right)}  - b_{R_q }^{\left( {{\rm{NC}}} \right)} } \right|} &\quad {\rm{if}}\quad \mu _{q - 1}  = 1 \\
 \end{cases}
\end{equation}
\normalsize \hrulefill \vspace*{0pt}
\end{figure*}
\begin{figure*}[!t]
\setcounter{equation}{16}
\begin{equation}
\label{Eq_16}
\mathcal{I}\left( {s;\mu _{q - 1} } \right) = \frac{4}{{\pi} }\int\nolimits_0^{{\pi  \mathord{\left/
 {\vphantom {\pi  2}} \right.
 \kern-\nulldelimiterspace} 2}} {\left( {\frac{1}{{\sin ^2 \left( \theta  \right)}} + 4\left( { - 1} \right)^{\mu _{q - 1} } s} \right)^{ - 1} d\theta } \mathop  = \limits^{\left( a \right)} \begin{cases}
   \left( {2s} \right)^{ - 1} \left[ {1 - \sqrt {\left( {1 + 4s} \right)^{ - 1} } } \right] & \; {\rm{if}}\; \mu _{q - 1}  = 0\; {\rm{and}}\; s >  - \left( {{1 \mathord{\left/
 {\vphantom {1 4}} \right.
 \kern-\nulldelimiterspace} 4}} \right) \hfill  \\
    - \left( {2s} \right)^{ - 1} \left[ {1 - \sqrt {\left( {1 - 4s} \right)^{ - 1} } } \right] & \; {\rm{if}}\; \mu _{q - 1}  = 1\; {\rm{and}}\; s < {1 \mathord{\left/
 {\vphantom {1 4}} \right.
 \kern-\nulldelimiterspace} 4} \hfill  \\
\end{cases}
\end{equation}
\normalsize \hrulefill \vspace*{0pt}
\end{figure*}
\begin{figure*}[!t]
\setcounter{equation}{17}
\begin{equation}
\label{Eq_17}
\begin{split}
 \mathcal{\bar M}_{\Delta ^{\left( {{\rm{C - MRC}}} \right)} } \left( {\left. s \right|{\bf{c}},{\bf{\tilde c}},\mathcal{E}_m } \right) & \approx \prod\limits_{t = 1}^{N_S } {\left[ {4\bar \gamma _{S_t D} s\left( {1 - s} \right)} \right]^{ - \left| {\tilde b_{S_t }  - b_{S_t } } \right|} }  \times \prod\limits_{q = 1,\;\mu _{q - 1}  = 1}^{N_R } {\left( {4\bar \gamma _{eqR_q } } \right)^{ - 1} \left[ {\mathcal{I}\left( {s;\mu _{q - 1}  = 1} \right)} \right]^{\left| {\tilde b_{R_q }^{\left( {{\rm{NC}}} \right)}  - b_{R_q }^{\left( {{\rm{NC}}} \right)} } \right|} }  \\
  &\times \prod\limits_{q = 1,\;\mu _{q - 1}  = 0}^{N_R } {\left[ {\left( {4\bar \gamma _{eqR_q } s} \right)^{ - 1}  + \left( {4\bar \gamma _{R_q D} s\left( {1 - s} \right)} \right)^{ - 1}  - \left( {4\bar \gamma _{eqR_q } } \right)^{ - 1} \mathcal{I}\left( {s;\mu _{q - 1}  = 0} \right)} \right]^{\left| {\tilde b_{R_q }^{\left( {{\rm{NC}}} \right)}  - b_{R_q }^{\left( {{\rm{NC}}} \right)} } \right|} }  \\
 \end{split}
\end{equation}
\normalsize \hrulefill \vspace*{0pt}
\end{figure*}
Finally, by inserting (\ref{Eq_12}) in (\ref{Eq_8}), the APEP can be re--written as shown in (\ref{Eq_13}) at the top of the next page, where ${\pmb{\gamma }}_{SR }$ and ${\pmb{\gamma }}_{RD }$ are short--hands to denote all channels from the $N_S$ sources to the $N_R$ relays, and from the $N_R$ relays to $D$, respectively. For high--SNR, $\mathcal{F}_t \left( \cdot, \cdot \right)$ and $\mathcal{G}\left( {\left. s \right|{\bf{c}},{\bf{\tilde c}},\mathcal{E}_m } \right) = \prod\nolimits_{q = 1}^{N_R } {\mathcal{G}_q \left( {\left. s \right|{\bf{c}},{\bf{\tilde c}},\mathcal{E}_m } \right)}$ can be computed in closed--form as shown in Appendix \ref{App_APEP}. The final result is:
\setcounter{equation}{14}
\begin{equation}
\label{Eq_14}
\mathcal{F}_t \left( {\left. s \right|{\bf{c}},{\bf{\tilde c}}} \right) \approx \left[ {4\bar \gamma _{S_t D} s\left( {1 - s} \right)} \right]^{ - \left| {\tilde b_{S_t }  - b_{S_t } } \right|}
\end{equation}
\noindent and $\mathcal{G}_q \left( {\left. \cdot \right|\cdot, \cdot, \cdot } \right)$ and $\mathcal{I}\left( {\cdot; \cdot } \right)$ are given in (\ref{Eq_15}) and (\ref{Eq_16}), respectively, at the top of the next page, and $\mathop  = \limits^{\left( a \right)}$ follows form \cite[Eq. (5A.9)]{Simon}.

From (\ref{Eq_14})--(\ref{Eq_16}), ${\mathcal{\bar M}_{\Delta ^{\left( {{\rm{C - MRC}}} \right)} } \left( {\left. \cdot \right| \cdot } \right)}$ in (\ref{Eq_13}) can be written as shown in (\ref{Eq_17}) at the top of the next page. From (\ref{Eq_14})--(\ref{Eq_16}), $\mathcal{L}_m \left( {{\bf{c}},{\bf{\tilde c}}} \right) = \left( {2\pi j} \right)^{ - 1} \int\nolimits_{\delta  - j\infty }^{\delta  + j\infty } {\mathcal{\bar M}_{\Delta ^{\left( {{\rm{C - MRC}}} \right)} } \left( {\left. s \right|{\bf{c}},{\bf{\tilde c}},\mathcal{E}_m } \right)s^{ - 1} ds}$ in (\ref{Eq_13}) can be computed by using either the residues theorem via the computation of the positive poles of ${\mathcal{\bar M}_{\Delta ^{\left( {{\rm{C - MRC}}} \right)} } \left( {\left. \cdot \right| \cdot } \right)}$ \cite[Eq. (6)]{Biglieri} or the Gauss--Chebyshev quadrature rule with $\delta$ being chosen equal to one--half of the smallest real part of the non--negative poles of ${\mathcal{\bar M}_{\Delta ^{\left( {{\rm{C - MRC}}} \right)} } \left( {\left. \cdot \right| \cdot } \right)}$ \cite[Sec. 9B.2]{Simon}, \cite[Eq. (10)]{Biglieri}. The residues theorem is to be preferred for simple network topologies (with one or two relays) and specific network codes that result in expressions of ${\mathcal{\bar M}_{\Delta ^{\left( {{\rm{C - MRC}}} \right)} } \left( {\left. \cdot \right| \cdot } \right)}$ with simple poles. In this case, the integral expression of $\mathcal{I}\left( {\cdot;\cdot } \right)$ in (\ref{Eq_16}) is more convenient to be used compared to its closed--form expression. The reason is that closed--form expressions of $\mathcal{L}_m \left( \cdot,\cdot\right)$ can be obtained by first applying the residues theorem and then computing the integral in (\ref{Eq_16}). An example is given in Appendix \ref{App_APEP_SingleRelay}, where it is shown that our framework coincides with \cite[Eq. (32)]{Nasri} for single--relay network--coded cooperative networks with encoding vector ${\bf{g}}_{R_1 }  = \left[ {1,1, \ldots ,1} \right]$. Another example is available in \cite[Table I]{Iezzi_ICC2012} for one--source two--relay networks. On the other hand, the Gauss--Chebyshev quadrature rule is an efficient single--integral numerical method that can be used for generic network topologies and binary encoding vectors. Its accuracy and convergence speed depend on $\delta$, which, from \cite[Sec. 9B.2]{Simon}, \cite[Eq. (10)]{Biglieri} and by direct inspection of (\ref{Eq_17}), can be chosen as follows:
\setcounter{equation}{18}
\begin{equation}
\label{Eq_18}
\delta  = \begin{cases}
   {1 \mathord{\left/
 {\vphantom {1 2}} \right.
 \kern-\nulldelimiterspace} 2} & \; {\rm{if}}\quad m = 0 \hfill  \\
   {1 \mathord{\left/
 {\vphantom {1 2}} \right.
 \kern-\nulldelimiterspace} 2} & \; {\rm{if}}\quad m \ne 0\quad {\rm{and}}\; \sum\limits_{q = 1 \atop \mu _{q - 1}  = 1}^{N_R } {\left| {\tilde b_{R_q }^{\left( {{\rm{NC}}} \right)}  - b_{R_q }^{\left( {{\rm{NC}}} \right)} } \right|}  = 0 \hfill  \\
   {1 \mathord{\left/
 {\vphantom {1 8}} \right.
 \kern-\nulldelimiterspace} 8} & \; {\rm{if}}\quad m \ne 0\quad {\rm{and}}\; \sum\limits_{q = 1 \atop \mu _{q - 1}  = 1}^{N_R } {\left| {\tilde b_{R_q }^{\left( {{\rm{NC}}} \right)}  - b_{R_q }^{\left( {{\rm{NC}}} \right)} } \right|}  \ne 0 \hfill  \\
\end{cases}
\end{equation}
\subsection{Computation of the Average Bit Error Probability of the Sources} \label{ABEP_Sources}
From (\ref{Eq_13}) and (\ref{Eq_17}), the ABEP of the sources can be computed by applying the Union--Bound method \cite[Sec. 13.1.3]{Simon}. Let ${\bf{b}} = \left[ {{\bf{b}}_S ,{\bf{b}}_R } \right]$ and ${\bf{\tilde b}} = \left[ {{\bf{\tilde b}}_S ,{\bf{\tilde b}}_R } \right]$ be $1 \times
\left( {N_S  + N_R^{\left( {{{PC}}} \right)} } \right)$ binary vectors defined according to Section \ref{ML_and_CMRC}. Then, by assuming equiprobable transmitted bits, the ABEP of $S_t$ is:
\setcounter{equation}{19}
\begin{equation}
\label{Eq_19} \begin{split}
{\rm{ABEP}}_{S_t }^{\left( {{\rm{UB}}} \right)} & \le 2^{ - \left( {N_S  + N_R^{\left( {{{PC}}} \right)} } \right)} \sum\limits_{\bf{b}} {\sum\limits_{{\bf{\tilde b}}} {\left| {d_{S_t } } \right|{\rm{APEP}}\left( {{\bf{c}} \to {\bf{\tilde c}}} \right)} } \\ & \mathop  = \limits^{\left( a \right)} \sum\limits_{{\bf{\tilde b}}} {\tilde b_{S_t } {\rm{APEP}}\left( {{\bf{0}} \to {\bf{\tilde c}}} \right)} \end{split}
\end{equation}
\noindent where $\left| {d_{S_t } } \right| = \left| {\tilde b_{S_t }  - b_{S_t } } \right| \in \left\{ {0,1} \right\}$, which is introduced in (\ref{Eq_11}), takes into account that a wrong decoded codeword, \emph{i.e.}, ${\bf{c}} \ne {\bf{\tilde c}}$, does not necessarily result in a bit error, \emph{i.e.}, ${b_{S_t }  \ne \tilde b_{S_t } }$, for $S_t$.

The equality in $\mathop  = \limits^{\left( a \right)}$, which greatly reduces the computational complexity, deserves some comments and a proof. First of all, $\mathop  = \limits^{\left( a \right)}$ holds for communication systems that satisfy the so--called uniform error property \cite{Raphaeli}, \emph{i.e.}, the error probability is independent of the transmitted codeword, and, thus, without loss of generality, the zero codeword, \emph{i.e.}, ${\bf{c}} = {\bf{0}}$, can be assumed to be transmitted. As described in \cite{Raphaeli}, the uniform error property does not depend only on the linearity of the network code, but it also depends on the code, the modulation, the demodulator, and the fading channel. The validity of the uniform error property for our system model can be proved by direct inspection of (\ref{Eq_17}). More specifically, for a given pair $\left( {{\bf{c}},{\bf{\tilde c}}} \right)$, (\ref{Eq_17}) uniquely depends on: i) the number of distinct binary digits between ${\bf{c}}$ and ${\bf{\tilde c}}$; and ii) the position where these binary digits are different. The latter property originates from the considered arbitrary network topology and the i.n.i.d assumption of channel fading. Let $\mathcal{J}_P\left( {{\bf{c}},{\bf{\tilde c}}} \right) = \left\{ {\left. p \right|\left| {\tilde c_p  - c_p } \right| = \tilde c_p  \oplus c_p  = 0,\;p = 1,2, \ldots ,N_S  + N_R } \right\}$ and $\mathcal{\bar J}_P\left( {{\bf{c}},{\bf{\tilde c}}} \right) = \left\{ {\left. p \right|\left| {\tilde c_p  - c_p } \right| = \tilde c_p  \oplus c_p  = 1,\;p = 1,2, \ldots ,N_S  + N_R } \right\}$ be the sets of $\mathcal{N}_P$ and $N_S  + N_R - \mathcal{N}_P$ positions where ${\bf{c}}$ and ${\bf{\tilde c}}$ are equal and different, respectively. By direct inspection and under the assumption of linear network codes, it can be verified that $\mathcal{J}\left( {{\bf{c}},{\bf{\tilde c}}} \right) = \mathcal{J}\left( {{\bf{0}},{\bf{\tilde c}} \oplus {\bf{c}}} \right)$ and $\mathcal{\bar J}\left( {{\bf{c}},{\bf{\tilde c}}} \right) = \mathcal{\bar J}\left( {{\bf{0}},{\bf{\tilde c}} \oplus {\bf{c}}} \right)$ for every pair $\left( {{\bf{c}},{\bf{\tilde c}}} \right)$. Furthermore, since the network code is assumed to be linear, ${{\bf{\tilde c}} \oplus {\bf{c}}}$ still belongs to the codebook. This allows us to affirm that the communication system under analysis satisfies the uniform error property and, thus, $\mathop  = \limits^{\left( a \right)}$ in (\ref{Eq_19}) can be used to compute the ABEP of the sources.
\begin{figure*}[!t]
\setcounter{equation}{21}
\begin{equation}
\label{Eq_20bis}
\Omega _q^{\left( {PC} \right)}  = \left\{ {\left. {{\bf{\tilde c}}} \right|\tilde b_{R_q }  = \tilde c_{N_S  + q}  \oplus \left( {g_{S_1 R_q } \tilde b_{S_1 }  \oplus g_{S_2 R_q } \tilde b_{S_2 }  \oplus  \ldots  \oplus g_{S_{N_S } R_q } \tilde b_{S_{N_S } } } \right) = 1,\;R_q  \in \mathcal{N}_R^{\left( {PC} \right)} ,\;w_\mathcal{H} \left( {{\bf{\tilde c}}} \right) = 1} \right\}
\end{equation}
\normalsize \hrulefill \vspace*{0pt}
\end{figure*}
\begin{figure*}[!t]
\setcounter{equation}{23}
\begin{equation} \footnotesize
\label{Eq_22}
\left\{ \begin{array}{l}
 \mathcal{\bar M}_{\Delta ^{\left( {{\rm{C - MRC}}} \right)} } \left( {\left. s \right|{\bf{0}},{\bf{\tilde c}},\mathcal{E}_m } \right) = \mathcal{\bar M}_{\Delta ^{\left( {{\rm{C - MRC}}} \right)} } \left( {\left. s \right|{\bf{\tilde c}},\mathcal{E}_m } \right) = \left( {{{E_m } \mathord{\left/
 {\vphantom {{E_m } {N_0 }}} \right.
 \kern-\nulldelimiterspace} {N_0 }}} \right)^{ - \left( {w_{\mathcal{H}}^{\left( S \right)} \left( {{\bf{\tilde c}}} \right) + w_{\mathcal{H}}^{\left( {R,\mathcal{E}_m^{\left( {{\rm{ok}}} \right)} } \right)} \left( {{\bf{\tilde c}}} \right) + {\rm{card}}\left( {\mathcal{E}_m^{\left( {{\rm{nok}}} \right)} } \right)} \right)} \Psi \left( {\left. s \right|{\bf{\tilde c}},\mathcal{E}_m } \right) \\
 \begin{split} \Psi \left( {\left. s \right|{\bf{\tilde c}},\mathcal{E}_m } \right) &= \left[ {s\left( {1 - s} \right)} \right]^{ - w_{\mathcal{H}}^{\left( S \right)} \left( {{\bf{\tilde c}}} \right)} \prod\limits_{t = 1}^{N_S } {\left( {4\chi _{S_t } \sigma _{S_t D}^2 } \right)^{ - \tilde b_{S_t } } }  \times \left[ { - \left( {2s} \right)^{ - 1} \left( {1 - \sqrt {\left( {1 - 4s} \right)^{ - 1} } } \right)} \right]^{w_{\mathcal{H}}^{\left( {R,\mathcal{E}_m^{\left( {{\rm{nok}}} \right)} } \right)}   {\left( {{\bf{\tilde c}}} \right)}   } \prod\limits_{q = 1 \atop \mu _{q - 1}  = 1}^{N_R } {\left( {4\sigma _{eqR_q }^2 } \right)^{ - 1} } \\ & \times \prod\limits_{q = 1 \atop \mu _{q - 1}  = 0}^{N_R } {\left[ {\left( {4\sigma _{eqR_q }^2 s} \right)^{ - 1}  + \left( {4\chi _{R_q } \sigma _{R_q D}^2 s\left( {1 - s} \right)} \right)^{ - 1}  - \left( {4\sigma _{eqR_q }^2 } \right)^{ - 1} \left( {2s} \right)^{ - 1} \left( {1 - \sqrt {\left( {1 + 4s} \right)^{ - 1} } } \right)} \right]^{\tilde b_{R_q }^{\left( {{\rm{NC}}} \right)} } }\end{split}  \\
 \end{array} \right.
\end{equation}
\normalsize \hrulefill \vspace*{0pt}
\end{figure*}
\subsection{Computation of the Average Bit Error Probability of the Partial--Cooperative Relays} \label{ABEP_Relays}
With arguments similar to (\ref{Eq_19}) and using the uniform error property, the ABEP of $R_q  \in \mathcal{N}_R^{\left( {PC} \right)}$ is:
\setcounter{equation}{20}
\begin{equation}
\label{Eq_20} \begin{split}
{\rm{ABEP}}_{R_q }^{\left( {{\rm{UB}}} \right)} & \le \sum\limits_{{\bf{\tilde b}}} {\tilde b_{R_q } {\rm{APEP}}\left( {{\bf{0}} \to {\bf{\tilde c}}} \right)} \\ & \approx \sum\limits_{{\bf{\tilde c}} \in \Omega _q^{\left( {PC} \right)} } {{\rm{APEP}}\left( {{\bf{0}} \to {\bf{\tilde c}}} \right)} \end{split}
\end{equation}
\noindent where $\Omega _q^{\left( {PC} \right)}$ is defined in (\ref{Eq_20bis}) shown at the top of this page, and $w_{\mathcal{H}} \left( {{\bf{\tilde c}}} \right) = \sum\nolimits_{p = 1}^{N_S  + N_R } {\tilde c_p }$ is the Hamming weight of ${{\bf{\tilde c}}}$. More specifically, $\Omega _q^{\left( {PC} \right)}$ is the set of possible codewords with unit Hamming weight and $\tilde b_{R_q }  = 1$ for $R_q  \in \mathcal{N}_R^{\left( {PC} \right)}$. The reason why only codewords with unit Hamming weight are considered is that the diversity order of partial--cooperative relays is equal to one. Even though this is intuitive from the working operation of our transmission protocol, a formal proof of the diversity order of partial--cooperative relays is postponed to Section \ref{NCdesign_for_Diversity}.

By construction, the set $\Omega _q^{\left( {PC} \right)}$ contains, at the most, the $N_S$ codewords ${\bf{\tilde c}}^{\left( t \right)}  = \left[ {0, \cdots ,0,1,0, \ldots ,0} \right]$ for $t=1,2,\ldots,N_S$ where only the $t$--th entry is non--zero, and the codeword ${\bf{\tilde c}}^{\left( {q} \right)}  = \left[ {0, \cdots ,0,1,0, \ldots ,0} \right]$ where only the $(N_S + q)$--th entry is non--zero. It is worth noticing that ${\bf{\tilde c}}^{\left( {q} \right)}$ is always present in $\Omega _q^{\left( {PC} \right)}$, while the presence of ${\bf{\tilde c}}^{\left( t \right)}$ for $t=1,2,\ldots,N_S$ depends on the codebook, \emph{i.e.}, the network code. To account for this, we introduce the indicator variable $\xi ^{\left( t \right)}$ for $t=1,2,\ldots,N_S$, such that  $\xi ^{\left( t \right)}  = 1$ if ${\bf{\tilde c}}^{\left( t \right)}$ is in the codebook and $\xi ^{\left( t \right)}  = 0$ if ${\bf{\tilde c}}^{\left( t \right)}$ is not in the codebook. Accordingly, (\ref{Eq_20}) can be re--written as follows:
\setcounter{equation}{22}
\begin{equation}
\label{Eq_21} \begin{split}
{\rm{ABEP}}_{R_q }^{\left( {{\rm{UB}}} \right)} & \le \sum\limits_{t = 1}^{N_S } {\xi ^{\left( t \right)} {\rm{APEP}}\left( {{\bf{0}} \to {\bf{\tilde c}}^{\left( t \right)} } \right)} \\ & + {\rm{APEP}}\left( {{\bf{0}} \to {\bf{\tilde c}}^{\left( q \right)} } \right) \\ & \mathop  \approx \limits^{\left( a \right)} \sum\limits_{t = 1}^{N_S } {\xi ^{\left( t \right)} \left( {4\bar \gamma _{S_t D} } \right)^{ - 1} } \\ & + \left[ {\left( {4\bar \gamma _{R_q D} } \right)^{ - 1}  + \left( {4\bar \gamma _{eqR_q } } \right)^{ - 1} } \right] \end{split}
\end{equation}
\noindent where $\mathop  \approx \limits^{\left( a \right)}$ is proved in Appendix \ref{App_ABEP_RelaysPC} by using the residues theorem to compute (\ref{Eq_13}).

From (\ref{Eq_21}), the ABEP of the partial--cooperative relays can be bounded as $\left( {4\bar \gamma _{R_q D} } \right)^{ - 1}  + \left( {4\bar \gamma _{eqR_q } } \right)^{ - 1}  \le {\rm{ABEP}}_{R_q }^{\left( {{\rm{UB}}} \right)}  \le \sum\nolimits_{t = 1}^{N_S } \left( {4\bar \gamma _{S_t D} } \right)^{ - 1}  + \left( {4\bar \gamma _{R_q D} } \right)^{ - 1}  + \left( {4\bar \gamma _{eqR_q } } \right)^{ - 1}$. Thus, two conclusions can be drawn: i) ${\rm{ABEP}}_{R_q }^{\left( {{\rm{UB}}} \right)}  \ge \left( {4\bar \gamma _{R_q D} } \right)^{ - 1}$ for every network topology, network codes, and fading channels. Since ${\rm{ABEP}}_0  \approx \left( {4\bar \gamma _{R_q D} } \right)^{ - 1}$ is the high--SNR approximation of the error probability of a single--hop transmission from $R_q$ to $D$ \cite{Simon}, it follows that the partial--cooperative relays undergo a performance loss, compared to single--hop transmission, which is the price they pay for helping the sources; and ii) this performance loss decreases for good source--to--relay channels, since in this case $\bar \gamma _{eqR_q }^{ - 1}  \to 0$, while it increases with the number of sources that are network--coded at the relay. In Section \ref{DiversityOrder} and in Section \ref{Insights_Framework}, we investigate whether this performance loss is offset by a better diversity order and/or coding gain for the sources.
\begin{figure*}[!t]
\setcounter{equation}{24}
\begin{equation}
\label{Eq_23}
\left\{ \begin{array}{l}
 G_d \left( {{\bf{\tilde c}},\mathcal{E}_m } \right)\mathop  = \limits^{\left( a \right)} w_{\mathcal{H}}^{\left( S \right)} \left( {{\bf{\tilde c}}} \right) + w_{\mathcal{H}}^{\left( {R,\mathcal{E}_m^{\left( {{\rm{ok}}} \right)} } \right)} \left( {{\bf{\tilde c}}} \right) + \left( {w_{\mathcal{H}}^{\left( {R,\mathcal{E}_m^{\left( {{\rm{nok}}} \right)} } \right)} \left( {{\bf{\tilde c}}} \right) + \bar w_{\mathcal{H}}^{\left( {R,\mathcal{E}_m^{\left( {{\rm{nok}}} \right)} } \right)} \left( {{\bf{\tilde c}}} \right)} \right)\mathop  = \limits^{\left( b \right)} w_{\mathcal{H}}^{\left( S \right)} \left( {{\bf{\tilde c}}} \right) + w_{\mathcal{H}}^{\left( R \right)} \left( {{\bf{\tilde c}}} \right) + \bar w_{\mathcal{H}}^{\left( {R,\mathcal{E}_m^{\left( {{\rm{nok}}} \right)} } \right)} \left( {{\bf{\tilde c}}} \right) \\
 G_d \left( {{\bf{\tilde c}}} \right)\mathop  = \limits^{\left( c \right)} \min _{\mathcal{E}_m } \left\{ {G_d \left( {{\bf{\tilde c}},\mathcal{E}_m } \right)} \right\} = \min _{\mathcal{E}_m } \left\{ {w_{\mathcal{H}}^{\left( S \right)} \left( {{\bf{\tilde c}}} \right) + w_{\mathcal{H}}^{\left( R \right)} \left( {{\bf{\tilde c}}} \right) + \bar w_{\mathcal{H}}^{\left( {R,\mathcal{E}_m^{\left( {{\rm{nok}}} \right)} } \right)} \left( {{\bf{\tilde c}}} \right)} \right\}\mathop  = \limits^{\left( d \right)} w_{\mathcal{H}}^{\left( S \right)} \left( {{\bf{\tilde c}}} \right) + w_{\mathcal{H}}^{\left( R \right)} \left( {{\bf{\tilde c}}} \right) \\
 \end{array} \right.
\end{equation}
\normalsize \hrulefill \vspace*{0pt}
\end{figure*}
\begin{figure*}[!t]
\setcounter{equation}{26}
\begin{equation}
\label{Eq_25}
{\rm{ABEP}}_{S_t }^{\left( {{\rm{UB}}} \right)}  \approx {\rm{ABEP}}_{S_t }^{\left( {{\rm{SV}}} \right)}  = \left( {{{E_m } \mathord{\left/
 {\vphantom {{E_m } {N_0 }}} \right.
 \kern-\nulldelimiterspace} {N_0 }}} \right)^{ - {\rm{SV}}^{\left( {S_t } \right)} } \sum\limits_{{\bf{\tilde b}},\;\tilde b_{S_t }  = 1,\;w_{\mathcal{H}} \left( {{\bf{\tilde c}}} \right) = {\rm{SV}}^{\left( {S_t } \right)} } {\sum\limits_{m = 0}^{2^{N_R }  - 1} {\left( {\left( {2\pi j} \right)^{ - 1} \int\nolimits_{\delta  - j\infty }^{\delta  + j\infty } {\Psi \left( {\left. s \right|{\bf{\tilde c}},\mathcal{E}_m } \right)s^{ - 1} ds} } \right)} }
\end{equation}
\normalsize \hrulefill \vspace*{0pt}
\end{figure*}
\section{Diversity Analysis and Network Code Design} \label{Diversity_NCdesign}
The objective of this section is threefold: i) to study the diversity order of sources and partial--cooperative relays; ii) to provide design criteria for binary network codes such that, in addition to having a solvable NC problem, the diversity order of each source can be assigned individually (unequal error robustness to multipath fading); and iii) to summarize relevant performance trends that can be inferred from our analytical framework. A byproduct of our diversity analysis is a simplified expression of the ABEP of the sources.
\subsection{Achievable Diversity Order} \label{DiversityOrder}
We introduce the notation $E_{S_t }  = \chi _{S_t } E_m$ and $E_{R_q }  = \chi _{R_q } E_m$. With the equal energy allocation policy described in Section \ref{SystemModel}, we have $\chi _{S_t }  = 1$, $\chi _{R_q }  = 1$ if $R_q  \in \mathcal{N}_{R_q }^{\left( {PC} \right)}$, and $\chi _{R_q }  = 1/2$ if $R_q  \in \mathcal{N}_{R_q }^{\left( {FC} \right)}$. By assuming ${\bf{c}} = {\bf{0}}$ according to the uniform error property in Section \ref{ABEP_Sources}, $\mathcal{\bar M}_{\Delta ^{\left( {{\rm{C - MRC}}} \right)} } \left( {\left.  \cdot  \right| \cdot } \right)$ in (\ref{Eq_17}) simplified to (\ref{Eq_22}) shown at the top of this page, where: i) $\sigma _{eqR_q }^2  = \left( {\sum\nolimits_{t = 1}^{N_S } {g_{S_t R_q } \chi _{S_t }^{ - 1} \sigma _{S_t R_q }^{ - 2} } } \right)^{ - 1}$; ii) $w_{\mathcal{H}}^{\left( S \right)} \left( {{\bf{\tilde c}}} \right) = \sum\nolimits_{t = 1}^{N_S } {\tilde b_{S_t } }$ is the Hamming weight of the systematic bits of ${{\bf{\tilde c}}}$, \emph{i.e.}, the bits transmitted from the sources; iii) $w_{\mathcal{H}}^{\left( {R,\mathcal{E}_m^{\left( {{\rm{ok}}} \right)} } \right)} \left( {{\bf{\tilde c}}} \right) = \sum\nolimits_{q = 1,\;\mu _{q - 1}  = 0}^{N_R } {\tilde b_{R_q }^{\left( {{\rm{NC}}} \right)} }$ and $w_{\mathcal{H}}^{\left( {R,\mathcal{E}_m^{\left( {{\rm{nok}}} \right)} } \right)} \left( {{\bf{\tilde c}}} \right) = \sum\nolimits_{q = 1,\;\mu _{q - 1}  = 1}^{N_R } {\tilde b_{R_q }^{\left( {{\rm{NC}}} \right)} }$ are the Hamming weights of the parity bits of ${{\bf{\tilde c}}}$, \emph{i.e.}, the bits transmitted from the relays forwarding a correct and an incorrect network--coded bit, respectively. Also, $w_{\mathcal{H}}^{\left( R \right)} \left( {{\bf{\tilde c}}} \right) = w_{\mathcal{H}}^{\left( {R,\mathcal{E}_m^{\left( {{\rm{nok}}} \right)} } \right)} \left( {{\bf{\tilde c}}} \right) + w_{\mathcal{H}}^{\left( {R,\mathcal{E}_m^{\left( {{\rm{ok}}} \right)} } \right)} \left( {{\bf{\tilde c}}} \right) = \sum\nolimits_{q = 1}^{N_R } {\tilde b_{R_q }^{\left( {{\rm{NC}}} \right)} }$ is the Hamming weights of all parity bits; and iv) ${\rm{card}}\left( {\mathcal{E}_m^{\left( {{\rm{nok}}} \right)} } \right)$ is the cardinality of ${\mathcal{E}_m^{\left( {{\rm{nok}}} \right)} }$, \emph{i.e.}, the number of relays that transmit an incorrect network--coded bit.

From (\ref{Eq_22}), it follows that the diversity order of the APEP linked to $\mathcal{\bar M}_{\Delta ^{\left( {{\rm{C - MRC}}} \right)} } \left( {\left. \cdot \right|{\bf{\tilde c}},\mathcal{E}_m } \right)$ depends on ${{\bf{\tilde c}}}$ and $\mathcal{E}_m$, and it is equal to $G_d \left( {{\bf{\tilde c}},\mathcal{E}_m } \right) = w_{\mathcal{H}}^{\left( S \right)} \left( {{\bf{\tilde c}}} \right) + w_{\mathcal{H}}^{\left( {R,\mathcal{E}_m^{\left( {{\rm{ok}}} \right)} } \right)} \left( {{\bf{\tilde c}}} \right) + {\rm{card}}\left( {\mathcal{E}_m^{\left( {{\rm{nok}}} \right)} } \right)$ \cite{Giannakis}. Since the diversity order, $G_d^{\left( {\rm Z} \right)}$, of node ${\rm Z} = \left\{ {S_t ,R_q  \in \mathcal{N}_{R_q }^{\left( {PC} \right)} } \right\}$ is determined by the APEP with the smallest decaying exponent, we have $G_d^{\left( {\rm Z} \right)}  = \min _{{\bf{\tilde c}},\mathcal{E}_m } \left\{ {\left. {G_d \left( {{\bf{\tilde c}},\mathcal{E}_m } \right)} \right|\tilde b_{\rm Z}  = 1} \right\}$. The conditioning upon ${\tilde b_{\rm Z}  = 1}$ originates from (\ref{Eq_19}) and (\ref{Eq_20}), which show that the APEP contributes to the ABEP of ${\rm Z}$ if and only if ${\tilde b_{\rm Z}  = 1}$. Let $\bar w_\mathcal{H}^{\left( {R,\mathcal{E}_m^{\left( {{\rm{nok}}} \right)} } \right)} \left( {{\bf{\tilde c}}} \right) = {\rm{card}}\left( {\mathcal{E}_m^{\left( {{\rm{nok}}} \right)} } \right) - w_{\mathcal{H}}^{\left( {R,\mathcal{E}_m^{\left( {{\rm{nok}}} \right)} } \right)} \left( {{\bf{\tilde c}}} \right)$ be the complement of $w_{\mathcal{H}}^{\left( {R,\mathcal{E}_m^{\left( {{\rm{nok}}} \right)} } \right)} \left( {{\bf{\tilde c}}} \right)$. Then,  $G_d \left( \cdot,\cdot  \right)$ simplifies to (\ref{Eq_23}) shown at the top of this page, where: i) $\mathop  = \limits^{\left( a \right)}$ follows from ${\rm{card}}\left( {\mathcal{E}_m^{\left( {{\rm{nok}}} \right)} } \right) = w_{\mathcal{H}}^{\left( {R,\mathcal{E}_m^{\left( {{\rm{nok}}} \right)} } \right)} \left( {{\bf{\tilde c}}} \right) + \bar w_{\mathcal{H}}^{\left( {R,\mathcal{E}_m^{\left( {{\rm{nok}}} \right)} } \right)} \left( {{\bf{\tilde c}}} \right)$; ii) $\mathop  = \limits^{\left( b \right)}$ follows from $w_{\mathcal{H}}^{\left( R \right)} \left( {{\bf{\tilde c}}} \right) = w_{\mathcal{H}}^{\left( {R,\mathcal{E}_m^{\left( {{\rm{nok}}} \right)} } \right)} \left( {{\bf{\tilde c}}} \right) + w_{\mathcal{H}}^{\left( {R,\mathcal{E}_m^{\left( {{\rm{ok}}} \right)} } \right)} \left( {{\bf{\tilde c}}} \right)$; iii) in $\mathop  = \limits^{\left( c \right)}$, the minimum is computed only with respect to $\mathcal{E}_m$ for a fixed codeword ${{\bf{\tilde c}}}$; and iv) $\mathop  = \limits^{\left( d \right)}$ follows from the fact that $\bar w_{\mathcal{H}}^{\left( {R,\mathcal{E}_m^{\left( {{\rm{nok}}} \right)} } \right)} \left( {{\bf{\tilde c}}} \right) \ge 0$ by definition. Thus, from (\ref{Eq_23}):
\setcounter{equation}{25}
\begin{equation}
\label{Eq_24} \begin{split}
G_d^{\left( {\rm Z} \right)}  & = \min_{{\bf{\tilde c}}} \left\{ {\left. {G_d \left( {{\bf{\tilde c}}} \right)} \right|\tilde b_{\rm Z}  = 1} \right\} \\ &= \min _{{\bf{\tilde c}}} \left\{ {\left. {w_{\mathcal{H}}^{\left( S \right)} \left( {{\bf{\tilde c}}} \right) + w_{\mathcal{H}}^{\left( R \right)} \left( {{\bf{\tilde c}}} \right)} \right|\tilde b_{\rm Z}  = 1} \right\} \\ & \mathop  = \limits^{\left( a \right)} \min _{{\bf{\tilde c}}} \left\{ {\left. {w_{\mathcal{H}} \left( {{\bf{\tilde c}}} \right)} \right|\tilde b_{\rm Z}  = 1} \right\}\mathop  = \limits^{\left( b \right)} {\rm{SV}}^{\left( {\rm Z} \right)} \end{split}
\end{equation}
\noindent where: i) $\mathop  = \limits^{\left( a \right)}$ originates from the definition of Hamming weight of ${{\bf{\tilde c}}}$, \emph{i.e.}, $w_{\mathcal{H}} \left( {{\bf{\tilde c}}} \right) = w_{\mathcal{H}}^{\left( S \right)} \left( {{\bf{\tilde c}}} \right) + w_{\mathcal{H}}^{\left( R \right)} \left( {{\bf{\tilde c}}} \right)$; and ii) $\mathop  = \limits^{\left( b \right)}$ follows from \cite[Def. 1]{Dunning} with ${\rm{SV}}^{\left( {\rm Z} \right)}$ denoting the so--called separation vector (SV) of ${\rm Z}$. Let a codebook $\mathcal{C} = \left\{ {{\bf{\tilde c}}} \right\}$, ${\rm{SV}}^{\left( {\rm Z} \right)}$ is the minimum Hamming weight among the codewords of the codebook with ${\tilde b_{\rm Z}  = 1}$.

In summary, from (\ref{Eq_24}) two main conclusions can be drawn: i) the C--MRC in (\ref{Eq_7}) provides node ${\rm Z}$ with a diversity order equal to ${\rm{SV}}^{\left( {\rm Z} \right)}$; and ii) by properly choosing the codebook and, thus, the network code, each node can achieve a different diversity order, which provides it with a different robustness to multipath fading and flexibility for improved energy efficiency \cite{RayLiu_Loc2009}. In Section \ref{NCdesign_for_Diversity}, we provide guidelines to design the network code such that the sources can achieve the desired diversity order.
\subsection{Simplified Average Bit Error Probability of the Sources} \label{Simplified_ABEP_Sources}
The analysis of the diversity order in Section \ref{DiversityOrder} allows us to simplify the ABEP in (\ref{Eq_19}). In fact, for networks with a large number of sources and partial--cooperative relays the number of codewords can be very large, which results in the computation of many APEPs. Since, for high--SNR, the ABEP (\ref{Eq_19}) is dominated by the APEPs with decaying exponent ${\rm{SV}}^{\left( {S_t } \right)}$, then (\ref{Eq_19}) can be simplified to (\ref{Eq_25}) shown at the top of this page, where only the codewords resulting in a bit error for the sources, \emph{i.e.}, ${\tilde b_{S_t }  = 1}$, and providing the smallest decaying exponent with the SNR, \emph{i.e.}, ${{\rm{SV}}^{\left( {S_t } \right)} }$, are retained.
\subsection{Network Code Design for Cooperative Diversity} \label{NCdesign_for_Diversity}
In Section \ref{DiversityOrder}, it is proved that sources and partial--cooperative relays achieve a diversity order equal to the separation vector of the codebook (network code). This result is relevant from the \emph{system analysis} point of view. In this section, we look into two important \emph{system design} issues: i) to construct the network code such that the sources have a desired separation vector; and ii) to understand which relays contribute to the diversity order of the sources. The first design problem is relevant because a bigger separation vector could be assigned to sources either subject to deeper fading conditions or having less residual energy \cite{Iezzi_ICC2011}, \cite{RayLiu_Loc2009}. The second design problem is important because nodes that contribute to the diversity order of none of the sources provide a marginal contribution to their end--to--end performance, and, thus, they may be excluded from the relaying phase to reduce the cooperation overhead. In our implementation, this decision is made during the initialization phase, before actual data transmission, and, thus, these excluded nodes will be allowed to transmit the data available in their own buffers in the next available time--slot. More specifically, three situations can arise for these nodes: i) they will act as sources and will transmit their data to the destination; ii) they will act as sources and will ask other nodes to act as relays on their behalf for performance improvement; and iii) they will act as relays for other sources for which they can effectively contribute to their performance. As far as this latter case is concerned, a possible scenario is as follows. To better understand, we anticipate that partial--cooperative relays do not contribute to the diversity order of the sources, while full--cooperative relays do contribute. Accordingly, some partial--cooperative relays may agree to act as full--cooperative relays for other sources, from which they may have received special incentives for cooperation. In other words, the relays may decide their level of cooperation depending on the sources that ask for cooperation.

Since the diversity order coincides with the separation vector of the codebook, it follows that it is uniquely determined by the generator matrix, ${\bf{G}}^{\left( {{\rm{NC}}} \right)}$, of the network code. Note that this is not true, in general, for the coding gain \cite{Giannakis}. In fact, it also depends on the demodulation errors at the relays and the fading channels. In particular, ${\bf{G}}^{\left( {{\rm{NC}}} \right)}$ is an $\left( {N_S  + N_R^{\left( {{{PC}}} \right)} } \right) \times \left( {N_S  + N_R } \right)$ matrix such that ${\bf{c}} = {\bf{bG}}^{\left( {{\rm{NC}}} \right)}$. By direct inspection of the transmission protocol in Section \ref{SystemModel}, ${\bf{G}}^{\left( {{\rm{NC}}} \right)}$ can be constructed as the block matrix as follows:
\setcounter{equation}{27}
\begin{equation}
\label{Eq_26} \begin{split}
{\bf{G}}^{\left( {{\rm{NC}}} \right)}  & = \left[ {\begin{array}{*{20}c}
   {{\bf{I}}_{N_S  \times N_S } } \hfill & {{\bf{G}}^{\left( {PC} \right)} } \hfill & {{\bf{G}}^{\left( {FC} \right)} } \hfill  \\
   {{\bf{0}}_{N_R^{\left( {PC} \right)}  \times N_S } } \hfill & {{\bf{I}}_{N_R^{\left( {PC} \right)}  \times N_R^{\left( {PC} \right)} } } \hfill & {{\bf{0}}_{N_R^{\left( {PC} \right)}  \times N_R^{\left( {FC} \right)} } } \hfill  \\
\end{array}} \right] \\ & \mathop  = \limits^{\left( a \right)} \left[ {\left. {{\bf{I}}_{\left( {N_S  + N_R^{\left( {{\rm{PC}}} \right)} } \right) \times \left( {N_S  + N_R^{\left( {{\rm{PC}}} \right)} } \right)} } \right|\begin{array}{*{20}c}
   {{\bf{G}}^{\left( {FC} \right)} } \hfill  \\
   {{\bf{0}}_{N_R^{\left( {PC} \right)}  \times N_R^{\left( {FC} \right)} } } \hfill  \\
\end{array}} \right] \end{split}
\end{equation}
\noindent where: i) ${\bf{I}}_{n \times n}$ is an $n \times n$ identity matrix; ii) ${\bf{0}}_{l \times n}$ is an $l \times n$ all--zero matrix; iii) $\left(  \cdot  \right)^T$ is the transpose operator; iv) ${{\bf{G}}^{\left( {PC} \right)} }$ is the $N_S  \times N_R^{\left( {{{PC}}} \right)}$ matrix whose $q$--th column for ${R_q  \in \mathcal{N}_{R_q }^{\left( {PC} \right)} }$ is the encoding vector ${\bf{g}}_{R_q }^T$; and v) ${{\bf{G}}^{\left( {FC} \right)} }$ is the $N_S  \times N_R^{\left( {{{FC}}} \right)}$ matrix whose $q$--th column for ${R_q  \in \mathcal{N}_{R_q }^{\left( {FC} \right)} }$ is the encoding vector ${\bf{g}}_{R_q }^T$. In (\ref{Eq_26}), it is implicitly assumed that partial--cooperative relays transmit in time--slots $T_{N_S  + q}$ for $q = 1,2, \ldots ,N_R^{\left( {PC} \right)}$, while full--cooperative relays transmit in time--slots $T_{N_S  + q}$ for $q = N_R^{\left( {PC} \right)}+1,N_R^{\left( {PC} \right)}+2, \ldots ,N_R$. This assumption is retained only for simplicity of writing. In fact, the performance does not change with the transmission order of the relays if the channels are quasi--static during the whole cooperation phase, \emph{i.e.}, for $N_S+N_R$ time--slots. By direct inspection of (\ref{Eq_26}), we notice that ${\bf{G}}^{\left( {{\rm{NC}}} \right)}$ is in row--echelon form. The identity in $\mathop  = \limits^{\left( a \right)}$ is obtained by applying elementary row operations to ${\bf{G}}^{\left( {{\rm{NC}}} \right)}$ in order to get a reduced row--echelon form.

From \cite[Lemma 4]{Boyarinov_Marc1981} and \cite[Eq. (4)]{Dunning}, it is known that the separation vector of ${\bf{G}}^{\left( {{\rm{NC}}} \right)}$ can be directly inferred from its related $N_R^{\left( {FC} \right)}  \times \left( {N_S  + N_R } \right)$ parity--check matrix ${\bf{H}}^{\left( {{\rm{NC}}} \right)}$. From (\ref{Eq_26}), ${\bf{H}}^{\left( {{\rm{NC}}} \right)}$ is:
\setcounter{equation}{28}
\begin{equation}
\label{Eq_27}
{\bf{H}}^{\left( {{\rm{NC}}} \right)}  = \left[ {\begin{array}{*{20}c}
   {\left( {{\bf{G}}^{\left( {FC} \right)} } \right)^T } & {{\bf{0}}_{N_R^{\left( {FC} \right)}  \times N_R^{\left( {PC} \right)} }^T } & {{\bf{I}}_{N_R^{\left( {FC} \right)}  \times N_R^{\left( {FC} \right)} } }  \\
\end{array}} \right]
\end{equation}
\noindent where the rows of ${\left( {{\bf{G}}^{\left( {FC} \right)} } \right)^T }$ are the encoding vectors of the full--cooperative relays.

By direct inspection of ${\bf{H}}^{\left( {{\rm{NC}}} \right)}$, and from \cite[Lemma 4]{Boyarinov_Marc1981} and \cite[Eq. (4)]{Dunning}, the following considerations for the design of the network code can be drawn:
\begin{list}{$\bullet$}{\leftmargin=0em \itemindent=1.5em}
\item The diversity order of $S_t$ is $G_d^{\left( {S_t } \right)}  = {\rm{SV}}^{\left( {S_t } \right)}  = {\rm{nc}}_t  + 1$, where ${\rm{nc}}_t$ is the least number of columns of ${\bf{H}}^{\left( {{\rm{NC}}} \right)}$ whose linear combination yields the $t$--th column of ${\bf{H}}^{\left( {{\rm{NC}}} \right)}$. In other words, $S_t$ has diversity order ${\rm{SV}}^{\left( {S_t } \right)}$ if and only if the $t$--th column of ${\bf{H}}^{\left( {{\rm{NC}}} \right)}$ is linearly dependent on no fewer than ${\rm{nc}}_t  = {\rm{SV}}^{\left( {S_t } \right)}  - 1$ columns of ${\bf{H}}^{\left( {{\rm{NC}}} \right)}$ \cite[Lemma 4]{Boyarinov_Marc1981}. This provides a clear criterion to choose the encoding vectors of the full--cooperative relays, and, thus, the matrix ${{\bf{G}}^{\left( {FC} \right)} }$ in (\ref{Eq_26}) and (\ref{Eq_27}).
\item Similar to the sources, the diversity order of $R_q  \in \mathcal{N}_R^{\left( {PC} \right)}$ is $G_d^{\left( {R_q } \right)}  = {\rm{SV}}^{\left( {R_q } \right)}  = {\rm{nc}}_{N_S+q}  + 1$, where ${\rm{nc}}_{N_S+q}$ is the least number of columns of ${\bf{H}}^{\left( {{\rm{NC}}} \right)}$ whose linear combination yields the $(N_S+q)$--th column of ${\bf{H}}^{\left( {{\rm{NC}}} \right)}$. Since the $(N_S+q)$--th column of ${\bf{H}}^{\left( {{\rm{NC}}} \right)}$ is zero for $q=1,2,\ldots,N_R^{\left( {PC} \right)}$, it follows that ${\rm{nc}}_{N_S+q} = 0$ and, thus, $G_d^{\left( {R_q } \right)}  = {\rm{SV}}^{\left( {R_q } \right)}  = 1$ for all partial--cooperative relays. This confirms the analysis in Section \ref{ABEP_Relays}.
\item From (\ref{Eq_27}), it follows that the partial--cooperative relays do not contribute to the diversity order of the sources. In fact, ${\bf{H}}^{\left( {{\rm{NC}}} \right)}$ is independent of the encoding vectors, \emph{i.e.}, the matrix ${{\bf{G}}^{\left( {PC} \right)} }$ in (\ref{Eq_26}). Only the full--cooperative relays contribute to the diversity order of the sources, and, by properly choosing the encoding vectors in ${{\bf{G}}^{\left( {FC} \right)} }$, the range of achievable diversity orders of $S_t$ is $1 \le {\rm{SV}}^{\left( {S_t } \right)}  \le N_R^{\left( {FC} \right)}  + 1$ \cite[Lemma 5]{Boyarinov_Marc1981}.
\item In addition to the analytical derivation, it is interesting to provide intuition about the result that partial--cooperative relays do not contribute to the diversity order of the sources. For simplicity, let us assume that there are no decoding errors at the relays (\emph{i.e.}, the SNR of the source--to--relay links goes to infinity). As far as diversity is concerned, this is the best--case scenario. If the diversity order is equal to one in this case, then it will be equal to one for every SNR of the source--to--relays links. However, this no longer holds if the diversity order is greater than one, since decoding errors at the relays may reduce the diversity order. To start with, let us consider the simplest case study with a single source and a single relay. The relay acts as partial--cooperative, and, thus, it network--codes the bit received from the source with its own bit. Accordingly, the codewords of the codebook are: ``00'', ``01'', ``10'', ``11'' where the first bit is sent by the source and the second bit is the XOR of the bits of source and relay. It is apparent that the codebook contains all the possible codewords, and, thus, the Hamming distance of the information bits can only be equal to one. More in general, if all the relays are partial--cooperative, \emph{i.e.}, the network rate is equal to one, the intuitive reason why partial--cooperative relays do not contribute to the diversity of the sources is that the codebook of the equivalent distributed code contains all possible codewords of the universe set. As a consequence, the Hamming distance of each information bit can only be one. If the network rate is less than one, \emph{i.e.}, some relays are full--cooperative and some others are partial--cooperative, then providing an intuitive explanation is more complicated and the equivalent codebook needs to be investigated by direct inspection. Our understanding is that when a partial--cooperative relay network--codes its own bit with the bits received from the sources the net effect is that the transmitted network--coded bit no longer acts as a parity bit because of the presence of the information bit of the partial--cooperative relay. On the other hand, the transmitted network--coded bit is an actual information bit, and, thus, the Hamming distance of the equivalent code is reduced compared with the case study where all the relays are full--cooperative. Our research is currently devoted to understand whether partial--cooperative relays can contribute to the diversity order of the sources by increasing the Galois field and/or for some modulation orders. We are particularly interested in system setups with network rates less than one. In fact, if all relays are partial--cooperative the equivalent codebooks may coincide with the universe set for many choices of Galois field and modulation order. If so, the diversity order of the sources will never be greater than one.
\end{list}

From these considerations, the following design criterion for the network code is proposed. Let us consider a network with $N_S$ sources requesting diversity order $G_d^{\left( {S_t } \right)}$ and ${N}_R^{\left( {FC} \right)} $ full--cooperative relays. Partial--cooperative relays can be neglected since they do not contribute to $G_d^{\left( {S_t } \right)}$. Then, ${\bf{G}}^{\left( {{{FC}}} \right)}$, \emph{i.e.}, the encoding vectors of the full--cooperative relays, can be chosen as the generator matrix of a systematic $(N_S+{N}_R^{\left( {FC} \right)}, N_S)$ linear block code with separation vector ${\rm{SV}}^{\left( {S_t } \right)} = G_d^{\left( {S_t } \right)}$. Known techniques for the construction of these codes, along with the conditions for their existence for a given choice of $N_S$ and $N_R$, are available in the literature \cite{Boyarinov_Marc1981}, \cite{MacWilliamsSloane}. Examples of such systematic codes are listed in \cite[Table I]{VanGils} for various network topologies and separation vectors. Among the possible choices, an important case study is when each source needs the largest separation vector, \emph{i.e.}, ${\rm{SV}}^{\left( {S_t } \right)} = N_R^{\left( {FC} \right)}  + 1$ for $t=1, 2, \ldots, N_S$. This case study corresponds to the design of the so--called Maximum Distance Separable (MDS) codes \cite[Ch. 11]{MacWilliamsSloane}, which are known to attain the Singleton bound \cite{Singleton1964}. From \cite[Ch. 11, Corollary 7]{MacWilliamsSloane}, it follows that for $N_S \ge 2$, binary modulations, and binary network codes the Singleton bound can be attained only for single--relay networks. If more relays are available, non--binary codes should be used to attain the Singleton bound. For a two--source two--relay network with binary NC, this result has been proved, using simulations, in \cite{Iezzi_ICC2011} by considering all possible choices of the encoding vectors. On the other hand, for single--relay networks the possibility to attain the Singleton bound for arbitrary Galois fields has been proved in \cite{Nasri}. In fact, the network code used in \cite{Nasri} is an example of the the so--called trivial MDS codes described in \cite[Ch. 11, Problem 1]{MacWilliamsSloane} with code type $(n, n-1, 2)$ and $n=N_S+1$.
\begin{figure*}[!t]
\setcounter{equation}{29}
\begin{equation} \footnotesize
\label{Eq_28}
\left\{ \begin{array}{l}
 {\rm{ABEP}}_{S_t }^{\left( {{\rm{SV}}} \right)}  \approx \left( {{{E_m } \mathord{\left/
 {\vphantom {{E_m } {N_0 }}} \right.
 \kern-\nulldelimiterspace} {N_0 }}} \right)^{ - {\rm{SV}}^{\left( {S_t } \right)} } \sum\limits_{{\bf{\tilde b}}_S ,\;\tilde b_{S_t }  = 1,\;w_{\mathcal{H}} \left( {{\bf{\tilde c}}^{(FC)} } \right) = {\rm{SV}}^{\left( {S_t } \right)} } {\sum\limits_{m = 0}^{2^{N_R }  - 1} {\left( {\left( {2\pi j} \right)^{ - 1} \int\nolimits_{\delta  - j\infty }^{\delta  + j\infty } {\Psi \left( {\left. s \right|{\bf{\tilde c}}^{(FC)} ,\mathcal{E}_m } \right)s^{ - 1} ds} } \right)} }  \\
 \begin{split}
 \Psi \left( {\left. s \right|{\bf{\tilde c}}^{(FC)} ,\mathcal{E}_m } \right) &= \left[ {s\left( {1 - s} \right)} \right]^{ - w_{\mathcal{H}}^{\left( S \right)} \left( {{\bf{\tilde c}}}^{(FC)} \right)} \prod\limits_{t = 1}^{N_S } {\left( {4\chi _{S_t } \sigma _{S_t D}^2 } \right)^{ - \tilde b_{S_t } } } \\  &\times \left[ { - \left( {2s} \right)^{ - 1} \left( {1 - \sqrt {\left( {1 - 4s} \right)^{ - 1} } } \right)} \right]^{w_{\mathcal{H}}^{\left( {R,\mathcal{E}_m^{\left( {{\rm{nok}}} \right)} } \right)} \left( {{\bf{\tilde c}}}^{(FC)} \right)} \prod\limits_{q = 1,\;R_q  \in \mathcal{N}_R^{\left( {FC} \right)} ,\;\mu _{q - 1}  = 1}^{N_R } {\left( {4\sigma _{eqR_q }^2 } \right)^{ - 1} }  \\
  &\times \prod\limits_{q = 1,\;R_q  \in \mathcal{N}_R^{\left( {FC} \right)} ,\;\mu _{q - 1}  = 0}^{N_R } {\left[ {\left( {4\sigma _{eqR_q }^2 s} \right)^{ - 1}  + \left( {4\chi _{R_q } \sigma _{R_q D}^2 s\left( {1 - s} \right)} \right)^{ - 1}  - \left( {4\sigma _{eqR_q }^2 } \right)^{ - 1} \left( {2s} \right)^{ - 1} \left( {1 - \sqrt {\left( {1 + 4s} \right)^{ - 1} } } \right)} \right]^{\tilde b_{R_q }^{\left( {{\rm{NC}}} \right)} } } \end{split}  \\
 \end{array} \right.
\end{equation}
\normalsize \hrulefill \vspace*{0pt}
\end{figure*}
\subsection{Insights From the Analytical Framework} \label{Insights_Framework}
We close this section by providing some conclusions that can be drawn from our analytical framework and can help the understanding, the analysis, and the design of network--coded cooperative networks.
\begin{list}{$\bullet$}{\leftmargin=0em \itemindent=1.5em}
\item[{\bf{C1)}}] From the definition of $\sigma _{eqR_q }^2$ in (\ref{Eq_22}), it follows, as expected, that the larger the number of network--coded sources at each relay, the worse the performance. This result is in agreement with \cite{Nasri} and \cite{DiRenzo}, as well as with intuition. Mathematically speaking, this originates from the fact that increasing the number of network--coded sources results in a larger probability, $P_{R_q }^{\left( {{\rm{NC}}} \right)}$ in (\ref{Eq_4}), of transmitting an incorrect network--coded bit. Thus, the matrix ${\left( {{\bf{G}}^{\left( {FC} \right)} } \right)^T }$ in (\ref{Eq_27}) should not only be designed to provide the desired diversity order, but its rows should be as sparse as possible in order to minimize the error accumulation problem at the relays, and, thus, to improve the coding gain \cite{Giannakis}. In general, the design criterion on the diversity order has a more pronounced impact on the end--to--end performance. Furthermore, the closed--form expression of $\sigma _{eqR_q }^2$ in (\ref{Eq_22}) confirms that the error accumulation problem at the relays is inversely proportional to the quality of the source--to--relay links, \emph{i.e.}, $\sigma _{S_t R_q }^2$. This result is well--known for DemF--based relaying protocols \cite{Laneman2007}, \cite{Morgado}, and it can be considered as a sanity check for our analytical derivation.
\item[{\bf{C2)}}] In Section \ref{NCdesign_for_Diversity}, it is shown that ${\rm{SV}}^{\left( {S_t } \right)}  \ge 1$ for every choice of ${\left( {{\bf{G}}^{\left( {FC} \right)} } \right)^T }$ in (\ref{Eq_27}). In particular, $S_t$ enjoys no diversity gain if and only if ${\rm{SV}}^{\left( {S_t } \right)}  = 1$, which implies ${\rm{SV}}^{\left( {S_t } \right)}  = {\rm{nc}}_t  + 1 = 1 \Rightarrow {\rm{nc}}_t  = 0$. This is possible if and only if the $t$--th column of ${\left( {{\bf{G}}^{\left( {FC} \right)} } \right)^T }$ is all--zero, which, in turn, implies that $S_t$ is network--coded by none of the full--cooperative relays. This result is expected and agrees with intuition. In fact, if $S_t$ is network--coded by none of the full--cooperative relays this implies that it is only present in the direct link, and, thus, the diversity order can only be equal to one. This sanity check confirms the consistency of our analytical derivation. In addition, and, more importantly, it provides an interesting conclusion. It is sufficient that at least one entry of all columns of ${\left( {{\bf{G}}^{\left( {FC} \right)} } \right)^T }$ is non--zero to guarantee that each source enjoys at least second--order diversity, \emph{i.e.}, $G_d^{\left( {S_t } \right)}  = {\rm{SV}}^{\left( {S_t } \right)} \ge 2$. In other words, if each source is network--coded by at least one full--cooperative relay, then its diversity order is at least equal to two regardless of the specific choice of the network code, which can also be randomly generated.
\item[{\bf{C3)}}] Let us assume that the ${N_R^{\left( {FC} \right)} }$ full--cooperative relays have the same encoding vector and that at least two of its entries are non--zero, \emph{i.e.}, at least two sources are network--coded. If the encoding vectors are all the same and only one of its entries is non--zero, the system reduces to a cooperative network without NC. This latter case study is considered in {\bf{C6)}} below. Under these assumptions, the sources that are network--coded have diversity order two, while the sources that are not network--coded have diversity order one. In fact, the matrix ${\left( {{\bf{G}}^{\left( {FC} \right)} } \right)^T }$ in (\ref{Eq_27}) has a number of all--one columns equal to the sources that are network--coded, and the other columns are all--zero. The all--zero columns are responsible for the first--order diversity of the sources that are not network--coded. On the other hand, the all--one columns, since linearly dependent, are responsible for the second--order diversity of the network--coded sources.
\item[{\bf{C4)}}] Let us consider that the ${N_R^{\left( {FC} \right)} }$ full--cooperative relays network--code the data received from all the sources. This results in having an all--ones matrix ${\left( {{\bf{G}}^{\left( {FC} \right)} } \right)^T }$ in (\ref{Eq_27}). Accordingly, its columns are linearly dependent and the diversity order of each source is $G_d^{\left( {S_t } \right)}  = {\rm{SV}}^{\left( {S_t } \right)}  = 2$.
\item[{\bf{C5)}}] In \cite{Nasri}, it is shown that one relay is sufficient for all network--coded sources to achieve second--order diversity. The diversity analysis in Section \ref{NCdesign_for_Diversity} reveals that this encoding is optimum from the diversity point of view. In fact, we have proved that $1 \le {\rm{SV}}^{\left( {S_t } \right)}  \le N_R^{\left( {FC} \right)}  + 1$. If $N_R^{\left( {FC} \right)} = 1$, then $1 \le {\rm{SV}}^{\left( {S_t } \right)}  \le 2$. In \cite{Nasri}, it is proved that $G_d^{\left( {S_t } \right)}  = {\rm{SV}}^{\left( {S_t } \right)}  = 2$, which is the best achievable diversity with one full--cooperative relay. From {\bf{C3)}}, it follows that multiple relays can be useless, from the diversity point of view, if their encoding vectors are all the same since diversity orders no greater than two can be achieved. In the presence of multiple relays, the encoding vectors should be carefully designed in order to enjoy the benefits of cooperative diversity. The design guidelines provided in Section \ref{NCdesign_for_Diversity} are useful to this end.
\item[{\bf{C6)}}] As mentioned in Section \ref{Introduction}, network--coded cooperation is a generalization of repetition--based relaying. In particular, the former reduces to the latter if there is only one non--zero entry in each encoding vector ${{\bf{g}}_{R_q } }$ for $R_q  \in \mathcal{N}_R^{\left( {FC} \right)}$. Thus, the frameworks developed in the present paper, \emph{e.g.}, (\ref{Eq_25}), and the diversity analysis in Section \ref{DiversityOrder} can be applied to repetition--based relaying as well. In this case, the $t$--th column of ${\left( {{\bf{G}}^{\left( {FC} \right)} } \right)^T }$ in (\ref{Eq_27}) has a number of non--zero entries equal to the full--cooperative relays that forward the data received from $S_t$. Let $0 \le N_R^{\left( {S_t } \right)}  \le N_R^{\left( {FC} \right)}$ denote this number of relays. Also, since each relay can forward the data of one and only one source, it follows that the first $N_S$ columns of ${\left( {{\bf{G}}^{\left( {FC} \right)} } \right)^T }$ are independent of each other and that ${\rm{nc}}_t = N_R^{\left( {S_t } \right)}$. Thus, $G_d^{\left( {S_t } \right)}  = {\rm{SV}}^{\left( {S_t } \right)}  = {\rm{nc}}_t  + 1 = N_R^{\left( {S_t } \right)}  + 1$. This result is in agreement with state--of--the--art diversity analysis of repetition--based relaying \cite{Laneman2007}. However, unlike \cite{Laneman2007}, our frameworks provide also an accurate estimate of the coding gain.
\item[{\bf{C7)}}] In Section \ref{ABEP_Relays}, it is proved that the partial--cooperative relays undergo a performance loss, compared to single--hop transmission, which can be interpreted as the price they pay for helping the sources. Furthermore, in Section \ref{NCdesign_for_Diversity} it is shown that the partial--cooperative relays do not contribute to the diversity order of the sources. Thus, a fundamental question is whether the partial--cooperative relays are useful and whether they should be exploited during the relaying phase. To answer to this question, it is important to investigate the impact that these nodes have on the coding gain of the sources. To this end, let us consider two network topologies: 1) the first one with $N_S$ sources, $N_R^{\left( {FC} \right)}$ full--cooperative relays, and $N_R^{\left( {PC} \right)}$ = 0 partial--cooperative relays; and 2) the second one with $N_S$ sources, $N_R^{\left( {FC} \right)}$ full--cooperative relays, and $N_R^{\left( {PC} \right)}$ partial--cooperative relays. In both networks, the full--cooperative relays have the same encoding vectors, while in the second network the $N_R^{\left( {PC} \right)}$ partial--cooperative relays have arbitrary encoding vectors. Let $\mathcal{C}_1$ and $\mathcal{C}_2$ denote the codebooks of the two network topologies. In general, $\mathcal{C}_1 \ne \mathcal{C}_2$. From Section \ref{NCdesign_for_Diversity}, it is known that, in both networks, the sources have the same diversity order, \emph{i.e.}, ${\rm{SV}}_1^{\left( {S_t } \right)}  = {\rm{SV}}_2^{\left( {S_t } \right)} = {\rm{SV}}^{\left( {S_t } \right)}$. From (\ref{Eq_25}), we know that for high--SNR the ABEP of $S_t$ depends only on the codewords ${\bf{\tilde c}}_l^*  \in \mathcal{C}_l$ with ${\rm{SV}}_l^{\left( {S_t } \right)}  = w_{\mathcal{H}} \left( {{\bf{\tilde c}}_l^* } \right) = w_{\mathcal{H}}^{\left( S \right)} \left( {{\bf{\tilde c}}_l^* } \right) + w_{\mathcal{H}}^{\left( R \right)} \left( {{\bf{\tilde c}}_l^* } \right)$ for $l=1, 2$. For these codewords, the constraint ${\rm{SV}}_1^{\left( {S_t } \right)}  = {\rm{SV}}_2^{\left( {S_t } \right)}$ implies $w_{\mathcal{H}}^{\left( S \right)} \left( {{\bf{\tilde c}}_1^* } \right) = w_{\mathcal{H}}^{\left( S \right)} \left( {{\bf{\tilde c}}_2^* } \right)$ and $w_{\mathcal{H}}^{\left( R \right)} \left( {{\bf{\tilde c}}_1^* } \right) = w_{\mathcal{H}}^{\left( R \right)} \left( {{\bf{\tilde c}}_2^* } \right)$, which stems from the fact that both codebooks have, by construction, the same systematic bits. In addition, we can write $w_{\mathcal{H}}^{\left( R \right)} \left( {{\bf{\tilde c}}_l^* } \right) = w_{\mathcal{H}}^{\left( {R,FC} \right)} \left( {{\bf{\tilde c}}_l^* } \right) + w_{\mathcal{H}}^{\left( {R,PC} \right)} \left( {{\bf{\tilde c}}_l^* } \right)$, where $w_{\mathcal{H}}^{\left( {R,FC} \right)} \left( {{\bf{\tilde c}}_l^* } \right)$ and $w_{\mathcal{H}}^{\left( {R,PC} \right)} \left( {{\bf{\tilde c}}_l^* } \right)$ are the contributions to the Hamming weight related to full-- and partial--cooperative relays, respectively. By construction: i) $w_{\mathcal{H}}^{\left( {R,FC} \right)} \left( {{\bf{\tilde c}}_1^* } \right) = w_{\mathcal{H}}^{\left( {R,FC} \right)} \left( {{\bf{\tilde c}}_2^* } \right)$, since in both networks the full--cooperative relays use the same encoding vectors; and ii) $w_{\mathcal{H}}^{\left( {R,PC} \right)} \left( {{\bf{\tilde c}}_1^* } \right) = 0$, since in the first network there are no partial--cooperative relays. In conclusion, the constraint $w_{\mathcal{H}}^{\left( R \right)} \left( {{\bf{\tilde c}}_1^* } \right) = w_{\mathcal{H}}^{\left( R \right)} \left( {{\bf{\tilde c}}_2^* } \right)$ implies $w_{\mathcal{H}}^{\left( {R,PC} \right)} \left( {{\bf{\tilde c}}_2^* } \right) = 0$, which, in turn, implies, for these codewords, $\tilde b_q^{\left( {{\rm{NC}}} \right)} = 0$ for $R_q  \in \mathcal{N}_R^{\left( {PC} \right)}$. In other words, in the second network, the partial--cooperative relays do not contribute to the Hamming weight of the codewords ${\bf{\tilde c}}_2^*  \in C_2$. From $\Psi \left( {\left.  \cdot  \right|{\bf{\tilde c}}_2^* , \cdot } \right)$ in (\ref{Eq_22}), we notice that the relays with $\tilde b_q^{\left( {{\rm{NC}}} \right)} = 0$ do not contribute to the coding gain of the ABEP. Thus, the partial--cooperative relays do not contribute to the coding gain of the sources.

    In summary, our analytical study clearly shows that the partial--cooperative relays contribute to neither the diversity order nor to the coding gain of the sources. In addition, these relays undergo a performance loss due to applying NC on the data received from the sources. In conclusion, our analysis reveals that only full--cooperative relays should be used during the relaying phase. This result allows us to further simplify the computation of $\Psi \left( {\left.  \cdot  \right|{\cdot , \cdot } } \right)$ in (\ref{Eq_22}) and ${\rm{ABEP}}_{S_t }^{\left( {{\rm{SV}}} \right)}$ in (\ref{Eq_25}). Since the partial--cooperative relays contribute to neither the diversity order nor to the coding gain, they can be completely neglected from the computation of $\Psi \left( {\left.  \cdot  \right|{\cdot , \cdot } }\right)$ and ${\rm{ABEP}}_{S_t }^{\left( {{\rm{SV}}} \right)}$. Thus, (\ref{Eq_22}) and (\ref{Eq_25}) simplify as shown in (\ref{Eq_28}) at the top of the previous page, where ${\bf{\tilde c}}^{\left( {FC} \right)}  = \left[ {\tilde b_{S_1 } ,\tilde b_{S_2 } , \ldots ,\tilde b_{S_{N_S } } ,\tilde b_{R_1 }^{\left( {{\rm{NC}}} \right)} ,\tilde b_{R_2 }^{\left( {{\rm{NC}}} \right)} , \ldots ,\tilde b_{R_{N_R^{\left( {FC} \right)} } }^{\left( {{\rm{NC}}} \right)} } \right]$ accounts only for sources and full--cooperative relays. In other words, in the high--SNR the performance of the sources is determined by the sub--network containing only the sources and the full--cooperative relays. The partial--cooperative relays can be neglected regardless of the binary encoding vectors they use.

    Finally, some remarks about the usefulness of partial--cooperative relays are needed. In the present paper, we have focused our attention on the contribution of these relays to diversity order and coding gain. Our results have clearly shown that these relays have a negligible contribution to these performance metrics. However, other papers have shown that partial--cooperative relays have many advantages, which include energy efficiency, low transmission delay, as well as the possibility to avoid dedicated network elements for data relaying \cite{Zorzi_Phoenix}. Furthermore, partial--cooperative relays are always present in the general NC framework \cite{Ahlswede}, \cite{Koetter}. As a consequence, our framework has highlighted important performance tradeoffs for different types of relays. Full--cooperative relays seem to be more useful for diversity and end--to--end performance. On the other hand, partial--cooperative relays seem to be more useful for energy efficiency, higher rate, and low transmission delay. How these tradeoffs are affected by non--binary modulation and non--binary Galois field is currently being investigated by the authors.
\item[{\bf{C8)}}] So far, the encoding vectors at the relays have been assumed to be fixed a priori, \emph{i.e.}, deterministic NC. This implies that the encoding vectors must be agreed by relays and destination before transmission. A different option is to use random NC, which allows the relays to generate at random the encoding vectors. This solution is suitable for distributed implementations \cite{Ho_TIT}, \cite{Fitzek_ICC2011}. However, random binary NC cannot guarantee that the sources achieve a given diversity order since there is no a priori structure on the network code. In fact, in random binary NC the worst case scenario for the diversity of source $S_t$ happens when none of the encoding vectors at the relays include $S_t$, \emph{i.e.}, when $g_{S_t Rq}  = 0$ for $R_q  \in \mathcal{N}_R^{\left( {FC} \right)}$ and $q=1,2,\ldots,N_R$. In {\bf{C2)}}, it is shown that in this case the sources have diversity order one. Since $S_t$ is network--coded in none of the relays, this implies that its error performance is equal to the error probability of a single--hop transmission multiplied by the probability that $g_{S_t Rq}  = 0$ for $R_q  \in \mathcal{N}_R^{\left( {FC} \right)}$ and $q=1,2,\ldots,N_R$, \emph{i.e.}, ${{1 \mathord{\left/ {\vphantom {1 {2^{N_R^{\left( {FC} \right)} } }}} \right. \kern-\nulldelimiterspace} {2^{N_R^{\left( {FC} \right)} } }}}$. Thus, for high--SNR, ${\rm{ABEP}}_{S_t }^{\left( {{\rm{SV}}{\rm{, random}}} \right)}  \approx \left( {{1 \mathord{\left/ {\vphantom {1 {2^{N_R^{\left( {FC} \right)} } }}} \right. \kern-\nulldelimiterspace} {2^{N_R^{\left( {FC} \right)} } }}} \right)\left[ {4\chi _{S_t } \sigma _{S_t D}^2 \left( {{{E_m } \mathord{\left/ {\vphantom {{E_m } {N_0 }}} \right. \kern-\nulldelimiterspace} {N_0 }}} \right)} \right]^{ - 1}$. As a result, random binary NC allows the sources to achieve better end--to--end error performance than single--hop transmission. Also, the performance gain increases with ${N_R^{\left( {FC} \right)} }$. However, it does not help increasing the diversity order.
\end{list}

Finally, we would like to emphasize that the conclusions drawn in the present paper depend in many cases on the assumptions of binary modulation and binary NC. The impact of non--binary modulation and non--binary NC on these conclusions is currently being investigated by the authors.
\subsection{Generalization to Non--Binary Modulation and Non--Binary Network Code: Thoughts and Conjectures} \label{NonBinary_Conjectures}
Diversity analysis and related network code design discussed in the previous sections are applicable only to the system setup with binary modulation and binary network codes. In the present paper, these assumptions are retained only for analytical tractability and to provide sound proofs about the achievable diversity for generic network topologies and realistic channel models over all the communications links. In this section, we provide some thoughts (or conjectures) about the impact of the Galois field on the expected diversity order. The departing point of our thoughts moves from the outcomes obtained in the binary case. We have shown that if a proper demodulator at the destination is used, then classical systematic linear block codes can be used as network codes. More specifically, the design of diversity--achieving network codes is equivalent to the design of linear systematic block codes over fully-- interleaved point--to–-point links (see Section \ref{NCdesign_for_Diversity} and \cite{DiRenzo}). In other words, if the demodulator is well designed to account for demodulation errors at the relays, then network codes for distributed diversity can be constructed by assuming ideal (error--free) source--to--relay links. In what follows, we call this demodulator, assuming that it exists for every system setup, genie--aided detector. The two--step demodulator in \cite{DiRenzo} and the C--MRC demodulator in Section \ref{ML_and_CMRC} are two practical examples of detectors for binary modulation and binary network codes. Likewise, the C--MRC demodulator in \cite{Nasri} for single--relay networks is another practical example under the assumption that modulation order and Galois field are the same, as well as that a MDS network code is used \cite[Ch. 11]{MacWilliamsSloane}. Let us now assume that such demodulator exists, regardless of its computational complexity and network CSI needed at the destination to counteract the error propagation problem. Under these assumptions and if modulation order and Galois field are the same, we expect that the results for the binary case can be generalized to the non--binary case. In particular, we conjecture that non--binary linear systematic block codes can be used as diversity--achieving network codes. This implies that existence conditions and code constructions for equal (\emph{e.g.}, MDS code design) and unequal (based on the separation vector) diversity apply. For example, from \cite[Ch. 11, Corollary 7]{MacWilliamsSloane} and \cite{Singleton1964} this would imply that the Singleton bound can be achieved for the design of MDS codes if the size, $q$, of the Galois field of the network code (and, thus, the modulation order) satisfies the inequalities $q \ge N_S  + 1$ and $q \ge N_R^{\left( {FC} \right)}  + 1$ if $N_S \ge 2$ and $N_R^{\left( {FC} \right)} \ge 2$. This conjecture is similar to the results recently obtained in \cite{XiaoDec2010}, \cite{Topakkaya}--\cite{Rebelatto} for the erasure channel model. On the other hand, the analysis of the system setup where Galois field and modulation order are different is more complicated to be conjectured since it would need further assumptions on how modulation and NC are mixed together. Formal proofs of these conjectures, as well as the design of practical demodulators to obtain these diversity gains for realistic source--to--relays links and for an arbitrary number of sources and relays are currently being investigated by the authors.
\section{Numerical and Simulation Results} \label{NumericalResults}
In this section, analytical framework and findings are validated through Monte Carlo simulations. For simplicity, numerical examples for i.i.d. fading channels are described. However, the framework has been verified for various channel conditions. In all analyzed scenarios, we have obtained a good accuracy for high--SNR. An example for i.n.i.d fading channels is shown in Fig. \ref{Fig5_NIID}. The encoding vectors are obtained, for a given network topology and SV, from \cite{VanGils}. As far as the analytical framework is concerned, the ABEP of the sources, ${\rm{ABEP}}_{S_t }^{\left( {{\rm{SV}}} \right)}$, is computed using (\ref{Eq_28}) and the Gauss--Chebyshev quadrature rule in \cite[Sec. 9B.2]{Simon}, \cite[Eq. (10)]{Biglieri} with $\delta$ given in (\ref{Eq_18}). For completeness, ${\rm{ABEP}}_{S_t }^{\left( {{\rm{UB}}} \right)}$ in (\ref{Eq_19}) is shown as well. It is obtained from (\ref{Eq_28}) by neglecting the condition on ${\rm{SV}}^{\left( {S_t } \right)}$.

\begin{figure}[!t]
\centering
\includegraphics [width=\columnwidth] {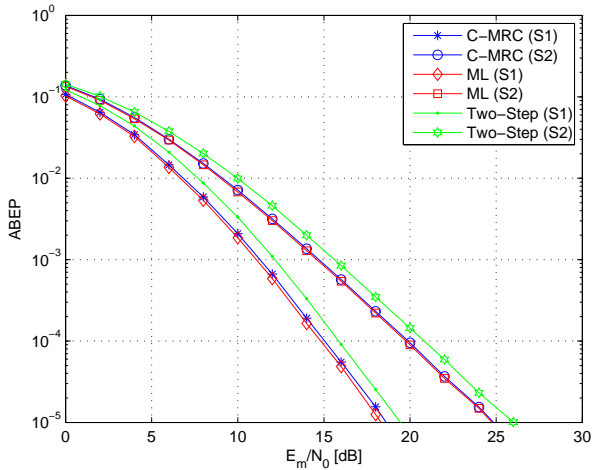}
\caption{ABEP of a 2--source 2--relay network. Both relays are full--cooperative. Performance comparison of three diversity combiners at the destination: i) the ML--optimum demodulator in (\ref{Eq_3}); ii) the two--step demodulator in \cite[Eq. (6)]{DiRenzo}; and iii) the C--MRC in (\ref{Eq_7}). Setup: i) channel fading is i.i.d. with $\sigma _0^2  = 1$; and ii) the encoding vectors are ${\bf{g}}_{R_1 }  = \left[ {1,0} \right]$, ${\bf{g}}_{R_2 }  = \left[ {1,1} \right]$, which yield ${\rm{SV}}^{\left( S_1 \right)}  = 3$ and ${\rm{SV}}^{\left( S_2 \right)}  = 2$. For clarity, only Monte Carlo simulations are shown.} \label{Fig1}
\end{figure}
In Fig. \ref{Fig1}, a network topology with two sources and two relays is studied, and the ABEP of three diversity combiners is compared. It can be observed that the C--MRC provides near--ML performance with reduced signal processing complexity. Furthermore, it provides, with lower implementation complexity, better performance than the two--step demodulator in \cite{DiRenzo}. In fact, the weighting factors of the C--MRC are simpler to be computed. Other network topologies with more sources and relays have been studied, and the same performance trend as in Fig. \ref{Fig1} has been observed in all analyzed case studies.

\begin{figure}[!t]
\centering
\includegraphics [width=\columnwidth] {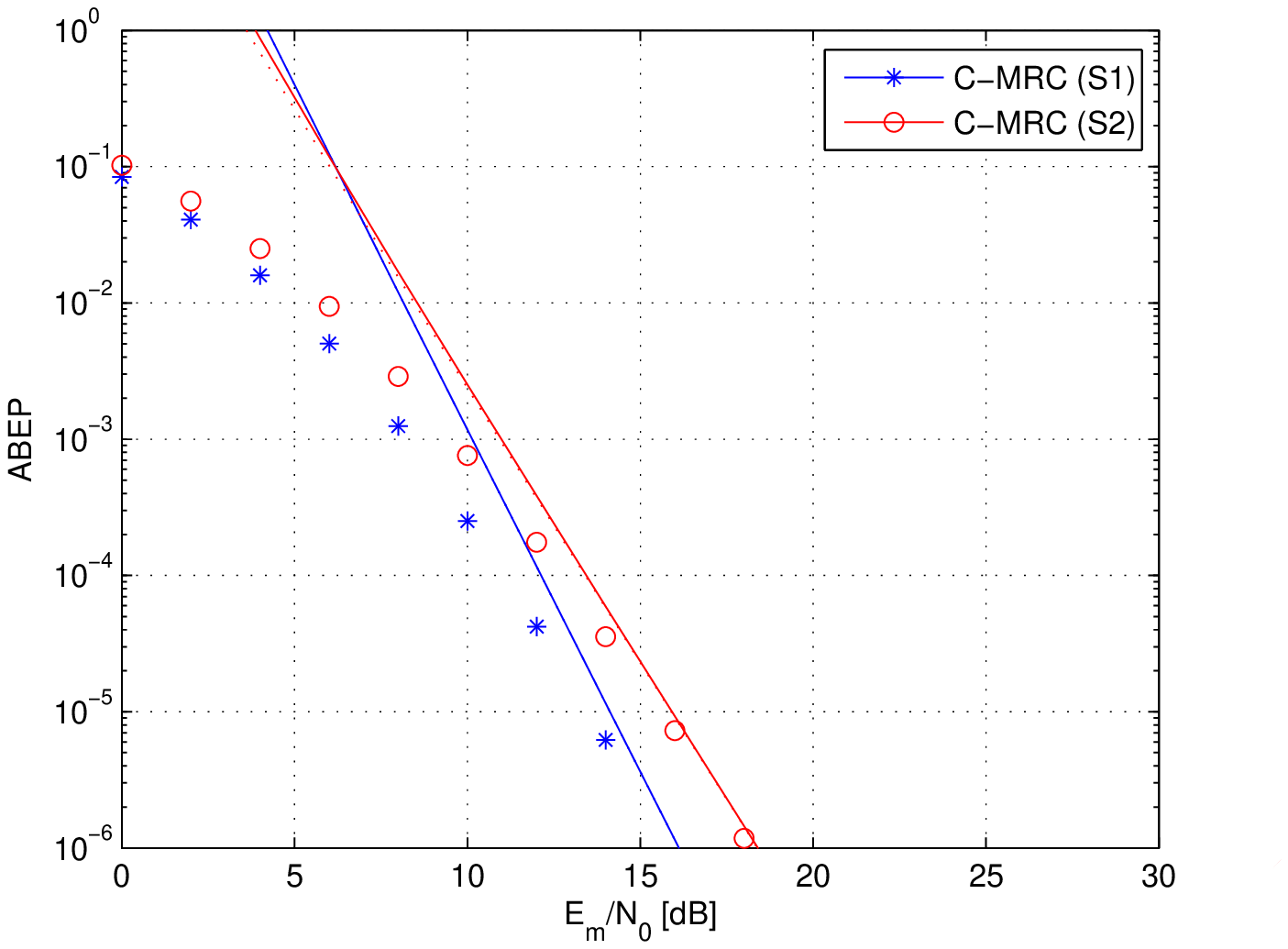}
\caption{ABEP of a 2--source 5--relay network. All relays are full--cooperative. The C--MRC in (\ref{Eq_7}) is used at the destination. Setup: i) channel fading is i.i.d. with $\sigma _0^2  = 1$; and ii) the encoding vectors are ${\bf{g}}_{R_1 }  = \left[ {1,0} \right]$, ${\bf{g}}_{R_2 }  = \left[ {1,0} \right]$, ${\bf{g}}_{R_3 }  = \left[ {1,1} \right]$, ${\bf{g}}_{R_4 }  = \left[ {1,1} \right]$, ${\bf{g}}_{R_5 }  = \left[ {0,1} \right]$, which yield ${\rm{SV}}^{\left( S_1 \right)} = 5$, and ${\rm{SV}}^{\left( S_2 \right)}  = 4$. Markers show Monte Carlo simulations, solid lines show ${\rm{ABEP}}_{S_t }^{\left( {{\rm{UB}}} \right)}$, and dotted lines show ${\rm{ABEP}}_{S_t }^{\left( {{\rm{SV}}} \right)}$.}
\label{Fig3}
\end{figure}

\begin{figure}[!t]
\centering
\includegraphics [width=\columnwidth] {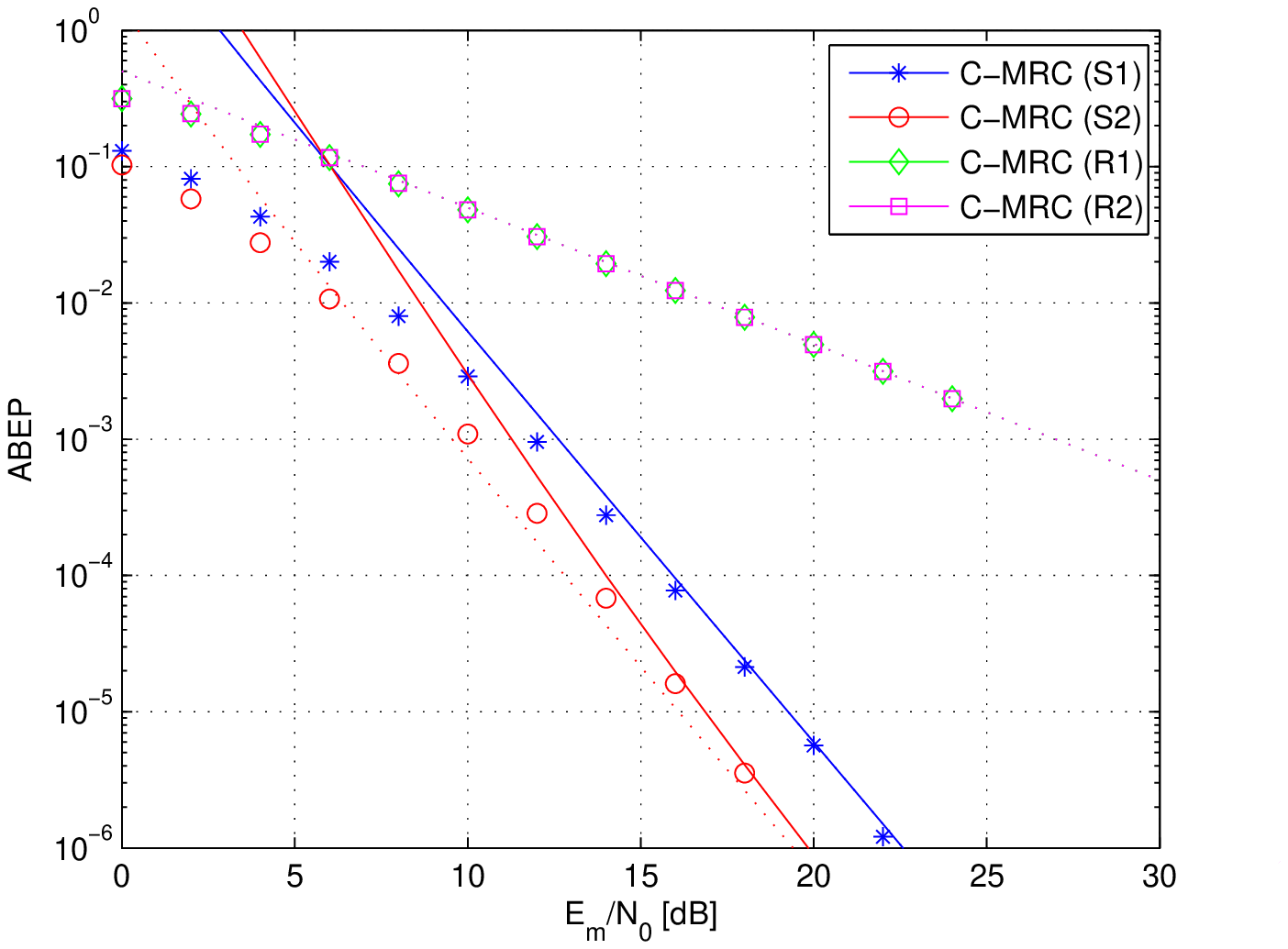}
\caption{ABEP of a 2--source 5--relay network. $R_1$ and $R_2$ are partial--cooperative relays, and $R_3$, $R_4$, and $R_5$ are full--cooperative relays. The C--MRC in (\ref{Eq_7}) is used at the destination. Setup: i) channel fading is i.i.d. with $\sigma _0^2  = 1$; and ii) the encoding vectors are ${\bf{g}}_{R_1 }  = \left[ {1,0} \right]$, ${\bf{g}}_{R_2 }  = \left[ {1,0} \right]$, ${\bf{g}}_{R_3 }  = \left[ {1,1} \right]$, ${\bf{g}}_{R_4 }  = \left[ {1,1} \right]$, ${\bf{g}}_{R_5 }  = \left[ {0,1} \right]$, which yield ${\rm{SV}}^{\left( S_1 \right)} = {\rm{SV}}^{\left( S_2 \right)} = 3$. As for the sources, markers show Monte Carlo simulations, solid lines show ${\rm{ABEP}}_{S_t }^{\left( {{\rm{UB}}} \right)}$, and dotted lines show ${\rm{ABEP}}_{S_t }^{\left( {{\rm{SV}}} \right)}$. More specifically, ${\rm{ABEP}}_{S_t }^{\left( {{\rm{UB}}} \right)}$ and ${\rm{ABEP}}_{S_t }^{\left( {{\rm{SV}}} \right)}$ are computed by considering only the sub--network with sources and full--cooperative relays. As for the partial--cooperative relays, markers show Monte Carlo simulations, and dotted lines are obtained from (\ref{Eq_21}).} \label{Fig4}
\end{figure}

\begin{figure}[!t]
\centering
\includegraphics [width=\columnwidth] {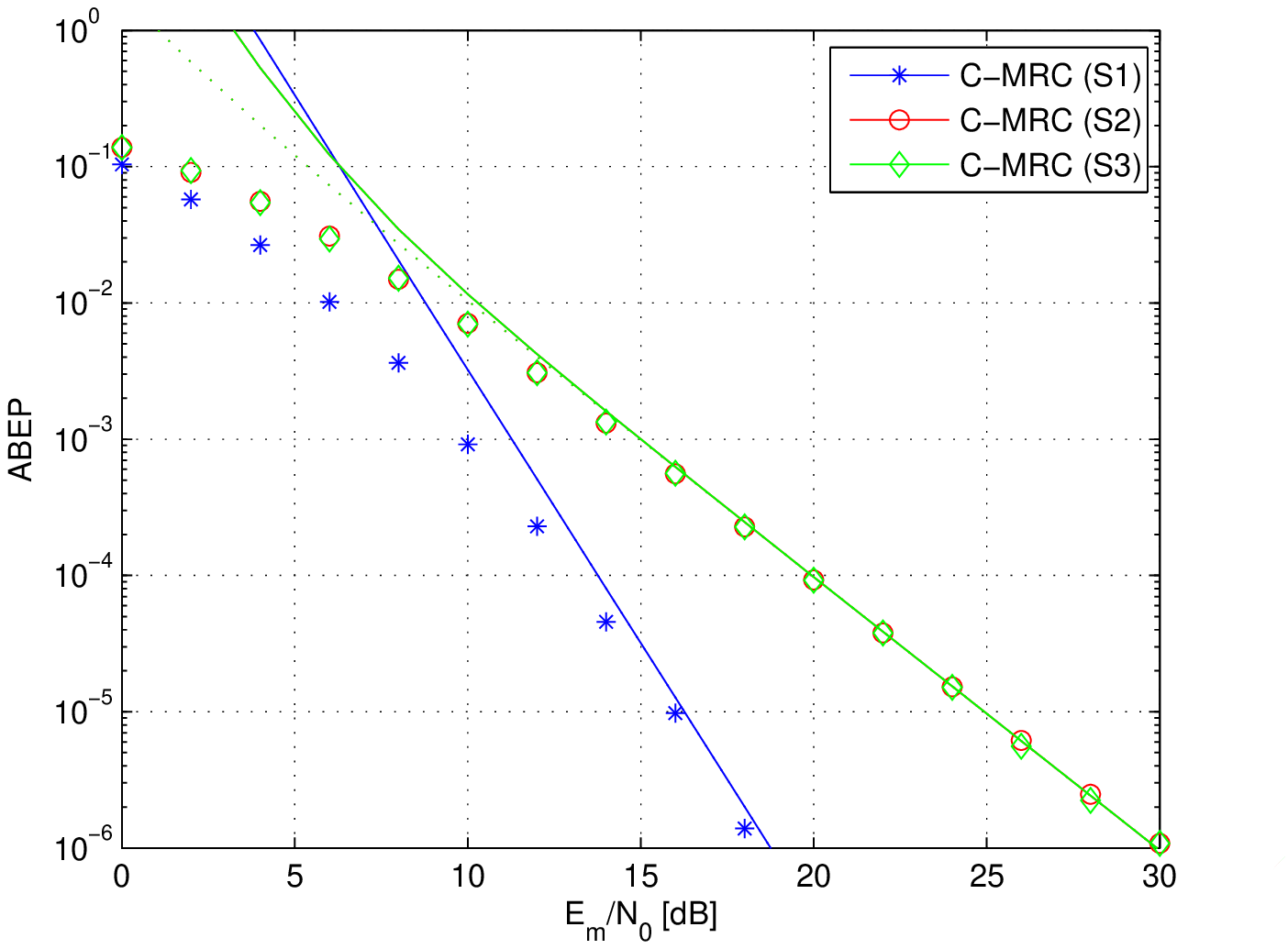}
\caption{ABEP of a 3--source 3--relay network. All relays are full--cooperative. The C--MRC in (\ref{Eq_7}) is used at the destination. Setup: i) channel fading is i.i.d. with $\sigma _0^2  = 1$; and ii) the encoding vectors are ${\bf{g}}_{R_1 }  = \left[ {1,0,1} \right]$, ${\bf{g}}_{R_2 }  = \left[ {1,1,0} \right]$, ${\bf{g}}_{R_3 }  = \left[ {1,0,0} \right]$, which yield ${\rm{SV}}^{\left( S_1 \right)} = 4$, and ${\rm{SV}}^{\left( S_2 \right)} = {\rm{SV}}^{\left( S_3 \right)}  = 2$. Markers show Monte Carlo simulations, solid lines show ${\rm{ABEP}}_{S_t }^{\left( {{\rm{UB}}} \right)}$, and dotted lines show ${\rm{ABEP}}_{S_t }^{\left( {{\rm{SV}}} \right)}$.} \label{Fig5}
\end{figure}

\begin{figure}[!t]
\centering
\includegraphics [width=\columnwidth] {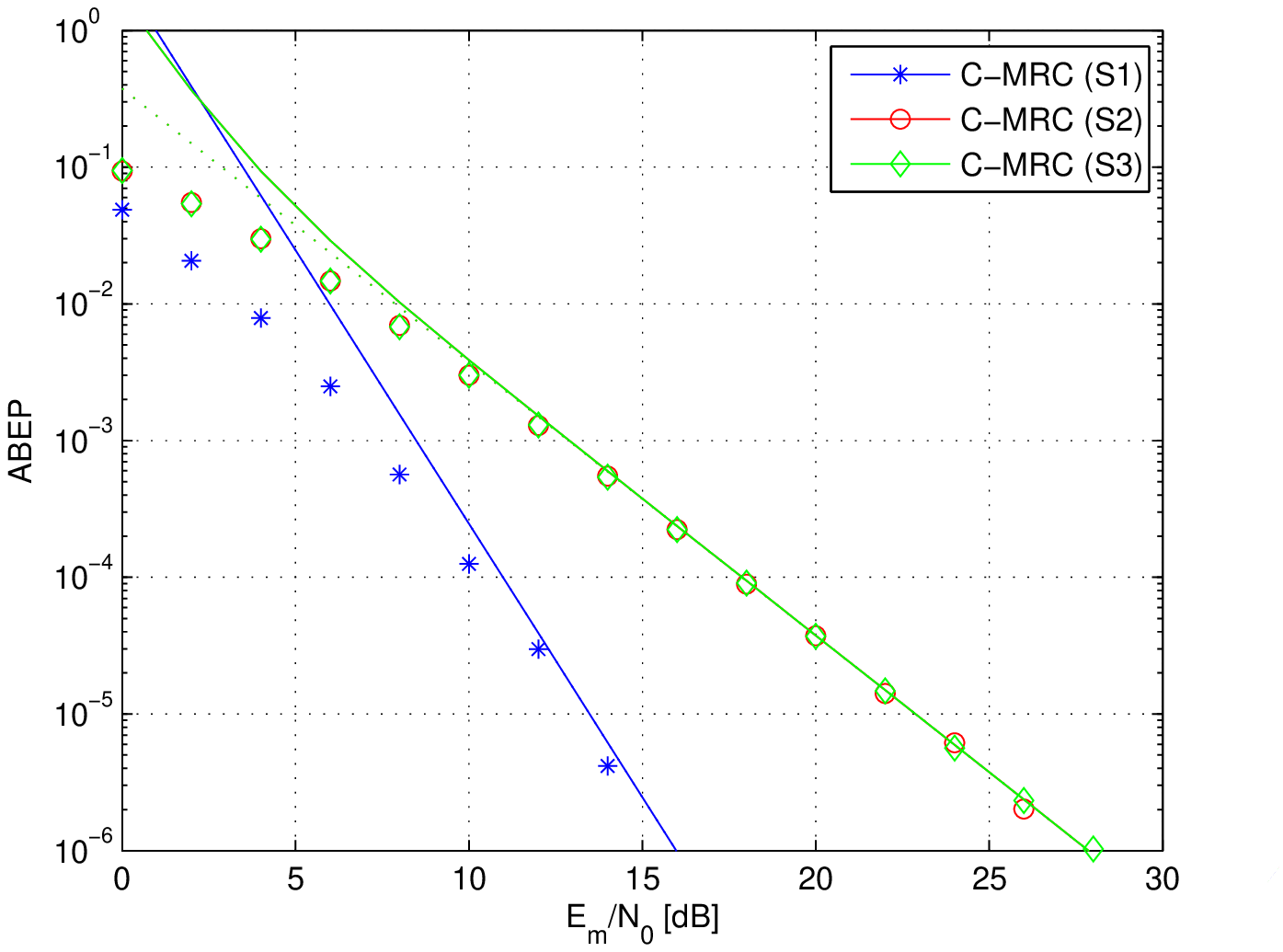}
\caption{ABEP of a 3--source 3--relay network. All relays are full--cooperative. The C--MRC in (\ref{Eq_7}) is used at the destination. Setup: i) channel fading is i.n.i.d. with $\sigma _0^2  = 1$, $\sigma _{S_t D}^2  = \sigma _{R_q D}^2  = \sigma _0^2$, and $\sigma _{S_t R_q }^2  = \left( {1000\sigma _0 } \right)^2$ for $t=1, 2, \ldots, N_S$, $q=1, 2, \ldots, N_R$, \emph{i.e.}, the source--to--relay links are better than the other links; and ii) the encoding vectors are ${\bf{g}}_{R_1 }  = \left[ {1,0,1} \right]$, ${\bf{g}}_{R_2 }  = \left[ {1,1,0} \right]$, ${\bf{g}}_{R_3 }  = \left[ {1,0,0} \right]$, which yield ${\rm{SV}}^{\left( S_1 \right)} = 4$, and ${\rm{SV}}^{\left( S_2 \right)} = {\rm{SV}}^{\left( S_3 \right)}  = 2$. Markers show Monte Carlo simulations, solid lines show ${\rm{ABEP}}_{S_t }^{\left( {{\rm{UB}}} \right)}$, and dotted lines show ${\rm{ABEP}}_{S_t }^{\left( {{\rm{SV}}} \right)}$.} \label{Fig5_NIID}
\end{figure}

\begin{figure}[!t]
\centering
\includegraphics [width=\columnwidth] {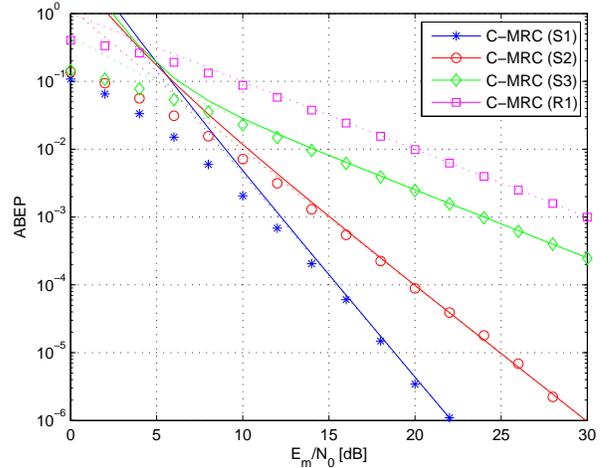}
\caption{ABEP of a 3--source 3--relay network. $R_1$ is a partial--cooperative relay, and $R_2$ and $R_3$ are full--cooperative relays. The C--MRC in (\ref{Eq_7}) is used at the destination. Setup: i) channel fading is i.i.d. with $\sigma _0^2  = 1$; and ii) the encoding vectors are ${\bf{g}}_{R_1 }  = \left[ {1,0,1} \right]$, ${\bf{g}}_{R_2 }  = \left[ {1,1,0} \right]$, ${\bf{g}}_{R_3 }  = \left[ {1,0,0} \right]$, which yield ${\rm{SV}}^{\left( S_1 \right)} = 3$, ${\rm{SV}}^{\left( S_2 \right)} = 2$, and ${\rm{SV}}^{\left( S_3 \right)} = 1$. As for the sources, markers show Monte Carlo simulations, solid lines show ${\rm{ABEP}}_{S_t }^{\left( {{\rm{UB}}} \right)}$, and dotted lines show ${\rm{ABEP}}_{S_t }^{\left( {{\rm{SV}}} \right)}$. More specifically, ${\rm{ABEP}}_{S_t }^{\left( {{\rm{UB}}} \right)}$ and ${\rm{ABEP}}_{S_t }^{\left( {{\rm{SV}}} \right)}$ are computed by considering only the sub--network with sources and full--cooperative relays. As for the partial--cooperative relay, markers show Monte Carlo simulations, and dotted lines are obtained from (\ref{Eq_21}).} \label{Fig6}
\end{figure}

In Fig. \ref{Fig3} and Fig. \ref{Fig4}, a network topology with two sources and five relays is considered, and the analytical frameworks are compared with Monte Carlo simulations. For high--SNR, we observe a good accuracy. It is worth mentioning that, unlike ${\rm{ABEP}}_{S_t }^{\left( {{\rm{UB}}} \right)}$, ${\rm{ABEP}}_{S_t }^{\left( {{\rm{SV}}} \right)}$ is not an upper--bound since some codewords are not considered in the computation. However, ${\rm{ABEP}}_{S_t }^{\left( {{\rm{SV}}} \right)}$ asymptotically (high--SNR) overlaps with ${\rm{ABEP}}_{S_t }^{\left( {{\rm{UB}}} \right)}$, as discussed in Section \ref{Simplified_ABEP_Sources}. The ABEP of the partial--cooperative relays obtained from (\ref{Eq_21}) is shown as well. The figures confirm the correctness of the diversity analysis in Section \ref{Diversity_NCdesign}. More specifically, Fig. \ref{Fig4} shows the diversity order degradation caused when two relays in Fig. \ref{Fig3} are no longer full--cooperative but become partial--cooperative. The ABEP of Fig. \ref{Fig4} coincides with that of the sub--network without partial--cooperative relays. In Fig. \ref{Fig5} and Fig. \ref{Fig6}, the same analysis is conducted for a network topology with three sources and three relays. Conclusions similar to Fig. \ref{Fig3} and Fig. \ref{Fig4} can be drawn. Furthermore, Fig. \ref{Fig5_NIID} shows an example for i.n.i.d. fading by assuming the same network topology and network code as in Fig. \ref{Fig5}. More specifically, the source--to--relay channels are assumed to be of better quality than the other channels. We observe a very good accuracy of our framework. In addition, by comparing Fig. \ref{Fig5} and Fig. \ref{Fig5_NIID} we observe, as expected, that the ABEP in Fig. \ref{Fig5_NIID} is better since error propagation is less pronounced for this setup.

\begin{figure}[!t]
\centering
\includegraphics [width=\columnwidth] {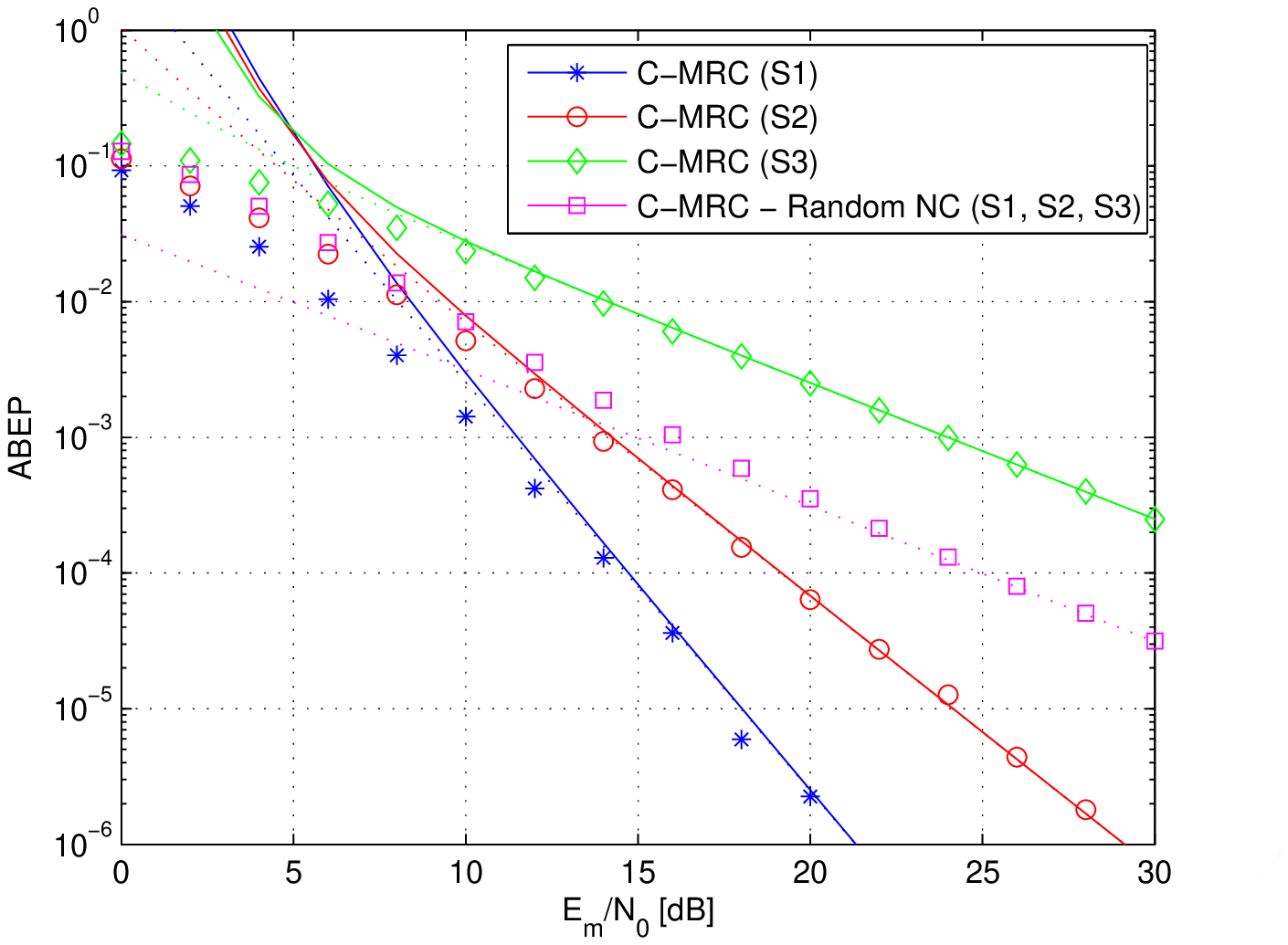}
\caption{ABEP of a 3--source 3--relay network. The relays use repetition--based cooperation. The C--MRC in (\ref{Eq_7}) is used at the destination. Setup: i) channel fading is i.i.d. with $\sigma _0^2  = 1$; and ii) the encoding vectors (only one source is relayed) are ${\bf{g}}_{R_1 }  = \left[ {1,0,0} \right]$, ${\bf{g}}_{R_2 }  = \left[ {1,0,0} \right]$, ${\bf{g}}_{R_3 }  = \left[ {0,1,0} \right]$, which yield ${\rm{SV}}^{\left( S_1 \right)} = 3$, ${\rm{SV}}^{\left( S_2 \right)} = 2$, and ${\rm{SV}}^{\left( S_3 \right)}  = 1$. Markers show Monte Carlo simulations, solid lines show ${\rm{ABEP}}_{S_t }^{\left( {{\rm{UB}}} \right)}$, and dotted lines show ${\rm{ABEP}}_{S_t }^{\left( {{\rm{SV}}} \right)}$. The figure shows also the ABEP when the relays generate at random the binary encoding vectors: random binary NC. In this case, markers show Monte Carlo simulations and dotted lines show ${\rm{ABEP}}_{S_t }^{\left( {{\rm{SV}}{\rm{, random}}} \right)}$ in Section \ref{Insights_Framework}.} \label{Fig7}
\end{figure}

In Fig. \ref{Fig7}, the ABEP of repetition--based cooperative relaying and random binary NC is shown. A network topology with three sources and three relays is investigated. Figure \ref{Fig7} confirms that repetition--based cooperative relaying is a special case of network--coded cooperation, and that our framework can be used to provide good estimates of both diversity order and coding gain. Also, Fig. \ref{Fig7} confirms the first--order diversity of random binary NC, as well as the accuracy of ${\rm{ABEP}}_{S_t }^{\left( {{\rm{SV}}{\rm{, random}}} \right)}$ computed in Section \ref{Insights_Framework}.

\begin{figure}[!t]
\centering
\includegraphics [width=\columnwidth] {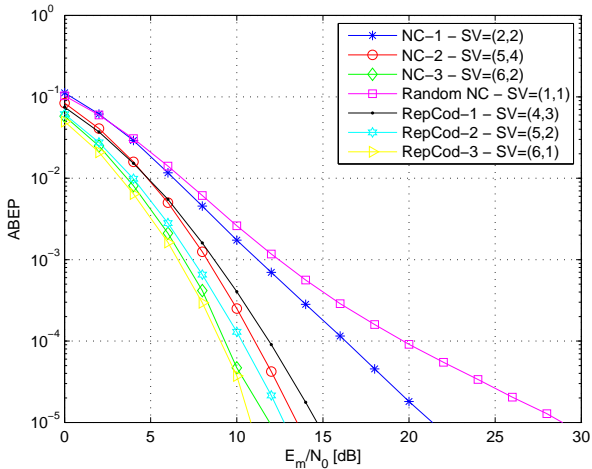}
\caption{ABEP of a 2--source 5--relay network (source $S_1$). All relays are full--cooperative. The C--MRC in (\ref{Eq_7}) is used at the destination. Channel fading is i.i.d. with $\sigma _0^2  = 1$. Various encoding vectors are considered to compare network--coded (NC) and repetition--based (RepCod) cooperative diversity: i) NC--1 with ${\bf{g}}_{R_1 } = {\bf{g}}_{R_2 } = {\bf{g}}_{R_3 } = {\bf{g}}_{R_4 } = {\bf{g}}_{R_5 }  = \left[ {1,0} \right]$; ii) NC--2 with ${\bf{g}}_{R_1 } = {\bf{g}}_{R_2 }  = \left[ {1,0} \right]$, ${\bf{g}}_{R_3 } = {\bf{g}}_{R_4 } = \left[ {1,1} \right]$, ${\bf{g}}_{R_5 }  = \left[ {0,1} \right]$; iii) NC--3 with ${\bf{g}}_{R_1 }  = \left[ {1,1} \right]$, ${\bf{g}}_{R_2 } = {\bf{g}}_{R_3 } = {\bf{g}}_{R_4 } = {\bf{g}}_{R_5 }  = \left[ {1,0} \right]$; iv) RepCod--1 with ${\bf{g}}_{R_1 } = {\bf{g}}_{R_2 } = {\bf{g}}_{R_3 }  = \left[ {1,0} \right]$, ${\bf{g}}_{R_4 } = {\bf{g}}_{R_5 }  = \left[ {0,1} \right]$; v) RepCod--2 with ${\bf{g}}_{R_1 } = {\bf{g}}_{R_2 } = {\bf{g}}_{R_3 }  = {\bf{g}}_{R_4 } = \left[ {1,0} \right]$, ${\bf{g}}_{R_5 }  = \left[ {0,1} \right]$; vi) RepCod--3 with ${\bf{g}}_{R_1 } = {\bf{g}}_{R_2 } = {\bf{g}}_{R_3 }  = {\bf{g}}_{R_4 } = {\bf{g}}_{R_5 } = \left[ {1,0} \right]$; and vii) Random NC with binary encoding vectors generated at random. The separation vectors of both sources are shown in the legend of the figure. For clarity, only Monte Carlo simulations are shown.} \label{Fig8}
\end{figure}

\begin{figure}[!t]
\centering
\includegraphics [width=\columnwidth] {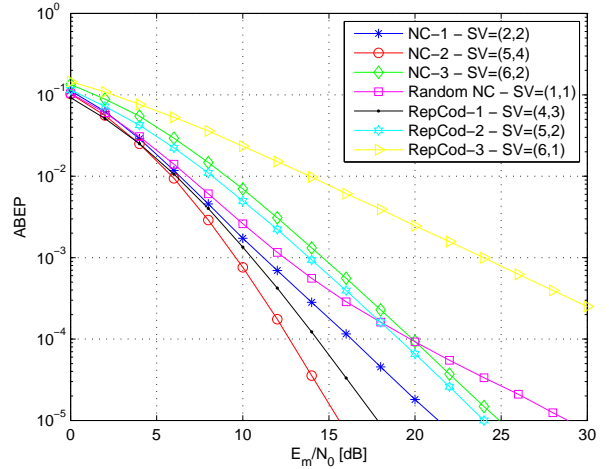}
\caption{ABEP of a 2--source 5--relay network (source $S_2$). All relays are full--cooperative. The C--MRC in (\ref{Eq_7}) is used at the destination. Channel fading is i.i.d. with $\sigma _0^2  = 1$. Various encoding vectors are considered to compare network--coded (NC) and repetition--based (RepCod) cooperative diversity, as described in the caption of Fig. \ref{Fig8}. The separation vectors of both sources are shown in the legend of the figure. For clarity, only Monte Carlo simulations are shown.} \label{Fig9}
\end{figure}

In Fig. \ref{Fig8} and Fig. \ref{Fig9}, the ABEP of network--coded (NC) and repetition--based (RepCod) cooperative diversity is compared by assuming the same network topology (two sources and five relays) and the same number of time--slots. Figure \ref{Fig8} and Fig. \ref{Fig9} show that, by properly choosing the encoding vectors, the transmission protocol described in Section \ref{SystemModel} is flexible enough to accommodate various performance tradeoffs for the sources. For example, the setup denoted by NC--3 outperforms the setup denoted by RepCod--3. In fact, the ABEP of $S_1$ is almost the same and the ABEP of $S_2$ is much better thanks to the higher diversity order. This choice of the encoding vectors is convenient when $S_1$ needs high robustness to multipath fading. On the other hand, the setup denoted by NC--2 appears to be useful when both sources require almost the same ABEP. In fact, compared with NC--3 and RepCod--3, the performance degradation of $S_1$ is tolerable, and the performance gain of $S_2$ is significant. Furthermore, NC--2 outperforms RepCod--1, and it provides a good alternative to RepCod--2. In fact, the big performance gain enjoyed by $S_2$ is obtained with a small performance loss for $S_1$. Finally, it is apparent that random binary NC is, in general, not a good solution, especially in the high--SNR, because of the single--order diversity order.

In addition to the numerical examples depicted in Fig. \ref{Fig1}--\ref{Fig9}, the considerations in Section \ref{Insights_Framework} have been verified through Monte Carlo simulations. An example is NC--1 shown in Fig. \ref{Fig8} and Fig. \ref{Fig9}, which confirms {\bf{C4)}} in Section \ref{Insights_Framework}. The other results are not shown due to space limitations.

Finally, we would like to emphasize that in the present paper many numerical results are given only as illustrative examples, which are aimed at substantiating analytical derivations and main findings. However, many other options are possible for network code design, which depend on the application--specific robustness to multipath fading requested by each source. In particular, it is interesting to comment on how close to the Singleton bound (full--diversity) the network codes considered in this section are. From Section \ref{NCdesign_for_Diversity} and Section \ref{NonBinary_Conjectures}, we know that for binary network codes full--diversity can be achieved by all the sources only for single--relay networks. However, full--diversity may be achieved by \emph{some} sources. For example, the maximum achievable diversity order is equal to three in Fig. \ref{Fig1}, six in Fig. \ref{Fig3}, and four in Fig. \ref{Fig4} and Fig. \ref{Fig5}. Similar comments apply to the other figures, but comments are here omitted for the sake of conciseness. The network code in Fig. \ref{Fig1} is an example where the Singleton bound can be achieved by at least one source. This network code can be considered to be near--optimum since the second source achieves, at least, second--order diversity. The network code in Fig. \ref{Fig3} is an example where the Singleton bound is not achieved by any sources, but they both have a quite high diversity order. From \cite[Table I, row 7]{VanGils}, it follows that other network codes are available to allow one source to achieve full--diversity order equal to six. However, the price to pay is that the diversity order of the other source is only equal to two. Thus, even though the network code in Fig. \ref{Fig3} is not strong enough to attain the Singleton bound for none of the sources, it provides a good diversity--multiplexing tradeoff for both sources. The network code in Fig. \ref{Fig4} offers a trade--off similar to the network code in Fig. \ref{Fig3}, where no source achieves full--diversity but both have diversity order equal to three. One source could achieve full--diversity equal to four by using the network code in \cite[Table I, row 2]{VanGils}. However, the diversity order of the other source is limited to two. Similar to Fig. \ref{Fig1}, the network code in Fig. \ref{Fig5} allows at least one source to obtain full--diversity. In conclusion, our numerical examples clearly show that using NC allows cooperative networks to achieve a wide range of diversity orders, which can be chosen in agreement with the robustness to multipath fading requested by each source. Furthermore, these diversity orders can be achieved with the same number of time--slots, and, thus, the same network rate.
\begin{figure*}[!t]
\setcounter{equation}{30}
\begin{equation}
\label{Eq_AppI__1}
\begin{split}
 \mathcal{F}_t \left( {\left. s \right|{\bf{c}},{\bf{\tilde c}}} \right) & = {\mathbb{E}}_{\gamma _{S_t D} } \left\{ {\exp \left( { - 4\gamma _{S_t D} d_{S_t }^2 s} \right)\exp \left( {4\gamma _{S_t D} d_{S_t }^2 s^2 } \right)} \right\} \\ & \mathop  = \limits^{\left( a \right)} \int\nolimits_0^{ + \infty } {\exp \left( { - 4\zeta d_{S_t }^2 s} \right)\exp \left( {4\zeta d_{S_t }^2 s^2 } \right)\left[ {\left( {{1 \mathord{\left/
 {\vphantom {1 {\bar \gamma _{S_t D} }}} \right.
 \kern-\nulldelimiterspace} {\bar \gamma _{S_t D} }}} \right)\exp \left( { - {\zeta  \mathord{\left/
 {\vphantom {\zeta  {\bar \gamma _{S_t D} }}} \right.
 \kern-\nulldelimiterspace} {\bar \gamma _{S_t D} }}} \right)} \right]d\zeta }  \\
 & \mathop  = \limits^{\left( b \right)} \left[ {1 + 4\bar \gamma _{S_t D} d_{S_t }^2 s\left( {1 - s} \right)} \right]^{ - 1} \mathop  \approx \limits^{\left( c \right)} \left[ {4\bar \gamma _{S_t D} s\left( {1 - s} \right)} \right]^{ - \left| {\tilde b_{S_t }  - b_{S_t } } \right|}  \\
 \end{split}
\end{equation}
\normalsize \hrulefill \vspace*{0pt}
\end{figure*}
\begin{figure*}[!t]
\setcounter{equation}{31}
\begin{equation} \footnotesize
\label{Eq_AppI__2}
\left\{ \begin{array}{l}
 \mathcal{G}\left( {\left. s \right|{\bf{c}},{\bf{\tilde c}},\mathcal{E}_m } \right) = \prod\limits_{q = 1,\;\mu _{q - 1}  = 0}^{N_R } {\mathcal{G}_q \left( {\left. s \right|{\bf{c}},{\bf{\tilde c}},\mathcal{E}_m ,\mu _{q - 1}  = 0} \right)}  \times \prod\limits_{q = 1,\;\mu _{q - 1}  = 1}^{N_R } {\mathcal{G}_q \left( {\left. s \right|{\bf{c}},{\bf{\tilde c}},\mathcal{E}_m ,\mu _{q - 1}  = 1} \right)}  \\
 \mathcal{G}_q \left( {\left. s \right|{\bf{c}},{\bf{\tilde c}},\mathcal{E}_m ,\mu _{q - 1}  = 0} \right) = {\mathbb{E}}_{\left\{ {{\pmb{\gamma }}_{SR_q } ,{{\gamma }}_{R_q D} } \right\}} \left\{ {\exp \left( { - 4\lambda _{R_q } \gamma _{R_q D} d_{R_q }^2 s} \right)\exp \left( {4\lambda _{R_q }^2 \gamma _{R_q D} d_{R_q }^2 s^2 } \right)\left( {1 - \sum\limits_{t = 1}^{N_S } {g_{S_t R_q } Q\left( {\sqrt {2\gamma _{S_t R_q } } } \right)} } \right)} \right\} \\
 \mathcal{G}_q \left( {\left. s \right|{\bf{c}},{\bf{\tilde c}},\mathcal{E}_m ,\mu _{q - 1}  = 1} \right) = {\mathbb{E}}_{\left\{ {{\pmb{\gamma }}_{SR_q } ,{{\gamma }}_{R_q D} } \right\}} \left\{ {\exp \left( {4\lambda _{R_q } \gamma _{R_q D} d_{R_q }^2 s} \right)\exp \left( {4\lambda _{R_q }^2 \gamma _{R_q D} d_{R_q }^2 s^2 } \right)\left( {\sum\limits_{t = 1}^{N_S } {g_{S_t R_q } Q\left( {\sqrt {2\gamma _{S_t R_q } } } \right)} } \right)} \right\} \\
 \end{array} \right.
\end{equation}
\normalsize \hrulefill \vspace*{0pt}
\end{figure*}
\begin{figure*}[!t]
\setcounter{equation}{32}
\begin{equation}
\label{Eq_AppI__3}
\left\{ \begin{array}{l}
 \Upsilon \left( s \right) = {\mathbb{E}}_{\left\{ {{\pmb{\gamma }}_{SR_q } ,\gamma _{R_q D} } \right\}} \left\{ {\exp \left( { - 4\min \left\{ {\gamma _{eqR_q } ,\gamma _{R_q D} } \right\}d_{R_q }^2 s} \right)\exp \left( {4\left( {\min \left\{ {\gamma _{eqR_q } ,\gamma _{R_q D} } \right\}} \right)^2 \gamma _{R_q D}^{ - 1} d_{R_q }^2 s^2 } \right)} \right\} \\
 \Gamma _t \left( s \right) = {\mathbb{E}}_{\left\{ {{\pmb{\gamma }}_{SR_q } ,\gamma _{R_q D} } \right\}} \left\{ {\exp \left( { - 4\min \left\{ {\gamma _{eqR_q } ,\gamma _{R_q D} } \right\}d_{R_q }^{\left( m \right)} s} \right)\exp \left( {4\left( {\min \left\{ {\gamma _{eqR_q } ,\gamma _{R_q D} } \right\}} \right)^2 \gamma _{R_q D}^{ - 1} d_{R_q }^2 s^2 } \right)Q\left( {\sqrt {2\gamma _{S_t R_q } } } \right)} \right\} \\
 \end{array} \right.
\end{equation}
\normalsize \hrulefill \vspace*{0pt}
\end{figure*}
\section{Conclusion} \label{Conclusion}
In this paper, we have contributed to the theoretical understanding of network--coded cooperative diversity protocols. We have shown that these networks generalize repetition--based cooperative relaying protocols, and that they can offer, by properly choosing the encoding vectors at the relays, good design flexibility to accommodate various performance and rate tradeoffs. For a fixed network topology and desired end--to--end diversity order, the encoding vectors can be constructed from linear UEP block codes. By assuming the C--MRC at the destination, an asymptotically--tight analytical framework for arbitrary network topologies and binary encoding vectors is developed, and its achievable diversity is studied analytically. It is shown that the diversity order of each source coincides with the separation vector of the distributed network code. The framework accounts for two classes of relays, \emph{i.e.}, partial-- and full--cooperative, and it is shown that the former class contributes to neither the diversity order nor to the diversity gain of the sources. Thus, partial--cooperative relays are of no use for binary network--coded cooperative diversity protocols. Ongoing research is concerned with the analysis of these networks with non--binary modulations and non--binary encoding vectors, as well as with the end--to--end code design over more general fading channels (\emph{e.g.}, block--fading channels) with channel coding \cite{Marium_CAMAD2012}, \cite{Caire}.
\appendices
\begin{figure*}[!t]
\setcounter{equation}{33}
\begin{equation} \footnotesize
\label{Eq_AppI__4}
\begin{split}
 \Upsilon _1 \left( s \right) & \mathop  = \limits^{\left( a \right)} \int\nolimits_0^{ + \infty } {\int\nolimits_0^{\zeta _{eqRq} } {\exp \left( { - 4\zeta _{R_q D} d_{R_q }^2 s} \right)\exp \left( {4\zeta _{R_q D} d_{R_q }^2 s^2 } \right)\left[ {\left( {{1 \mathord{\left/
 {\vphantom {1 {\bar \gamma _{R_q D} }}} \right.
 \kern-\nulldelimiterspace} {\bar \gamma _{R_q D} }}} \right)\exp \left( { - {{\zeta _{R_q D} } \mathord{\left/
 {\vphantom {{\zeta _{R_q D} } {\bar \gamma _{R_q D} }}} \right.
 \kern-\nulldelimiterspace} {\bar \gamma _{R_q D} }}} \right)} \right]\left[ {\left( {{1 \mathord{\left/
 {\vphantom {1 {\bar \gamma _{eqR_q } }}} \right.
 \kern-\nulldelimiterspace} {\bar \gamma _{eqR_q } }}} \right)\exp \left( { - {{\zeta _{eqRq} } \mathord{\left/
 {\vphantom {{\zeta _{eqRq} } {\bar \gamma _{eqR_q } }}} \right.
 \kern-\nulldelimiterspace} {\bar \gamma _{eqR_q } }}} \right)} \right]d\zeta _{R_q D} d\zeta _{eqRq} } }  \\
 &\mathop  = \limits^{\left( b \right)} \left[ {1 + 4\bar \gamma _{R_q D} d_{R_q }^2 s\left( {1 - s} \right)} \right]^{ - 1} \left[ {1 - {{\bar \gamma _{R_q D} } \mathord{\left/
 {\vphantom {{\bar \gamma _{R_q D} } {\left( {4\bar \gamma _{R_q D} \bar \gamma _{eqR_q } d_{R_q }^2 s\left( {1 - s} \right) + \bar \gamma _{eqR_q }  + \bar \gamma _{R_q D} } \right)}}} \right.
 \kern-\nulldelimiterspace} {\left( {4\bar \gamma _{R_q D} \bar \gamma _{eqR_q } d_{R_q }^2 s\left( {1 - s} \right) + \bar \gamma _{eqR_q }  + \bar \gamma _{R_q D} } \right)}}} \right]\mathop  \approx \limits^{\left( c \right)} \left[ {4\bar \gamma _{R_q D} s\left( {1 - s} \right)} \right]^{ - \left| {\tilde b_{R_q }^{\left( {{\rm{NC}}} \right)}  - b_{R_q }^{\left( {{\rm{NC}}} \right)} } \right|}
 \end{split}
\end{equation}
\normalsize \hrulefill \vspace*{0pt}
\end{figure*}
\begin{figure*}[!t]
\setcounter{equation}{34}
\begin{equation} \footnotesize
\label{Eq_AppI__5}
\begin{split}
 \Upsilon _2 \left( s \right) & \mathop  = \limits^{\left( a_1 \right)} \int\nolimits_0^{ + \infty } {\Upsilon _2 \left( {s;\zeta _{eqRq} } \right)\exp \left( { - 4\zeta _{eqRq} d_{R_q }^2 s} \right)\left[ {\left( {{1 \mathord{\left/
 {\vphantom {1 {\bar \gamma _{eqR_q } }}} \right.
 \kern-\nulldelimiterspace} {\bar \gamma _{eqR_q } }}} \right)\exp \left( { - {{\zeta _{eqRq} } \mathord{\left/
 {\vphantom {{\zeta _{eqRq} } {\bar \gamma _{eqR_q } }}} \right.
 \kern-\nulldelimiterspace} {\bar \gamma _{eqR_q } }}} \right)} \right]d\zeta _{eqRq} }  \\
 & \mathop  \approx \limits^{\left( a_2 \right)} \int\nolimits_0^{ + \infty } {\left\{ {\int\nolimits_{\zeta _{eqRq} }^{ + \infty } {\left( {{1 \mathord{\left/
 {\vphantom {1 {\bar \gamma _{R_q D} }}} \right.
 \kern-\nulldelimiterspace} {\bar \gamma _{R_q D} }}} \right)\exp \left( { - {{\zeta _{R_q D} } \mathord{\left/
 {\vphantom {{\zeta _{R_q D} } {\bar \gamma _{R_q D} }}} \right.
 \kern-\nulldelimiterspace} {\bar \gamma _{R_q D} }}} \right)d\zeta _{R_q D} } } \right\}\exp \left( { - 4\zeta _{eqRq} d_{R_q }^2 s} \right)\left( {{1 \mathord{\left/
 {\vphantom {1 {\bar \gamma _{eqR_q } }}} \right.
 \kern-\nulldelimiterspace} {\bar \gamma _{eqR_q } }}} \right)\exp \left( { - {{\zeta _{eqRq} } \mathord{\left/
 {\vphantom {{\zeta _{eqRq} } {\bar \gamma _{eqR_q } }}} \right.
 \kern-\nulldelimiterspace} {\bar \gamma _{eqR_q } }}} \right)d\zeta _{eqRq} }  \\
 & \mathop  = \limits^{\left( b \right)} \left[ {1 + 4\bar \gamma _{eqR_q } d_{R_q }^2 s + \left( {{{\bar \gamma _{eqR_q } } \mathord{\left/
 {\vphantom {{\bar \gamma _{eqR_q } } {\bar \gamma _{R_q D} }}} \right.
 \kern-\nulldelimiterspace} {\bar \gamma _{R_q D} }}} \right)} \right]^{ - 1} \mathop  \approx \limits^{\left( c \right)} \left[ {4\bar \gamma _{eqR_q }s} \right]^{ - \left| {\tilde b_{R_q }^{\left( {{\rm{NC}}} \right)}  - b_{R_q }^{\left( {{\rm{NC}}} \right)} } \right|}  \\
 \end{split}
\end{equation}
\normalsize \hrulefill \vspace*{0pt}
\end{figure*}
\begin{figure*}[!t]
\setcounter{equation}{35}
\begin{equation}
\label{Eq_AppI__6}
\Upsilon _2 \left( {s;\zeta _{eqRq} } \right) = \int\nolimits_{\zeta _{eqRq} }^{ + \infty } {\exp \left[ {4\left( {{{\zeta _{eqRq}^2 } \mathord{\left/ {\vphantom {{\zeta _{eqRq}^2 } {\zeta _{R_q D} }}} \right. \kern-\nulldelimiterspace} {\zeta _{R_q D} }}} \right)d_{R_q }^2 s^2 } \right]\left( {{1 \mathord{\left/ {\vphantom {1 {\bar \gamma _{R_q D} }}} \right. \kern-\nulldelimiterspace} {\bar \gamma _{R_q D} }}} \right)\exp \left( { - {{\zeta _{R_q D} } \mathord{\left/ {\vphantom {{\zeta _{R_q D} } {\bar \gamma _{R_q D} }}} \right. \kern-\nulldelimiterspace} {\bar \gamma _{R_q D} }}} \right)d\zeta _{R_q D} }  = \int\nolimits_0^{{1 \mathord{\left/ {\vphantom {1 {\zeta _{eqRq} }}} \right. \kern-\nulldelimiterspace} {\zeta _{eqRq} }}} {g\left( x \right)f\left( x \right)dx}
\end{equation}
\normalsize \hrulefill \vspace*{0pt}
\end{figure*}
\begin{figure*}[!t]
\setcounter{equation}{36}
\begin{equation} \footnotesize
\label{Eq_AppI__7}
\begin{split}
 \Gamma _t \left( {s; \theta } \right) &= {\mathbb{E}}_{\left\{ {{\pmb{\gamma }}_{SR_q } ,\gamma _{R_q D} } \right\}} \left\{ {\exp \left( { - 4\min \left\{ {\gamma _{eqR_q } ,\gamma _{R_q D} } \right\}d_{R_q }^{\left( m \right)} s} \right)\exp \left( {4\left( {\min \left\{ {\gamma _{eqR_q } ,\gamma _{R_q D} } \right\}} \right)^2 \gamma _{R_q D}^{ - 1} d_{R_q }^2 s^2 } \right)\exp \left( { - {{\gamma _{S_t R_q } } \mathord{\left/
 {\vphantom {{\gamma _{S_t R_q } } {\sin ^2 \left( \theta  \right)}}} \right.
 \kern-\nulldelimiterspace} {\sin ^2 \left( \theta  \right)}}} \right)} \right\} \\
 &\mathop  = \limits^{\left( a \right)} {\mathbb{E}}_{\left\{ {{\pmb{\gamma }}_{SR_q } ,\gamma _{R_q D} } \right\}} \left\{ {\exp \left( { - 4\min \left\{ {\gamma _{S_t R_q } ,\gamma _{eqR_q }^{\left( {\backslash t} \right)} ,\gamma _{R_q D} } \right\}d_{R_q }^{\left( m \right)} s} \right)\exp \left( {4\left( {\min \left\{ {\gamma _{S_t R_q } ,\gamma _{eqR_q }^{\left( {\backslash t} \right)} ,\gamma _{R_q D} } \right\}} \right)^2 \gamma _{R_q D}^{ - 1} d_{R_q }^2 s^2 } \right)  \exp \left(- \frac{\gamma _{S_t R_q }}{{\sin ^2 \left( \theta  \right)}} \right)    }       \right\} \\
 &\mathop  = \limits^{\left( b \right)} \Gamma _t^{\left( 1 \right)} \left( {s;\theta } \right) + \Gamma _t^{\left( 2 \right)} \left( {s;\theta } \right) + \Gamma _t^{\left( 3 \right)} \left( {s;\theta } \right) \\
 \end{split}
\end{equation}
\normalsize \hrulefill \vspace*{0pt}
\end{figure*}
\section{High--SNR Computation of $\mathcal{F}_t \left( {\left. \cdot \right| \cdot } \right)$ and $\mathcal{G} \left( {\left. \cdot \right|\cdot } \right)$ in (\ref{Eq_13})} \label{App_APEP}
For high--SNR, $\mathcal{F}_t \left( {\left. \cdot \right| \cdot } \right)$ in (\ref{Eq_13}) can be computed in closed--form as shown in (\ref{Eq_AppI__1}) at the top of this page, where: i) in $\mathop  = \limits^{\left( a \right)}$, it is taken into account that ${\gamma _{S_t D} }$ is an exponential RV with parameter $\bar \gamma _{S_t R_q }$; ii) $\mathop  = \limits^{\left( b \right)}$ follows from \cite[Eq. (3.310)]{GradshteynRyzhik}; and iii) $\mathop  = \limits^{\left( c \right)}$ holds for high--SNR, \emph{i.e.}, $\bar \gamma _{S_t D}  \gg 1$. This concludes the proof of (\ref{Eq_14}).

In (\ref{Eq_13}), let us consider the generic event $\mathcal{E}_m$ for ${0 < m < 2^{N_R }  - 1}$. The cases $m=0$ and $m=2^{N_R }  - 1$ can be obtained with similar analytical steps. Then, $\mathcal{G} \left( {\left. \cdot \right|\cdot } \right)$ in (\ref{Eq_13}) can be re--written as shown in (\ref{Eq_AppI__2}) at the top of this page, where: i) ${\pmb{\gamma }}_{SR_q }$ is a short--hand to denote all channels from the $N_S$ sources to $R_q$; and ii) it is taken into account that $d_{R_q }^{\left( m \right)}  = d_{R_q }^2$ if ${\mu _{q - 1}  = 0}$ and $d_{R_q }^{\left( m \right)}  = - d_{R_q }^2$ if ${\mu _{q - 1}  = 1}$. To compute (\ref{Eq_AppI__2}) in closed--form, the expectations $\Upsilon \left( \cdot \right)$ and $\Gamma \left( s \right) = \sum\nolimits_{t = 1}^{N_S } {g_{S_t R_q } \Gamma _t \left( s \right)}$ shown in (\ref{Eq_AppI__3}) at the top of this page need to be studied. When computing ${\Gamma _t \left( \cdot \right)}$, we assume ${g_{S_t R_q } } = 1$ as only in this case it contributes to $\Gamma \left( \cdot \right)$.

$\Upsilon \left( \cdot \right)$ can be computed in closed--form as $\Upsilon \left( s \right) = \Upsilon _1 \left( s \right) + \Upsilon _2 \left( s \right)$. $\Upsilon _1 \left( \cdot \right) $ is shown in (\ref{Eq_AppI__4}) at the top of the next page, where: i) in $\mathop  = \limits^{\left( a \right)}$, it is taken into account that ${\gamma _{R_q D} }$ and ${\gamma _{eq R_q} }$ are exponential RVs with parameters $\bar \gamma _{R_q D }$ and $\bar \gamma _{eq R_q}$; ii) $\mathop  = \limits^{\left( b \right)}$ follows from \cite[Eq. (3.310)]{GradshteynRyzhik}; and iii) $\mathop  = \limits^{\left( c \right)}$ holds for high--SNR, \emph{i.e.}, ${{E_m } \mathord{\left/ {\vphantom {{E_m } {N_0 }}} \right. \kern-\nulldelimiterspace} {N_0 }} \gg 1$. $\Upsilon _2 \left( \cdot \right) $ is shown in (\ref{Eq_AppI__5}) at the top of the next page, where $\mathop  = \limits^{\left( b \right)}$ and $\mathop  = \limits^{\left( c \right)}$ follow from considerations similar to (\ref{Eq_AppI__4}). On the other hand, $\mathop  = \limits^{\left( a_1 \right)}$ and $\mathop  = \limits^{\left( a_2 \right)}$ deserve further clarifications. In $\mathop  = \limits^{\left( a_1 \right)}$, we define $f\left( x \right) = \left( {\bar \gamma _{R_q D} x^2 } \right)^{ - 1} \exp \left[ { - {1 \mathord{\left/ {\vphantom {1 {\left( {x\bar \gamma _{R_q D} } \right)}}} \right. \kern-\nulldelimiterspace} {\left( {x\bar \gamma _{R_q D} } \right)}}} \right]$, $g\left( x \right) = \exp \left( {4\zeta _{eqRq}^2 d_{R_q }^2 s^2 x} \right)$, and $\Upsilon _2 \left( {\cdot;\cdot } \right)$ is given in (\ref{Eq_AppI__6}) shown at the top of this page. From these definitions, the high--SNR approximation in $\mathop  = \limits^{\left( a_2 \right)}$ follows by noticing, with arguments similar to \cite[Fig. 1]{Giannakis}, that the behavior of $g\left( x \right)$ around $x \to 0$ mainly determines the high--SNR behavior of $\Upsilon _2 \left( \cdot \right)$. With this in mind, $\mathop  = \limits^{\left( a_2 \right)}$ is obtained by replacing $g\left( \cdot \right)$ with its first--order Taylor expansion, \emph{i.e.}, $g\left( x \to 0 \right) \approx 1$.

Using the Craig representation of the Q--function \cite[Eq. (4.2)]{Simon}, we have $\Gamma _t \left( s \right) = \int\nolimits_0^{{\pi  \mathord{\left/{\vphantom {\pi  2}} \right. \kern-\nulldelimiterspace} 2}} {\Gamma _t \left( {s;\theta } \right)d\theta }$, where $\Gamma _t \left( {\cdot; \cdot } \right)$ is defined in (\ref{Eq_AppI__7}) at the top of this page, and: i) $\mathop  = \limits^{\left( a \right)}$ is obtained by introducing $\gamma _{eqR_q }^{\left( {\backslash t} \right)}  = \min _{\tau  \ne t = 1,2, \ldots ,N_S } \left\{ {g_{S_\tau  R_q }^{ - 1} \gamma _{S_\tau  R_q } } \right\}$, which is an exponential RV with parameter $\bar \gamma _{eqR_q }^{\left( {\backslash t} \right)}  = \left( {\sum\nolimits_{\tau  \ne t = 1}^{N_S } {g_{S_\tau  R_q } \bar \gamma _{S_\tau  R_q }^{ - 1} } } \right)^{ - 1}$; and ii) $\mathop  = \limits^{\left( b \right)}$ follows by decomposing $\Gamma _t \left( {\cdot;\cdot } \right)$ into the summation of three terms, \emph{i.e.}, $\Gamma _t^{\left( 1 \right)} \left( {\cdot;\cdot } \right)$ for $\min \left\{ {\gamma _{S_t R_q }, \gamma _{eqR_q }^{\left( {\backslash t} \right)} ,\gamma _{R_q D} } \right\} = \gamma _{S_t R_q }$, $\Gamma _t^{\left( 2 \right)} \left( {\cdot;\cdot } \right)$ for $\min \left\{ {\gamma _{S_t R_q }, \gamma _{eqR_q }^{\left( {\backslash t} \right)} ,\gamma _{R_q D} } \right\} = \gamma _{eqR_q }^{\left( {\backslash t} \right)}$, and $\Gamma _t^{\left( 3 \right)} \left( {\cdot;\cdot } \right)$ for $\min \left\{ {\gamma _{S_t R_q }, \gamma _{eqR_q }^{\left( {\backslash t} \right)} ,\gamma _{R_q D} } \right\} = \gamma _{R_q D}$. Similar to the computation of $\Upsilon \left( \cdot \right)$, each of the three summands in $\mathop  = \limits^{\left( b \right)}$ can be decomposed into the summation of two terms, and each of them can be computed in closed--form and for high--SNR by resorting the same analytical steps and approximations used to compute $\Upsilon _1 \left( \cdot \right)$ and $\Upsilon _2 \left( \cdot \right)$. The details of the derivation are here omitted due to space limitations. In particular, our analysis has shown that, for high--SNR, $\Gamma _t^{\left( 2 \right)} \left( {\cdot;\cdot } \right)$ and $\Gamma _t^{\left( 3 \right)} \left( {\cdot;\cdot } \right)$ decay as $\left( {{{E_m } \mathord{\left/ {\vphantom {{E_m } {N_0 }}} \right. \kern-\nulldelimiterspace} {N_0 }}} \right)^{ - 2}$, while $\Gamma _t^{\left( 1 \right)} \left( {\cdot;\cdot } \right)$ decays as $\left( {{{E_m } \mathord{\left/ {\vphantom {{E_m } {N_0 }}} \right. \kern-\nulldelimiterspace} {N_0 }}} \right)^{ - 1}$. Thus, $\Gamma _t^{\left( 1 \right)} \left( {\cdot;\cdot } \right)$ is the dominant term for high--SNR. In conclusion, we have:
\setcounter{equation}{37}
\begin{equation}
\label{Eq_AppI__8}
\Gamma _t \left( {s;\theta } \right) \approx \Gamma _t^{\left( 1 \right)} \left( {s;\theta } \right) \approx \left[ {\bar \gamma _{S_t R_q } \left( {{1 \mathord{\left/
 {\vphantom {1 {\sin ^2 \left( \theta  \right)}}} \right.
 \kern-\nulldelimiterspace} {\sin ^2 \left( \theta  \right)}} + d_{R_q }^{\left( m \right)} s} \right)} \right]^{-1}
\end{equation}

Finally, by inserting (\ref{Eq_AppI__3})--(\ref{Eq_AppI__5}), (\ref{Eq_AppI__7}), and (\ref{Eq_AppI__8}) in (\ref{Eq_AppI__2}), $\mathcal{G}\left( {\left. \cdot \right| \cdot} \right)$ in (\ref{Eq_13}) can be obtained after some algebra and by taking into account that $\prod\nolimits_{q = 1,\;\mu _{q - 1}  = 0}^{N_R } {\left[ {1 - \left( {4\bar \gamma _{eqR_q } } \right)^{ - 1} } \right]}  \to 1$ for high--SNR. This concludes the proof.
\begin{figure*}[!t]
\setcounter{equation}{38}
\begin{equation}
\label{Eq_AppII__1}
{\rm{APEP}}\left( {{\bf{0}} \to {\bf{\tilde c}}} \right) = \left( {2\pi j} \right)^{ - 1} \int\nolimits_{\delta  - j\infty }^{\delta  + j\infty } {\mathcal{\bar M}_1 \left( {\left. s \right|{\bf{\tilde c}}} \right)\mathcal{\bar M}_2 \left( {\left. s \right|{\bf{\tilde c}},\mathcal{E}_0 } \right)s^{ - 1} ds}  + \left( {2\pi j} \right)^{ - 1} \int\nolimits_{\delta  - j\infty }^{\delta  + j\infty } {\mathcal{\bar M}_1 \left( {\left. s \right|{\bf{\tilde c}}} \right)\mathcal{\bar M}_3 \left( {\left. s \right|{\bf{\tilde c}},\mathcal{E}_1 } \right)s^{ - 1} ds}
\end{equation}
\normalsize \hrulefill \vspace*{0pt}
\end{figure*}
\begin{figure*}[!t]
\setcounter{equation}{42}
\begin{equation}
\label{Eq_AppII__5}
\left\{ \begin{array}{l}
 \begin{split} & \left( {{4 \mathord{\left/
 {\vphantom {4 \pi }} \right.
 \kern-\nulldelimiterspace} \pi }} \right)\int\nolimits_0^{{\pi  \mathord{\left/
 {\vphantom {\pi  2}} \right.
 \kern-\nulldelimiterspace} 2}} {\left[ {\left( {2\pi j} \right)^{ - 1} \int\nolimits_{\delta  - j\infty }^{\delta  + j\infty } {s^{ - 2} \left( {1 - s} \right)\left( {\sin ^{ - 2} \left( \theta  \right) + 4s} \right)^{ - 1} ds} } \right]d\theta } \\ & \hspace{5cm} \mathop  = \limits^{\left( a \right)} \left( {{4 \mathord{\left/
 {\vphantom {4 \pi }} \right.
 \kern-\nulldelimiterspace} \pi }} \right)\int\nolimits_0^{{\pi  \mathord{\left/
 {\vphantom {\pi  2}} \right.
 \kern-\nulldelimiterspace} 2}} {\left( {4 + \sin ^{ - 2} \left( \theta  \right)} \right)^{ - 1} d\theta }  = {{\left( {5 - \sqrt 5 } \right)} \mathord{\left/
 {\vphantom {{\left( {5 - \sqrt 5 } \right)} {10}}} \right.
 \kern-\nulldelimiterspace} {10}} \end{split} \\
 \begin{split} & \left( {{4 \mathord{\left/
 {\vphantom {4 \pi }} \right.
 \kern-\nulldelimiterspace} \pi }} \right)\int\nolimits_0^{{\pi  \mathord{\left/
 {\vphantom {\pi  2}} \right.
 \kern-\nulldelimiterspace} 2}} {\left[ {\left( {2\pi j} \right)^{ - 1} \int\nolimits_{\delta  - j\infty }^{\delta  + j\infty } {s^{ - 2} \left( {1 - s} \right)\left( {\sin ^{ - 2} \left( \theta  \right) - 4s} \right)^{ - 1} ds} } \right]d\theta } \\ & \hspace{5cm} \mathop  = \limits^{\left( b \right)} \left( {{4 \mathord{\left/
 {\vphantom {4 \pi }} \right.
 \kern-\nulldelimiterspace} \pi }} \right)\int\nolimits_0^{{\pi  \mathord{\left/
 {\vphantom {\pi  2}} \right.
 \kern-\nulldelimiterspace} 2}} {\left[ {\left( { - 4 + \sin ^{ - 2} \left( \theta  \right)} \right)^{ - 1}  + 16\sin ^4 \left( \theta  \right)\left( {4 - \sin ^{ - 2} \left( \theta  \right)} \right)^{ - 1} } \right]d\theta }  = 1 \end{split} \\
 \end{array} \right.
\end{equation}
\normalsize \hrulefill \vspace*{0pt}
\end{figure*}
\begin{figure*}[!t]
\setcounter{equation}{43}
\begin{equation}
\label{Eq_AppII__6}
{\rm{ABEP}}_{S_t }^{\left( {{\rm{SV}}} \right)}  = \frac{1}{{\bar \gamma _{S_t D} }}\left[ {\frac{1}{{16}}\left( {\frac{1}{{\bar \gamma _{R_1 D} }} + \sum\limits_{\tau  \ne t = 1}^{N_S } {\frac{1}{{\bar \gamma _{S_\tau  D} }}} } \right) + \frac{{45 + \sqrt 5 }}{{160}}\sum\limits_{\tau  = 1}^{N_S } {\frac{1}{{\bar \gamma _{S_\tau  R_1 } }}} } \right]
\end{equation}
\normalsize \hrulefill \vspace*{0pt}
\end{figure*}
\begin{figure*}[!t]
\setcounter{equation}{44}
\begin{equation}
\label{Eq_AppIII__1}
{\rm{APEP}}\left( {{\bf{0}} \to {\bf{\tilde c}}} \right) = \sum\limits_{m = 0}^{2^{N_R }  - 1} {\left( {\left( {2\pi j} \right)^{ - 1} \int\nolimits_{\delta  - j\infty }^{\delta  + j\infty } {\mathcal{\bar M}_1 \left( {\left. s \right|{\bf{\tilde c}}} \right)\mathcal{\bar M}_2 \left( {\left. s \right|{\bf{\tilde c}},\mathcal{E}_m } \right)\mathcal{\bar M}_3 \left( {\left. s \right|{\bf{\tilde c}},\mathcal{E}_m } \right)s^{ - 1} ds} } \right)}
\end{equation}
\normalsize \hrulefill \vspace*{0pt}
\end{figure*}
\section{High--SNR ABEP of Single--Relay Networks} \label{App_APEP_SingleRelay}
Let us consider the same system model as in \cite{Nasri} with $N_S$ sources and one full--cooperative relay ($N_R=1$) with encoding vector ${\bf{g}}_{R_1 }  = \left[ {1,1, \ldots ,1} \right]$. By direct inspection of the codebook and from Section \ref{Diversity_NCdesign}, we have $G_d^{\left( {S_t } \right)}  = {\rm{SV}}^{\left( {S_t } \right)}  = 2$ for $t=1,2,\ldots,N_S$. Since there is just one relay, by using $\mathcal{G}\left( {\left. \cdot \right| \cdot} \right)$  in (\ref{Eq_AppI__2}), the ${\rm{APEP}}\left( {{\bf{0}} \to {\bf{\tilde c}}} \right)$ in (\ref{Eq_13}) simplifies as shown in (\ref{Eq_AppII__1}) at the top of the next page, where: i) $\mathcal{\bar M}_1 \left( {\left. s \right|{\bf{\tilde c}}} \right) = \prod\nolimits_{t = 1}^{N_S } {\mathcal{F}_t \left( {\left. s \right|{\bf{\tilde c}}} \right)}$; ii) $\mathcal{\bar M}_2 \left( {\left. s \right|{\bf{\tilde c}},\mathcal{E}_0 } \right) = \mathcal{G}_1 \left( {\left. s \right|{\bf{\tilde c}},\mathcal{E}_0 ,\mu _0  = 0} \right)$; and iii) $\mathcal{\bar M}_3 \left( {\left. s \right|{\bf{\tilde c}},\mathcal{E}_1 } \right) = \mathcal{G}_1 \left( {\left. s \right|{\bf{\tilde c}},\mathcal{E}_1 ,\mu _0  = 1} \right)$.

From (\ref{Eq_25}), ${\rm{ABEP}}_{S_t }^{\left( {{\rm{SV}}} \right)}$ can be explicitly written as follows:
\setcounter{equation}{39}
\begin{equation}
\label{Eq_AppII__2}
{\rm{ABEP}}_{S_t }^{\left( {{\rm{SV}}} \right)}  = \sum\limits_{\tau  = 1}^{N_S  + 1} {{\rm{APEP}}\left( {{\bf{0}} \to {\bf{\tilde c}}^{\left( {t,\tau } \right)} } \right)}
\end{equation}
\noindent where ${{\bf{\tilde c}}^{\left( {t,\tau } \right)} }$ denotes the codeword whose entries are $\tilde c_t  = \tilde c_\tau   = 1$, and $\tilde c_p  = 0$ if $p \ne t$ and $p \ne \tau$ for $p=1, 2, \ldots, N_S+1$. This simplified expression of ${\rm{ABEP}}_{S_t }^{\left( {{\rm{SV}}} \right)}$ originates from the fact that ${{\bf{\tilde c}}^{\left( {t,\tau } \right)} }$ are the only codewords of the codebook with $w_{\mathcal{H}} \left( {{\bf{\tilde c}}^{\left( {t,\tau } \right)} } \right) = 2$.

By direct inspection, from (\ref{Eq_14})--(\ref{Eq_16}) we can readily obtain:
\setcounter{equation}{40}
\begin{equation}
\label{Eq_AppII__3}
\left\{ \begin{array}{l}
 \mathcal{\bar M}_1 \left( {\left. s \right|{\bf{\tilde c}}^{\left( {t,\tau } \right)} } \right) = \left( {\bar \gamma _{S_t D} \bar \gamma _{S_\tau  D} } \right)^{ - 1} \left[ {4s\left( {1 - s} \right)} \right]^{ - 2}  \\
 \mathcal{\bar M}_2 \left( {\left. s \right|{\bf{\tilde c}}^{\left( {t,\tau } \right)} ,\mathcal{E}_0 } \right) = 1 \\
 \mathcal{\bar M}_3 \left( {\left. s \right|{\bf{\tilde c}}^{\left( {t,\tau } \right)} ,\mathcal{E}_1 } \right) = \left( {4\bar \gamma _{eqR_1 } } \right)^{ - 1}  \\
 \end{array} \right.
\end{equation}
\noindent if $1 \le \tau  \le N_S$, and:
\setcounter{equation}{41}
\begin{equation}
\label{Eq_AppII__4}
\left\{ \begin{array}{l}
 \mathcal{\bar M}_1 \left( {\left. s \right|{\bf{\tilde c}}^{\left( {t,N_S  + 1} \right)} } \right) = \left[ {4\bar \gamma _{S_t D} s\left( {1 - s} \right)} \right]^{ - 1}  \\
 \begin{split} \mathcal{\bar M}_2 \left( {\left. s \right|{\bf{\tilde c}}^{\left( {t,N_S  + 1} \right)} ,\mathcal{E}_0 } \right) & = \left( {4\bar \gamma _{eqR_1 } s} \right)^{ - 1} \\ & \hspace{-2.5cm}  + \left( {4\bar \gamma _{R_1 D} s\left( {1 - s} \right)} \right)^{ - 1}  \\ & \hspace{-2.5cm} - \left( {4\bar \gamma _{eqR_1 } } \right)^{ - 1} \left[ {\left( {{4 \mathord{\left/
 {\vphantom {4 \pi }} \right.
 \kern-\nulldelimiterspace} \pi }} \right)\int\nolimits_0^{{\pi  \mathord{\left/
 {\vphantom {\pi  2}} \right.
 \kern-\nulldelimiterspace} 2}} {\left( {\sin ^{ - 2} \left( \theta  \right) + 4s} \right)^{ - 1} d\theta } } \right] \end{split} \\
 \begin{split} \mathcal{\bar M}_3 \left( {\left. s \right|{\bf{\tilde c}}^{\left( {t,N_S  + 1} \right)} ,\mathcal{E}_1 } \right) & = \left( {4\bar \gamma _{eqR_1 } } \right)^{ - 1} \\ & \hspace{-2.5cm} \times \left[ {\left( {{4 \mathord{\left/
 {\vphantom {4 \pi }} \right.
 \kern-\nulldelimiterspace} \pi }} \right)\int\nolimits_0^{{\pi  \mathord{\left/
 {\vphantom {\pi  2}} \right.
 \kern-\nulldelimiterspace} 2}} {\left( {\sin ^{ - 2} \left( \theta  \right) - 4s} \right)^{ - 1} d\theta } } \right] \end{split} \\
 \end{array} \right.
\end{equation}
\noindent if $\tau = N_S+1$.

By inserting (\ref{Eq_AppII__3}) and (\ref{Eq_AppII__4}) in (\ref{Eq_AppII__1}), the APEPs can be computed in closed--form. Due to space limitations, the details of the derivation are omitted. However, the adopted methodology is as follows: i) contour integrals are solved by using the residues theorem in \cite[Eq. (6)]{Biglieri}; ii) the integrals in (\ref{Eq_AppII__3}) and (\ref{Eq_AppII__4}) are computed only after solving the contour integrals in (\ref{Eq_AppII__1}), \emph{i.e.}, the order of integration is swapped; and iii) the identities in (\ref{Eq_AppII__5}) shown at the top of this page are used, where the equalities in $\mathop  = \limits^{\left( a \right)}$ and $\mathop  = \limits^{\left( b \right)}$ follow from the residues theorem in \cite[Eq. (6)]{Biglieri}.

From (\ref{Eq_AppII__5}) and some algebra, (\ref{Eq_AppII__2}) simplifies to (\ref{Eq_AppII__6}) shown at the top of this page, which coincides with \cite[Eq. (32)]{Nasri}. This concludes the proof.
\begin{figure*}[!t]
\setcounter{equation}{46}
\begin{equation} \footnotesize
\label{Eq_AppIII__3}
{\rm{APEP}}\left( {{\bf{0}} \to {\bf{\tilde c}}^{\left( {q^* } \right)} } \right) \approx \left( {2\pi j} \right)^{ - 1} \int\nolimits_{\delta  - j\infty }^{\delta  + j\infty } {\mathcal{\bar M}_2 \left( {\left. s \right|{\bf{\tilde c}}^{\left( {q^* } \right)} ,\mathcal{E}_0 } \right)s^{ - 1} ds}  + \left( {2\pi j} \right)^{ - 1} \int\nolimits_{\delta  - j\infty }^{\delta  + j\infty } {\mathcal{\bar M}_2 \left( {\left. s \right|{\bf{\tilde c}}^{\left( {q^* } \right)} ,\mathcal{E}_{m^* } } \right)\mathcal{\bar M}_3 \left( {\left. s \right|{\bf{\tilde c}}^{\left( {q^* } \right)} ,\mathcal{E}_{m^* } } \right)s^{ - 1} ds}
\end{equation}
\normalsize \hrulefill \vspace*{0pt}
\end{figure*}
\begin{figure*}[!t]
\setcounter{equation}{48}
\begin{equation}
\label{Eq_AppIII__5}
\begin{split}
 {\rm{APEP}}\left( {{\bf{0}} \to {\bf{\tilde c}}^{\left( {q^* } \right)} } \right) & \mathop  = \limits^{\left( a \right)} \left( {2\pi j} \right)^{ - 1} \int\nolimits_{\delta  - j\infty }^{\delta  + j\infty } {\left[ {4\bar \gamma _{R_{q^* } D} s^2 \left( {1 - s} \right)} \right]^{ - 1} ds} \\ &  + \left( {2\pi j} \right)^{ - 1} \int\nolimits_{\delta  - j\infty }^{\delta  + j\infty } {\left( {4\bar \gamma _{eqR_{q^* } } } \right)^{ - 1} \left[ {\left( {{4 \mathord{\left/
 {\vphantom {1 \pi }} \right.
 \kern-\nulldelimiterspace} \pi }} \right)\int\nolimits_0^{{\pi  \mathord{\left/
 {\vphantom {\pi  2}} \right.
 \kern-\nulldelimiterspace} 2}} {\left( {\sin ^{ - 2} \left( \theta  \right) - 4s} \right)^{ - 1} d\theta } } \right]s^{ - 1} ds}  \\
 & \mathop  = \limits^{\left( b \right)} \left( {4\bar \gamma _{R_{q^* } D} } \right)^{ - 1}  + \left( {4\bar \gamma _{eqR_{q^* } } } \right)^{ - 1} \\
 \end{split}
\end{equation}
\normalsize \hrulefill \vspace*{0pt}
\end{figure*}
\section{High--SNR ABEP of Partial--Cooperative Relays} \label{App_ABEP_RelaysPC}
By using $\mathcal{G}\left( {\left. \cdot \right| \cdot} \right)$  in (\ref{Eq_AppI__2}), the ${\rm{APEP}}\left( {{\bf{0}} \to {\bf{\tilde c}}} \right)$ in (\ref{Eq_13}) can be re--written as shown in (\ref{Eq_AppIII__1}) at the top of this page, where: i) $\mathcal{\bar M}_1 \left( {\left. s \right|{\bf{\tilde c}}} \right) = \prod\nolimits_{t = 1}^{N_S } {\mathcal{F}_t \left( {\left. s \right|{\bf{\tilde c}}} \right)}$; ii) $\mathcal{\bar M}_2 \left( {\left. s \right|{\bf{\tilde c}},\mathcal{E}_m } \right) = \prod\nolimits_{q = 1,\;\mu _{q - 1}  = 0}^{N_R } {\mathcal{G}_q \left( {\left. s \right|{\bf{\tilde c}},\mathcal{E}_m ,\mu _{q - 1}  = 0} \right)}$; and iii) $\mathcal{\bar M}_3 \left( {\left. s \right|{\bf{\tilde c}},\mathcal{E}_m } \right) = \prod\nolimits_{q = 1,\;\mu _{q - 1}  = 1}^{N_R } {\mathcal{G}_q \left( {\left. s \right|{\bf{\tilde c}},\mathcal{E}_m ,\mu _{q - 1}  = 1} \right)}$.

From (\ref{Eq_AppIII__1}), ${{\rm{APEP}}\left( {{\bf{0}} \to {\bf{\tilde c}}^{\left( t \right)} } \right)}$ in (\ref{Eq_21}) can be computed as:
\setcounter{equation}{45}
\begin{equation}
\label{Eq_AppIII__2}
\begin{split} & {\rm{APEP}}\left( {{\bf{0}} \to {\bf{\tilde c}}^{\left( t \right)} } \right) \\ & \hspace{1cm} \mathop  \approx \limits^{\left( a \right)} \left( {2\pi j} \right)^{ - 1} \int\nolimits_{\delta  - j\infty }^{\delta  + j\infty } {\mathcal{\bar M}_1 \left( {\left. s \right|{\bf{\tilde c}}^{\left( t \right)} } \right)s^{ - 1} ds} \\ & \hspace{1cm} \mathop  \approx \limits^{\left( b \right)} \left( {2\pi j} \right)^{ - 1} \int\nolimits_{\delta  - j\infty }^{\delta  + j\infty } {\left[ {4\bar \gamma _{S_t D} s^2 \left( {1 - s} \right)} \right]^{ - 1} ds} \\ & \hspace{1cm} \mathop  = \limits^{\left( c \right)} \left( {4\bar \gamma _{S_t D} } \right)^{ - 1} \end{split}
\end{equation}
\noindent where: i) $\mathop  \approx \limits^{\left( a \right)}$ follows by noticing that for high--SNR the dominant addend in (\ref{Eq_AppIII__1}) is $m=0$, and that $\mathcal{\bar M}_2 \left( {\left. s \right|{\bf{\tilde c}},\mathcal{E}_m } \right) = 1$ for every $m$ since $w_{\mathcal{H}}^{\left( {R,\mathcal{E}_m^{\left( {{\rm{ok}}} \right)} } \right)} \left( {{\bf{\tilde c}}^{\left( t \right)} } \right) = 0$ for every $m$; ii) $\mathop  \approx \limits^{\left( b \right)}$ follows from $\mathcal{\bar M}_1 \left( {\left. s \right|{\bf{\tilde c}}^{\left( t \right)} } \right) = \prod\nolimits_{t = 1}^{N_S } {\mathcal{F}_t \left( {\left. s \right|{\bf{\tilde c}}^{\left( t \right)} } \right)}  \approx \left[ {4\bar \gamma _{S_t D} s\left( {1 - s} \right)} \right]^{ - 1}$ with the last approximation coming from (\ref{Eq_14}); and iii) $\mathop  \approx \limits^{\left( c \right)}$ is obtained by applying the residues theorem \cite[Eq. (6)]{Biglieri}.

Let $q^*$ be the partial--cooperative relay of interest. From (\ref{Eq_AppIII__1}), ${{\rm{APEP}}\left( {{\bf{0}} \to {\bf{\tilde c}}^{\left( q^* \right)} } \right)}$ in (\ref{Eq_21}) can be computed as shown in (\ref{Eq_AppIII__3}) at the top of the next page because for high--SNR there are only two dominant addends in (\ref{Eq_AppIII__1}): i) $m=0$; and ii) $m=m^*$ with ${\mathcal{E}_{m^* } }$ being the event that $R_{q^* }$ is the only relay that forwards a wrong estimate of its network--coded bit to $D$.

From (\ref{Eq_15}), we have $\mathcal{\bar M}_2 \left( {\left. s \right|{\bf{\tilde c}}^{\left( {q^* } \right)} ,\mathcal{E}_{m^* } } \right) = 1$ and:
\setcounter{equation}{47}
\begin{equation}
\label{Eq_AppIII__4}
\left\{ \begin{array}{l}
 \begin{split} \mathcal{\bar M}_2 \left( {\left. s \right|{\bf{\tilde c}}^{\left( {q^* } \right)} ,\mathcal{E}_0 } \right) & = \left( {4\bar \gamma _{eqR_{q^*} } s} \right)^{ - 1}  + \left( {4\bar \gamma _{R_{q^*} D} s\left( {1 - s} \right)} \right)^{ - 1} \\ & \hspace{-2.25cm}  - \left( {4\bar \gamma _{eqR_{q^*} } } \right)^{ - 1} \left[ {\left( {{4 \mathord{\left/
 {\vphantom { \pi }} \right.
 \kern-\nulldelimiterspace} \pi }} \right)\int\nolimits_0^{{\pi  \mathord{\left/
 {\vphantom {\pi  2}} \right.
 \kern-\nulldelimiterspace} 2}} {\left( {\sin ^{ - 2} \left( \theta  \right) + 4s} \right)^{ - 1} d\theta } } \right] \end{split} \\
 \begin{split} \mathcal{\bar M}_3 \left( {\left. s \right|{\bf{\tilde c}}^{\left( {q^* } \right)} ,\mathcal{E}_{m^* } } \right) & = \left( {4\bar \gamma _{eqR_{q^*} } } \right)^{ - 1} \\ & \hspace{-2.25cm}  \times \left[ {\left( {{4 \mathord{\left/
 {\vphantom {1 \pi }} \right.
 \kern-\nulldelimiterspace} \pi }} \right)\int\nolimits_0^{{\pi  \mathord{\left/
 {\vphantom {\pi  2}} \right.
 \kern-\nulldelimiterspace} 2}} {\left( {\sin ^{ - 2} \left( \theta  \right) - 4s} \right)^{ - 1} d\theta } } \right] \end{split} \\
 \end{array} \right.
\end{equation}

Then, by inserting (\ref{Eq_AppIII__4}) in (\ref{Eq_AppIII__3}) we obtain ${\rm{APEP}}\left( {{\bf{0}} \to {\bf{\tilde c}}^{\left( {q^* } \right)} } \right)$ shown in (\ref{Eq_AppIII__5}) at the top of this page, where: i) $\mathop  = \limits^{\left( a \right)}$ takes into account that when applying the residues theorem only the addends containing positive poles have a non--zero contribution, while the others can be neglected \cite[Eq. (6)]{Biglieri}; and ii) $\mathop  = \limits^{\left( b \right)}$ can be obtained by solving the first integral as in (\ref{Eq_AppIII__2}), and the second integral by solving first the contour integral. In particular, we have used the identities: 1) $\left( {2\pi j} \right)^{ - 1} \int\nolimits_{\delta  - j\infty }^{\delta  + j\infty } {\left( {\sin ^{ - 2} \left( \theta  \right) - 4s} \right)^{ - 1} s^{ - 1} ds}  = \sin ^2 \left( \theta  \right)$, which originates from the application of the residues theorem \cite[Eq. (6)]{Biglieri}; and 2) $\left( {{4 \mathord{\left/ {\vphantom {4 \pi }} \right. \kern-\nulldelimiterspace} \pi }} \right)\int\nolimits_0^{{\pi  \mathord{\left/ {\vphantom {\pi  2}} \right. \kern-\nulldelimiterspace} 2}} {\sin ^2 \left( \theta  \right)d\theta }  = 1$. From (\ref{Eq_AppIII__2}) and (\ref{Eq_AppIII__5}), the proof is complete.
\begin{biography} {Marco Di Renzo}
(S'05--AM'07--M'09) was born in L'Aquila, Italy, in 1978. He received the Laurea (cum laude) and the Ph.D. degrees in Electrical and
Information Engineering from the Department of Electrical and Information Engineering, University of L'Aquila, Italy, in April 2003 and in January
2007, respectively.

From August 2002 to January 2008, he was with the Center of Excellence for Research DEWS, University of L'Aquila, Italy.
From February 2008 to April 2009, he was a Research Associate with the Telecommunications Technological Center of Catalonia (CTTC), Barcelona,
Spain. From May 2009 to December 2009, he was an EPSRC Research Fellow with the Institute for Digital Communications (IDCOM), The University of
Edinburgh, Edinburgh, United Kingdom (UK).

Since January 2010, he has been a Tenured Researcher (``Charg\'e de Recherche Titulaire'') with the French National Center for Scientific
Research (CNRS), as well as a faculty member of the Laboratory of Signals and Systems (L2S), a joint research laboratory of the CNRS,
the \'Ecole Sup\'erieure d'\'Electricit\'e (SUP\'ELEC), and the University of Paris--Sud XI, Paris, France. His main research interests are in
the area of wireless communications theory. He is a Principal Investigator of three European--funded research projects (Marie Curie ITN--GREENET, Marie Curie IAPP--WSN4QoL, and Marie Curie ITN--CROSSFIRE).

Dr. Di Renzo is the recipient of the special mention for the outstanding five--year (1997--2003) academic career, University of L'Aquila, Italy;
the THALES Communications fellowship for doctoral studies (2003--2006), University of L'Aquila, Italy; the Best Spin--Off Company Award (2004), Abruzzo Region, Italy; the Torres Quevedo award for research on ultra wide band systems and cooperative localization for wireless networks (2008--2009), Ministry of Science and Innovation, Spain; the ``D\'erogation pour l'Encadrement de Th\`ese'' (2010), University of Paris--Sud XI, France; the 2012 IEEE CAMAD Best Paper Award from the IEEE Communications Society; and the 2012 Exemplary Reviewer Award from the IEEE WIRELESS COMMUNICATIONS LETTERS of the IEEE Communications Society. He currently serves as an Editor of the IEEE COMMUNICATIONS LETTERS.
\end{biography}
\begin{biography} {Michela Iezzi}
(S'10) was born in Pescara, Italy, in 1983. She received the Bachelor and the Master degrees in Telecommunications Engineering from the Department of Electrical and Information Engineering, University of L'Aquila, Italy, in April 2006 and in September 2009, respectively. Since November 2009, she has been a Ph.D. candidate at the same University.
In 2010 and 2011, she visited the Laboratoire des Signaux et Syst\`emes (L2S), Centre National de la Recherche Scientifique--\'Ecole
Sup\'erieure d'\'Electricit\'e--Universit\'e Paris--Sud XI, where she conducted research on network--coded cooperative wireless networks.
Her main research interests are in the area of wireless communications.
\end{biography}
\begin{biography} {Fabio Graziosi}
(S'96–-M'97) was born in L'Aquila, Italy, in 1968. He received the Laurea degree (cum laude) and the Ph.D. degree in electronic engineering
from the University of L'Aquila, Italy, in 1993 and in 1997, respectively.
Since February 1997, he has been with the Department of Engineering, Computer Science, and Mathematics, University of L'Aquila, where he is currently an Associate Professor. He is a member of the Executive Committee, Center of Excellence Design methodologies for Embedded controllers, Wireless interconnect
and System--on--chip (DEWS), University of L'Aquila, and the Executive Committee, Consorzio Nazionale Interuniversitario per le Telecomunicazioni
(CNIT). He is also the Chairman of the Board of Directors of WEST Aquila s.r.l., a spin--off R\&D company of the University of L'Aquila, and the
Center of Excellence DEWS.
His current research interests are mainly focused on wireless communications systems with emphasis on wireless sensor networks, cognitive radio, and cooperative communications.
\end{biography}
\end{document}